\documentclass[12pt]{article}
\pdfoutput=1

\usepackage{putex}
\usepackage{amsmath,amssymb,amsfonts}
\usepackage{psfrag}
\usepackage{footmisc}
\usepackage{bbold}
\usepackage{url}
\usepackage{mathtools}
\usepackage{color}
\usepackage{braket}
\usepackage{float}
\usepackage[dvipsnames]{xcolor}
\usepackage{comment}

\usepackage{tikz}
    \usepackage{amssymb,amsfonts,amsmath}
    \usepackage{tkz-euclide}
        \usetikzlibrary{arrows,calc,patterns}
\usepackage{pgfplots}

\definecolor{darkblue}{rgb}{0.1,0.1,.7}
\definecolor{purple}{rgb}{0.6,0,0.6}
\definecolor{orange}{rgb}{0.9,0.6,0}
\definecolor{llgray}{rgb}{0.9,0.9,1}
\definecolor{dgreen}{rgb}{0,0.5,0}
\usepackage[colorlinks, linkcolor=darkblue, citecolor=darkblue, urlcolor=darkblue, linktocpage]{hyperref} 
\usepackage[square, comma, compress,numbers]{natbib}
\usepackage[]{amsmath}
\usepackage[]{graphicx}
\usepackage[]{latexsym}
\usepackage[utf8]{inputenc}
\usepackage{geometry}
\usepackage{amscd}
\usepackage[all,cmtip]{xy}
\usepackage{mathrsfs}

\usepackage[margin=10pt,font=small,labelfont=bf]{caption}
\usepackage{dsdshorthand}
\usepackage{changepage}
\usepackage{setspace}
\usepackage{tikz}
\usepackage{subcaption}
\usepackage{ marvosym }
\usepackage{tensor}
\usepackage{CJKutf8}
\usepackage{empheq}

\usepackage{amsmath, amssymb, amscd, amsthm, amsfonts}
\usepackage{physics}
\usepackage{color}
\usepackage{bm}
\usepackage{graphicx}
\usepackage{multicol}
\usepackage{ marvosym }
\usepackage{tensor}

\usepackage[titles]{tocloft}
\usetikzlibrary{arrows,positioning}

\def\SL2{\widetilde{SL}(2,\mathbb R)}

\newcommand\mR{\mathbb{R}}

\newcommand\mC{\mathbb{C}}

\newcommand{\id}{\mathbb{1}}

\numberwithin{equation}{section}

\newcommand{\grp}[1]{\mathrm{#1}}

\newcommand {\bes} {\begin {equation*}}
\newcommand {\ees} {\end {equation*}}
\newcommand {\beq} {\begin {equation}}
\newcommand {\eeq} {\end {equation}}
\newcommand {\bea} {\begin {eqnarray}}
\newcommand {\ea} {\end {eqnarray}}
\newcommand {\eea} {\end {eqnarray}}

\numberwithin{equation}{section}

\newcommand{\hilb}{\mathcal{H}}
\newcommand{\hilbM}{\hilb^{\mathrm{matter}}}
 \newcommand{\HD}{{\put(4,4){\oval(8,8)[b]}\put(0,4){\line(1,0){8}}\phantom{\subset\,}}}

\def\crosscap at (#1,#2){  \draw [thick] (#1,#2) circle (0.25);    \draw [thick] (#1-0.70710678/4,#2+0.70710678/4) -- (#1+0.70710678/4,#2-0.70710678/4);    \draw [thick] (#1+0.70710678/4,#2+0.70710678/4) -- (#1-0.70710678/4,#2-0.70710678/4)}

\newcommand\Star[3][]{%
\path[#1] (0  :#3) -- ( 36:#2) 
       -- (72 :#3) -- (108:#2)
       -- (144:#3) -- (180:#2)
       -- (216:#3) -- (252:#2)
       -- (288:#3) -- (324:#2)--cycle;}

\def\<{\langle}
\def\>{\rangle}

\tikzset{
    >=stealth',
    punkt/.style={
           rectangle,
           rounded corners,
           draw=black, very thick,
           text width=15em,
           minimum height=2em,
           text centered},
    pil/.style={
           ->,
           thick,
           shorten <=2pt,
           shorten >=2pt,}
}

 \def\ie{\begin{equation}\begin{aligned}}
\def\fe{\end{aligned}\end{equation}}

\tikzset{zigzag/.style={decorate,decoration=zigzag}}

\pgfplotsset{compat=1.18}

\begin{document}

\thispagestyle{plain}
\begin{center}
\vspace*{1cm}
\vspace*{1cm}
\LARGE
 \textbf{On the non-perturbative bulk Hilbert space of JT gravity}
\vspace*{1cm}
\normalsize
        
\vspace{0.4cm}

\textbf{ Luca V.~Iliesiu${}^{1,2}$, Adam Levine${}^3$, Henry W. Lin${}^2$, Henry Maxfield${}^2$,  M\'ark Mezei${}^{4}$}
\,\vspace{0.2cm}\\
{${}^1$  Department of Physics, University of California, Berkeley, CA 94720, USA}
\,\vspace{0.2cm}\\
{${}^2$ Stanford Institute for Theoretical Physics, Stanford University, Stanford, CA 94305, USA}
\,\vspace{0.2cm}\\
{${}^3$ Center for Theoretical Physics, MIT, Cambridge, MA 02139, USA
}
\,\vspace{0.2cm}\\
{${}^4$ Mathematical Institute, University of Oxford, Woodstock Road, Oxford, OX2 6GG, UK}    
\vspace{0.4cm}\\
 \textbf{Abstract}\\\vspace{-0.2cm}
 \end{center}

What is the bulk Hilbert space of quantum gravity?
In this paper, we resolve this problem in $2d$ JT gravity, both with and without matter, providing %
an explicit definition of a non-perturbative Hilbert space specified in terms of metric variables.
The states are wavefunctions of the length and matter state, but with a non-trivial and highly degenerate inner product. We explicitly identify the null states, and discuss their importance for defining operators non-perturbatively.
To highlight the power of the formalism we developed, we study the non-perturbative effects for two bulk linear operators that may serve as proxies for the experience of an observer falling into a two-sided black hole: one captures the length of an Einstein-Rosen bridge and the other captures the center-of-mass collision energy between two particles falling from opposite sides. We track the behavior of these operators up to times of order $e^{S_\text{BH}}$, at which point the wavefunction spreads to the complete set of eigenstates of these operators. If these observables are indeed good proxies for the experience of an infalling observer, our results indicate an $O(1)$  probability of detecting a firewall at late times that is self-averaging and universal.  %

\vspace{0.4cm}

\newpage
\tableofcontents

\newpage

\section{Introduction \& summary}

In recent years, it has become increasingly apparent that non-perturbative gravitational effects have remarkable explanatory power. Highlights include quantifying the information content of Hawking radiation \cite{Penington:2019npb, Almheiri:2019psf,  Penington:2019kki, Almheiri:2019qdq}, aspects of spectral statistics of black hole microstates \cite{Cotler:2016fpe, Saad:2018bqo, Saad:2019lba, Stanford:2019vob, Cotler:2020ugk}, and counting states of supersymmetric black holes using the supergravity path integral \cite{Dabholkar:2011ec, Dabholkar:2014ema, Iliesiu:2022kny}. These calculations can all be described in the framework of general relativity as an effective quantum theory, without using microscopic ideas such as string theory or holographic duality. Einstein's geometric theory of dynamical spacetime and quantum mechanics thus have far more unity and scope than previously envisaged.

However, the progress described above involves quantities that can be defined asymptotically, in a region where gravity is absent or weak. Concretely, the calculations involve boundary conditions that fix the asymptotic spacetime geometry and fields (for example, at the boundary of an asymptotically AdS spacetime), even if there are large quantum fluctuations in the interior. This becomes an important limitation for addressing more inherently `bulk' questions in contexts like the black hole interior (specifically, the experience of an infalling observer at late times) or cosmological spacetimes. We would therefore like a better understanding of non-perturbative gravitational effects in bulk language, with states described in terms of wavefunctions of spatial metrics and matter fields. In particular, such a language seems necessary to properly define bulk linear operators and describe their physics.

In this paper, we make progress towards this goal in the simple two-dimensional theory of JT gravity (coupled to matter). Specifically, we take the existing definition of the perturbative bulk Hilbert space defined on geodesic slices and describe how it is modified by non-perturbative effects. We then use this construction to study bulk operators non-perturbatively.

A prominent idea about the Hilbert space of gravity is that of redundancy, or null states. Historically, effective quantum gravity has been regarded as inconsistent due to the black hole information paradox; this is underpinned by the fact that a black hole can apparently have far too many interior states (as defined by quantizing fluctuations around a classical background spacetime, taking perturbative gravitational constraints into account). A possible resolution is that the geometric description of the Hilbert space is perfectly consistent but that apparently different states are not, in fact, linearly independent. States that are orthogonal in perturbation theory can obtain small overlaps from non-perturbative effects; if these overlaps are finely tuned, the resulting matrix of inner products can be degenerate, with many `null states' of zero norm. We will see this idea implemented very concretely in our study of JT gravity, obtaining explicit expressions for the degenerate inner product and null state wavefunctions. This phenomenon has important consequences for operators on the Hilbert space and for the associated observables in any theory of quantum gravity.

In the remainder of the introduction, we summarise our main results before outlining the structure of the paper.
 
\subsection{The bulk Hilbert space on geodesics}\label{ssec:introHilb}

A simplifying feature of JT gravity is that there is a tunable parameter $S_0$ which suppresses non-perturbative effects without altering perturbative gravity. In the limit $S_0\to \infty$, we recover a self-consistent `perturbative' theory, which includes all-loop quantum fluctuations but not topology-changing processes. We begin by describing the Hilbert space of this theory (which we call $\hilb_0$), before addressing the non-perturbative effects. Here, we restrict to pure JT gravity, leaving the generalisation to include matter to the main text.

We are interested in the `two-sided' Hilbert space, with two asymptotic spatial boundaries. We can describe states in $\hilb_0$ as wavefunctions $\psi(\ell)$ of the renormalized geodesic length $\ell$ between these boundaries, with the usual $L^2$ inner product (denoted $\langle \cdot|\cdot\rangle_0$ to distinguish it from the full inner product introduced later):
\begin{equation}\label{eq:H0intro}
    \hilb_0\simeq L^2(\mR), \qquad |\psi\rangle = \int d\ell\,  \psi(\ell) |\ell\rangle, \qquad \langle \psi|\psi\rangle_0 = \int  | \psi(\ell)|^2 d\ell \qquad (S_0\to\infty).
\end{equation}
We get this simple local inner product because every allowed geometry in this $S_0\to\infty$ theory (which must have the topology of a disc) has a unique geodesic between given boundary points. Time evolution is generated by a simple Hamiltonian $H$ acting on the wavefunction $\psi(\ell)$, of a non-relativistic particle in an exponential potential \cite{Bagrets:2017pwq,Harlow:2018tqv}:
\begin{equation}\label{eq:Hintro}
    H= H_L=H_R = -\tfrac{1}{2}\partial_\ell^2 + 2e^{-\ell}.
\end{equation}
This evolves either on the left or right boundary, since the left and right Hamiltonians $H_{L,R}$ are equal for pure JT gravity (this no longer holds with matter). This operator has a continuous spectrum (the positive reals $E>0$), so there is an infinite-dimensional space of states in any window of energies; this is a version of the information problem.

What happens when $S_0$ is finite, and we include non-perturbative effects? A complication is that we no longer have a unique geodesic on every geometry, so we might expect that a Hilbert space in terms of geodesic lengths no longer makes sense.\footnote{To avoid confusion, note that this is different from the baby universe Hilbert space \cite{Marolf:2020xie}. } Nonetheless, we can continue to define bulk `basis' states $|\ell\rangle$ by geodesic boundary conditions in the path integral. The inner product $\langle \ell'|\ell\rangle$ now becomes non-trivial, computed by the path integral over all geometries bounded by a pair of geodesics of lengths $\ell,\ell'$; this idea was articulated before in \cite{Gao:2021uro}. In fact, this is the standard way to define a Hilbert space when preparing states at the path integral level. 

Our first main result is an expression for this inner product in terms of the microscopic data of the theory (which in AdS/CFT language we can think of as the dual quantum mechanics). In pure JT gravity, this data is simply the discrete spectrum $E_i$ ($i=0,1,2,\ldots$) of the Hamiltonian in the complete non-perturbative theory. For a superposition $|\psi\rangle = \int d\ell \,\psi(\ell) |\ell\rangle$ of geodesic states as above, we find
\begin{equation}\label{eq:IPintro}
    \langle \psi|\psi\rangle = e^{-S_0} \sum_i \left|\int\phi_{E_i}(\ell)\psi(\ell)\,d\ell\right|^2,
\end{equation}
where $\phi_E(\ell)$ are eigenfunctions of the Schr\"odinger Hamiltonian \eqref{eq:Hintro} (given explicitly in \eqref{eq:phiEl}). Thus, for example, the inner product between two states with fixed length is given by 
\be
 \langle \ell|\ell'\rangle = e^{-S_0}\sum_i \phi_{E_i}(\ell) \phi_{E_i}(\ell')\,.
\ee

Normally, JT gravity is regarded as being dual to an ensemble of quantum systems with a random Hamiltonian, so we should explain what we mean by the spectrum $E_i$. There are two possible (but equivalent) perspectives we can take. First, we could take our definition of the non-perturbative theory to include not only sums over topologies, but also additional effects which `fix the member of the ensemble' (for example, additional asymptotic boundaries to fix a single `$\alpha$-state' of baby universes, a non-perturbatively small non-local interaction or eigenbranes). This leaves us with some specific non-perturbative spectrum (and a factorizing theory), the precise nature of which will depend on the details. Alternatively, we can think of the matrix of inner products $\langle \ell'|\ell\rangle$ as a random variable, with $E_i$ the eigenvalues of a random matrix. In that case, we will demonstrate the equality between random variables \eqref{eq:IPintro} holds for all moments of the inner product to all orders in a genus expansion, which is explicitly computable. Specifically, the average of this random variable takes the form, 
\begin{align}
\overline{\braket{\ell'}{\ell}} \quad  = \quad \delta(\ell-\ell') +  \quad 
\begin{tikzpicture}[rotate=90,scale=0.5,baseline={([yshift=-.6cm]current bounding box.center)}]
\node at (1,1.5) {$\ell$};
\node at (3,2) {$\ell'$};
\draw [thick, blue, fill = llgray,rotate=90,yshift=-2.2cm,scale=2] plot [smooth] coordinates { (-2,0)(-1,0.1)(0,0.7)(1,0.1)(2,0)} -- plot [smooth] coordinates {(2,0)(1,-0.1)(0,-0.7)(-1,-0.1)(-2,0)} ;
\clip (2.25,-1.51) rectangle (5.6,1.51);
\draw[smooth, thick, blue, fill = llgray] (0,0.5) to[out=0,in=0] (2,0.5) to[out=0,in=210] (3,1) to[out=30,in=150] (5,1) to[out=-30,in=30] (5,-1) to[out=210,in=-30] (3,-1) to[out=150,in=0] (2,-0.5) to[out=0,in=0] (0,-0.5) to[out=150,in=-150] (0,0.5);
\fill[smooth, white] (3.52,0) .. controls (3.8,-0.2) and (4.2,-0.2) .. (4.48,0);
\draw[smooth, blue, fill = white] (3.5,0) .. controls (3.8,0.2) and (4.2,0.2) .. (4.5,0);
\draw[smooth, blue] (3.4,0.1) .. controls (3.8,-0.25) and (4.2,-0.25) .. (4.6,0.1);
\end{tikzpicture}
\quad + \quad
O(e^{-4S_0})\,.
\end{align}
In either case, we use only some simple properties of the bulk theory and not the full details of its non-perturbative definition. This is particularly useful once we add matter since, in that case, the precise definition of the non-perturbative corrections is more ambiguous due to divergences on spacetimes with nontrivial topology. 

The modified norm \eqref{eq:IPintro} is highly degenerate. For any wavefunction $\psi(\ell)$, we can construct the `transform' $\hat{\psi}(E) =\int d\ell\, \phi_E(\ell) \psi(\ell)$, and the norm depends only on the values $\hat{\psi}(E_i)$ at the discrete energies giving the non-perturbative spectrum. In particular, if $\hat{\psi}(E_i)=0$ for all $i$, then we have a null state (and there will be many such states for which the function $\psi(\ell)$ is nonzero). Nonetheless, \eqref{eq:IPintro} is positive semi-definite. Note that we reserve the term `wavefunction' for the function $\psi(\ell)$ in the decomposition of a state in terms of geodesic states; because the inner product is modified and states $\ket{\ell}$ with different $\ell$ are no longer orthogonal, this is \emph{not} equal to the overlap $\langle \ell|\psi\rangle$ (though this does have an alternative interpretation as a  non-perturbative `Wheeler-DeWitt wavefunction', discussed in section \ref{ssec:NPWDW}). 

Despite the modified norm, the Hamiltonian \eqref{eq:Hintro} we used for the perturbative theory continues to apply in the exact theory. More precisely, acting with the differential operator \eqref{eq:H0intro} on a wavefunction $\psi(\ell)$ for the state $|\psi\rangle$ produces a wavefunction for $H|\psi\rangle$. Nonetheless, due to the redundancy of the geodesic length description, there are many other ways to represent the same operator.

\subsection{Defining bulk operators}\label{ssec:operatorsIntro}

Now that we have a description of the bulk Hilbert space $\hilb$, we can begin to discuss operators. Since the underlying vector space of wavefunctions $\psi(\ell)$ is unchanged by the non-perturbative effects, one might think that an operator on the perturbative Hilbert space $\hilb_0$ straightforwardly gives us an operator on $\hilb$.  However, there is an important subtlety: a generic operator on $\hilb_0$ will not map null states to null states, so does not give a well-defined operator on $\hilb$. (A second problem is that a Hermitian operator on $\hilb_0$ will not typically be Hermitian with respect to the modified inner product.) Given this, the non-perturbative definition of a perturbatively-defined bulk operator is inherently ambiguous.

Take the simple example of a `length operator' $\hat{\ell}$ for which geodesic states are eigenstates: $\hat{\ell}|\ell\rangle=\ell|\ell\rangle$. This is the usual position operator on the perturbative Hilbert space \eqref{eq:H0intro}, but it is ill-defined with the non-perturbative inner product \eqref{eq:IPintro} for the reasons explained above. However, for any invertible function $\mathcal{F}(\ell)$ of the length, we can make an alternative definition that does not suffer from the same problems:
\begin{align}
\widehat{\ell_\mathcal{F}} := \mathcal F^{-1} \left(\int d\ell \,\mathcal F(\ell) |\ell\rangle\langle \ell| \right).
    \label{eq:define-length-operator-intro}
\end{align}
With a trivial inner product, geodesic states $|\ell\rangle$ would be eigenstates of this operator with eigenvalue $\ell$, but this is not true once $\langle\ell'|\ell\rangle$ is non-diagonal. We get a different operator for different functions $\mathcal{F}$, so there is an ambiguity in defining operators non-perturbatively. This ambiguity can also be seen at a path integral level when summing over geometries with higher topologies. As we have already emphasized, higher-topology AdS$_2$ geometries have multiple geodesics between any two boundary points. Thus, when defining the geodesic length operator there is an ambiguity in which geodesic we choose (or the relative weight we lend to each geodesic). The relation between the Hilbert space ambiguity and the geometric ambiguity can be best seen by looking at the operator $\mathcal F(\widehat{\ell_\mathcal{F}})$, which has the same eigenfunctions as and related eigenvalues to $\widehat{\ell_\mathcal{F}}$. By using \eqref{eq:IPintro}, we shall show that matrix elements of this operator have a path integral interpretation as a weighted sum  of $\mathcal F(\ell)$ over all possible geodesics:
\be 
\bra{\psi'}\mathcal F(\widehat{\ell_\mathcal{F}})\ket{\psi} =\int%
\frac{Dg_{\mu\nu} D\Phi}{\text{Diffs}} \left(\sum_{\gamma} \mathcal F(\ell_\gamma)\right) e^{-I_\text{JT}}\,,
\label{eq:path-integral-expression-sum-over-geodesics}
 \ee
 where the boundary conditions are determined by the states $|\psi\rangle$ and $|\psi'\rangle$, and the sum runs over all non self-intersecting geodesic slices $\gamma$ between specified boundary points. In other words, for each relative weight of geodesics, there is an associated length operator that we can define at the bulk Hilbert space level.   

To highlight the usefulness of our formalism, we shall focus on two examples of bulk operators, both of which shall serve as useful probes for the interior of two-sided black holes. 

\subsection{Case study I: length operators}

\begin{center}
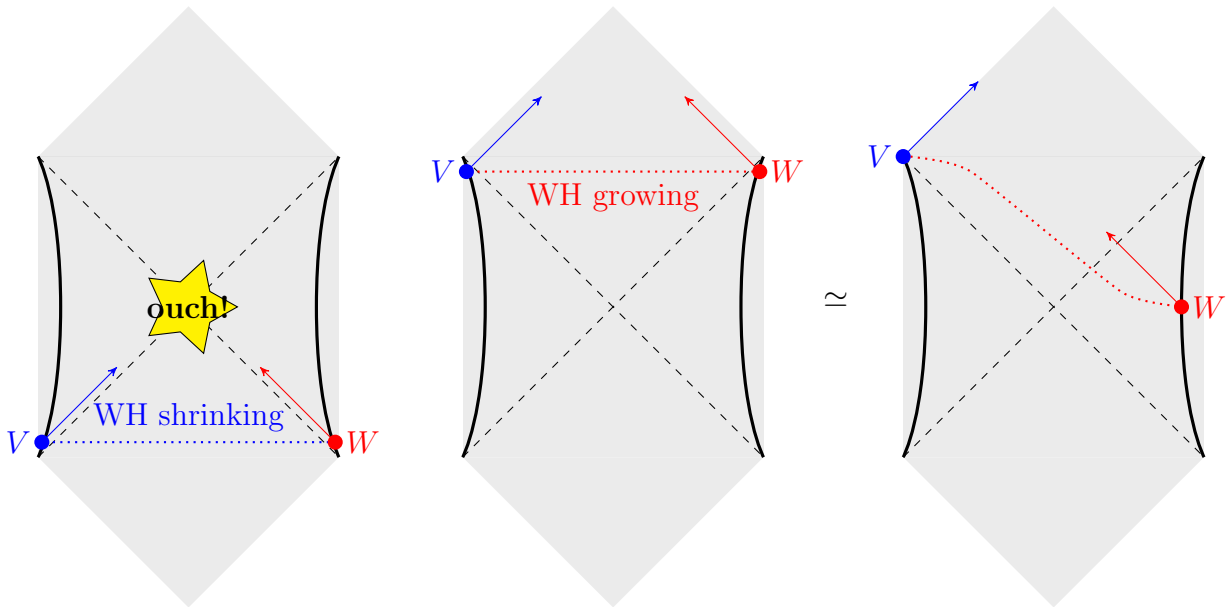
\begin{figure}[h!]
\begin{center}
\[\begin{tikzpicture}[baseline={([yshift=-0cm]current bounding box.center)}]
\fill[lightgray, opacity=0.3] (-2,-2) rectangle (2,2);
\fill[lightgray, opacity=0.3] (-2,-2) -- (0,-4) -- (2,-2);
\fill[lightgray, opacity=0.3] (-2,2) -- (0,4) -- (2,2);
\draw[dashed] (-2,-2) -- (2,2);
\draw[dashed] (-2,2) -- (2,-2);
\draw[black,very thick] plot [smooth, tension=1.3] coordinates {(-2,-2) (-1.7,0) (-2,2)};
\draw[black,very thick] plot [smooth, tension=1.3] coordinates {(2,-2) (1.7,0) (2,2)};
\draw[blue, thick, dotted] (-2,-1.8) -- node[above] {WH shrinking} (2,-1.8) ;
\fill[blue] (-1.95,-1.8) circle (0.1) node[left] {$V$};
\draw[blue,->] (-1.95,-1.8) -- (-1.95+1,-1.8+1);
\fill[red] (1.95,-1.8) circle (0.1) node[right] {$W$};
\draw[red,->] (1.95,-1.8) -- (1.95-1,-1.8+1);
\Star[fill=yellow,draw]{.35}{.65};
\node at (0,0) {{\bf ouch!}};
\end{tikzpicture} 
\hspace{0.4cm}
\begin{tikzpicture}[baseline={([yshift=-0cm]current bounding box.center)}]
\fill[lightgray, opacity=0.3] (-2,-2) rectangle (2,2);
\fill[lightgray, opacity=0.3] (-2,-2) -- (0,-4) -- (2,-2);
\fill[lightgray, opacity=0.3] (-2,2) -- (0,4) -- (2,2);
\draw[dashed] (-2,-2) -- (2,2);
\draw[dashed] (-2,2) -- (2,-2);
\draw[black,very thick] plot [smooth, tension=1.3] coordinates {(-2,-2) (-1.7,0) (-2,2)};
\draw[black,very thick] plot [smooth, tension=1.3] coordinates {(2,-2) (1.7,0) (2,2)};
\draw[red, thick, dotted] (-2,1.8) -- node[below] {WH growing} (2,1.8) ;
\fill[blue] (-1.95,1.8) circle (0.1) node[left] {$V$};
\draw[blue,->] (-1.95,1.8) -- (-1.95+1,1.8+1);
\fill[red] (1.95,1.8) circle (0.1) node[right] {$W$};
\draw[red,->] (1.95,1.8) -- (1.95-1,1.8+1);
\end{tikzpicture} \simeq
\begin{tikzpicture}[baseline={([yshift=-0cm]current bounding box.center)}]
\fill[lightgray, opacity=0.3] (-2,-2) rectangle (2,2);
\fill[lightgray, opacity=0.3] (-2,-2) -- (0,-4) -- (2,-2);
\fill[lightgray, opacity=0.3] (-2,2) -- (0,4) -- (2,2);
\draw[dashed] (-2,-2) -- (2,2);
\draw[dashed] (-2,2) -- (2,-2);
\draw[black,very thick] plot [smooth, tension=1.3] coordinates {(-2,-2) (-1.7,0) (-2,2)};
\draw[black,very thick] plot [smooth, tension=1.3] coordinates {(2,-2) (1.7,0) (2,2)};
\draw[red, thick, dotted] plot [smooth, tension=.3] coordinates {(-2,2) (-1.9,2) (-1.1,1.8) (0.9,0.2) (1.7,0) (1.8,0)};
\fill[blue] (-2,2) circle (0.1) node[left] {$V$};
\draw[blue,->] (-2,2) -- (-2+1,2+1);
\fill[red] (1.95-0.25,0) circle (0.1) node[right] {$W$};
\draw[red,->] (1.95-0.25,0) -- (1.95-1-0.25,0+1);
\end{tikzpicture} \]
\end{center}
\caption{\label{fig:collision} Two-sided near-extremal black hole setup in which we shall study the collision energy between an observer and a particle, infalling from opposite sides. In the left-most diagram, an infalling $\color{red} W$ particle encounters a collision almost immediately after crossing the horizon. By applying a boost to the middle figure, we get the right-most diagram. We see that  ${\color{red} W}$ does not experience any collision with $\color{blue} V$ until ${\color{red} W}$ gets extremely close to the future inner horizon (which is expected to be singular in any realistic near-extremal black hole.) In other words, while there is still a high-energy collision in the growing wormhole, the $W$ particle enjoys a potentially large amount of proper time in the interior of the black hole before the collision. In this sense, a growing wormhole is safer than a shrinking wormhole. 
}
\end{figure}
\end{center}
Studying the behavior of length operators non-perturbatively can help us understand whether an infalling observer sees an energetic particle when crossing the horizon of a black hole, a firewall \cite{Almheiri:2012rt,Stanford:2014jda}. Consider the setup presented in figure~\ref{fig:collision}. An observer falls into a two-sided black hole from the right at point $W$, and a particle is sent in from the left at point $V$.  The existence of a firewall is correlated with whether the wormhole between $V$ and $W$  is growing or shrinking \cite{Susskind:2015toa, Susskind:2020wwe, Stanford:2022fdt}. A shrinking wormhole leads to an exponential blue-shift factor which will lead to a violent collision between an infalling observer and the particle sent in from the opposite side. The bulk state, in this case, is usually referred to as a white hole. A growing wormhole leads to an exponential red-shift factor which makes the collision peaceful. The bulk state, in this case, is a black hole. Thus, the probability distribution for the "velocities" of the wormhole length can offer a proxy for the probability of hitting a firewall. Perturbatively, if we fix the energies on both sides to be close to $E$, this velocity operator should only have two possible eigenvalues ($+\sqrt{2 E}$ and $-\sqrt{2E}$), the probability of seeing a negative eigenvalue approximating that of seeing a firewall. While at early times, we expect the state of the black hole to have approximately zero probability of detecting a shrinking wormhole, at very late times, the black hole could be in a complicated linear combination of velocity eigenstates. At such times, what is the probability of detecting a shrinking wormhole?

To understand such probabilities we should thus start by having a well-defined length operator that acts on the bulk Hilbert space discussed above. A length operator $\hat \ell_\Delta$  that we will study in detail will be obtained by taking $\mathcal{F}(\ell) = e^{-\Delta \ell}$ in \eqref{eq:define-length-operator-intro}. This operator can be viewed as computing the length from the matrix elements of the bi-local operator $O_L O_R$, each operator of scaling dimension $\Delta$ being inserted on opposite sides of the black hole. In other words, this length operator is directly defined in terms of simple boundary operators. As we shall explain, the matrix elements of this length operator can be explicitly computed for any member of the JT ensemble, and from this, given a wavefunction, we can compute the probability of detecting any eigenvalue of the length operator or any eigenvalue of its associated velocity operator.\footnote{We will take the velocity operator to be defined as $\hat v_{\Delta} = i [H, \ell_{\Delta}]$, where $H = H_L = H_R$ is the ADM Hamiltonian operator on either boundary.} By numerically sampling from the ensemble dual to JT gravity, we can obtain probability distributions for the observed lengths and velocities that capture all possible corrections in $e^{-S_0}$.

 Starting with a two-sided black hole in a microcanonical window,\footnote{With $H_L = H_R$.} we find that the length of the wormhole between V and W starts by growing linearly with time, after which it plateaus at a time and value $~e^{S_\text{BH}}$.\footnote{This result is consistent with previous proposals for the length operator in the literature, for example \cite{Iliesiu:2021ari, Stanford:2022fdt}. However, as far as we are aware, the Hilbert space interpretation for all length operators discussed in the past was unclear. } Consequently, the expectation value of the associated wormhole velocity vanishes at late times. To better understand this result,  we compute the probability distribution of detecting any given wormhole velocity. We find that the probability of detecting a negative velocity is $1/2$ at late times. At first sight, this suggests that if the velocity operator is a good proxy for firewalls, the probability of encountering a firewall at late times is also $1/2$. This probability is self-averaging and universal: it is independent of the choice of ensemble and of the value of $\Delta$ that we choose in order to define $\hat \ell_\Delta$. This is consistent with the findings of \cite{Stanford:2022fdt}, who computed the probability for the wormhole to shrink to first non-trivial order in $e^{-S_0}$. However, we find that an $O(1)$ fraction of this probability is concentrated on eigenvalues of the velocity operator that do not have a good perturbative description, i.e. they are not close to $\pm \sqrt{2 E}$. In other words, a more refined description of our results is that there are equal $O(1)$ probabilities of detecting a black hole/white hole state, and there is an $O(1)$ probability of detecting a state that does not have any good perturbative description; we will refer to such states as gray holes.\footnote{This is slightly different than what is known as the gray hole scenario put forward in \cite{Susskind:2015toa}. There, the state at late times was believed to be in a linear combination of black hole states and white hole states. Here, we see the appearance of a third type of states that do not have a good semi-classical description, i.e.~a gray hole state. } This suggests that even if we start with a perturbatively well-defined firewall operator, whose eigenvalues are $\pm 1$ for detecting or not detecting a firewall, its non-perturbative definition will have some eigenvalues that could be very different from $\pm 1$ and associated eigenstates that do not have a good perturbative description.

\subsection{Case study II: center-of-mass collision energy}

A second operator that serves as a more direct proxy for the experience of an infalling observer is the center-of-mass collision energy between the observer and the infalling particle from the opposite side, as in figure~\ref{fig:collision}. We will thus define and study the operator associated with this energy. In our AdS$_2$ geometry, the CM collision energy can be obtained by studying the Casimir operator associated with the $\mathfrak{sl}(2, \mathbb{R})$ isometry of the spacetime \cite{Maldacena:2016upp,Lin:2019qwu}. While the perturbative properties of the length operator were previously understood, the collision energy operator or the related Casimir operator has not been previously studied in the perturbative regime. For the two-particle setup we are considering in figure~\ref{fig:collision}, the eigenvalues of such an operator are quantized. This operator is also conserved at the perturbative level: thus, we can analyze the two-particle wavefunction in the collision energy basis is equal on all time slices past the insertion points of $V$ and $W$. We will thus analyze this two-particle wavefunction in the Casimir basis on some slice to the future of $V$ and $W$ and track its behavior as we change the time at which $V$ and $W$ are inserted. Perturbatively, from this Casimir operator, we find that the likeliest collision energy that an observer would detect grows exponentially with time, and the wavefunction spread also grows exponentially with time. 

 Because it quickly spreads with time, the wavefunction in the collision basis sees non-perturbative corrections at relatively early times, $t \sim S_0$. The origin of such corrections comes from wormhole contributions in which the conservation of the Casimir is violated, a phenomenon similar to the violation of global symmetry charge conservation discussed in \cite{Harlow:2020bee, Chen:2020ojn, Hsin:2020mfa}. Even at such early times, an observer should see that the eigenvalues of the Casimir no longer have the quantized spacing predicted by the perturbative calculation. %
 Instead, we show that at large times the behavior of the Casimir operator mimics that of an exponential of the length operator that we discussed above. Thus, the same ambiguities that appeared in the non-perturbative definition of the length reappear when studying the behavior of the collision energy at late times. We thus expect that the universal behavior seen for the length operator to also be present for the collision energy operator, whose expectation value should thus plateau at a time and values $\sim e^{S_0}$.

\subsection{Outline}

The outline of the paper is as follows. In section \ref{sec:pureJT}, we review the perturbative properties of JT gravity, after which we derive the non-perturbative modification to the inner product and explore its effect on Hamiltonian evolution. In section \ref{sec:null}, we explicitly derive all null-states that exist in our construction that need to be quotiented out from our vector space in order to have a well-defined Hilbert space. In section \ref{sec:defining-non-perturbative-bulk-ops}, we explain in further detail why there is an ambiguity in the non-perturbative definition of any bulk operator both within the Hilbert space and at the path integral level. In section \ref{sec:a-length-operator}, we focus on a specific choice of wormhole length and velocity operators and study them numerically in great detail. In section \ref{sec:JT-with-matter}, we generalize our construction of a non-perturbative bulk Hilbert space to JT gravity with matter. We then use this construction in section \ref{sec:CM-collision-energy} to explicitly study the CM collision energy operator both at a perturbative and non-perturbative level. Finally, in section \ref{sec:discussion}, we discuss several consequences and interpretations of our results.\vspace{0.5cm}\\
\textit{Note:} During the development of these results, we became aware of the upcoming work of \cite{Blommaert:2024ftn}, who present related results (to all orders in $e^{-S_0}$) for the probability of detecting a firewall.

\section{The non-perturbative geodesic Hilbert space}
\label{sec:pureJT}

\subsection{Perturbative Hilbert space of pure JT: review}

We begin by reviewing the relevant results of quantization of pure JT gravity on spacetimes of disc topology, which can be obtained from many different perspectives \cite{Maldacena:2016upp, Mertens:2017mtv, Saad:2019pqd, Yang:2018gdb, Kitaev:2017awl, Harlow:2018tqv, Iliesiu:2019xuh, Penington:2023dql}.%

We study the Hilbert space with two asymptotic boundaries, defined by Dirichlet boundary conditions for the proper length of the boundary $L=\beta/\epsilon$ and for the dilaton $\Phi|_\partial = \frac{\Phi_b}{\epsilon}$, where we take the cutoff $\epsilon\to 0$. The value of $\phi_b$ simply sets the units of energy and time, and we work in units with $\phi_b=1$. The physical time $t$ on this boundary is defined (according to the usual holographic prescription) as $\epsilon$ times the proper time. The theory has a single degree of freedom, which we can take to be the renormalized geodesic length $\ell$ between the two boundaries: $\ell$ equals the proper length minus $2\log\epsilon^{-1}$, which gives a finite (but not necessarily positive) quantity when we take $\epsilon\to 0$.

The resulting `perturbative Hilbert space' $\hilb_0 = L^2(\mR)$ consists of wavefunctions of the length, with the usual inner product as given in \eqref{eq:H0intro}. We can define states directly in this description using geodesic boundary conditions, giving $\delta$-function states of definite length $|\ell\rangle$,  normalised so that $\langle \ell'|\ell\rangle_0 = \delta(\ell-\ell')$. Time evolution on the left or right side is generated by a Hamiltonian $H_L$ or $H_R$; these operators are in fact equal for pure JT, and act on  wavefunctions $\psi(\ell)$ as
\begin{equation}\label{eq:Hpure}
    H=H_L=H_R = -\frac{1}{2}\partial_\ell^2 + 2e^{-\ell}.
\end{equation}
The Hamiltonian is diagonalised by scattering states $|E\rangle$ for $E>0$, with wavefunctions
\begin{equation}
    \label{eq:phiEl}
\phi_E(\ell)= \langle \ell | E\rangle_0 =  4 K_{i \sqrt{8E}}\left(4 e^{-\ell/2}\right).
\end{equation}
We have defined these states with a convenient normalisation, which is singled out (up to an overall constant) by the fact that the wavefunction becomes independent of $E$ when we take $\ell\to -\infty$. (Note also that the Bessel function is real.) This gives the following inner products and completeness relation on $\hilb_0$:
\begin{equation}
    \langle E'|E\rangle_0 = \frac{\delta(E-E')}{\rho_0(E)}\,, \qquad \int_0^\infty dE \,\rho_0(E) |E\rangle\langle E| = \id_0\,.
    \label{eq:orth-lengths-to-energy}
\end{equation}
Here we have defined
\begin{equation}
\rho_0(E) = \frac{1}{4\pi^2} \sinh\left(2\pi \sqrt{2E}\right)\,
\end{equation}
which (up to a factor of $e^{S_0}$) is familiar as the density of states coming from the disk in JT: that is $Z_0(\beta) = e^{S_0}\int dE\, \rho_0(E) e^{-\beta E}$, where the left-hand side is defined as a path integral over Euclidean spacetimes with disk topology and renormalized boundary length $\beta$.

This disk partition function can also be interpreted as an inner product $Z_0(\tau+\tau') = \langle \tau'|\tau\rangle_0$ (computed at infinite $S_0$) by cutting it into two half-disks, with $|\tau\rangle$ the two-sided state defined by an asymptotic Euclidean boundary of renormalized length $\tau>0$. These states will be useful for us because they have a simple definition as thermofield double (TFD) states in the non-perturbative theory (defined by a dual boundary quantum mechanics as described in a moment). We can write $|\tau\rangle$ in terms of the energy eigenstates above, with
\begin{equation}\label{eq:TFD}
    |\tau\rangle = e^{S_0/2} \int dE\, \rho_0(E) e^{-\tau E} |E\rangle;
\end{equation}
this is fixed (up to an $E$-dependent phase) by the requirement that the overlaps give  $Z_0$. We can also write $|\tau\rangle$ as a superposition of geodesic states,
\begin{equation}\label{eq:ZHD}
    \begin{gathered}
    |\tau\rangle = e^{S_0/2}\int d\ell\, Z_{\HD}(\tau;\ell) |\ell\rangle, \\
    Z_\HD(\tau;\ell) =  \int dE \,\rho_0(E) e^{-\tau E} \phi_E(\ell).
\end{gathered}
\end{equation}
This expression is obtained by taking the inner product with $|\ell\rangle$ using \eqref{eq:phiEl}, but can also be computed directly from a path-integral on the half-disk geometry bounded by an asymptotic boundary and a geodesic boundary \cite{Yang:2018gdb}:
\begin{align}\label{eq:half-disc}
   e^{S_0/2}Z_{\HD}(\tau;\ell) = \langle \ell|\tau\rangle_0 =  \begin{tikzpicture}[baseline={([yshift=0cm]current bounding box.center)}]
    \clip (-2.2,0.1) rectangle (2.2,-2.2);
        \draw[thick,fill=lightgray, fill opacity=0.25] (0,0) circle (2);
        \draw[thick,blue] (-2,0.1) -- (2,0.1) node[midway,below] {$\ell$};
        \node at (0.4,-1.75) {$\tau$};
    \end{tikzpicture}\,.
\end{align}

\subsection{Defining the non-perturbative theory}

To study a similar geodesic Hilbert space in the non-perturbative theory, we must first give a precise definition of what we mean by that theory. Foreshadowing the key idea, we will identify a relation between geodesic states $|\ell\rangle$ and the asymptotic `TFD' states $|\tau\rangle$ which is independent of the full details of the bulk theory. Using this, we can define the inner products $\langle \ell'|\ell\rangle$ (and matrix elements like $\langle \ell'|e^{-\tau H}|\ell\rangle$) by first rewriting them in terms of asymptotically-defined inner products $\langle \tau'|\tau\rangle$. We can write these overlaps as $Z(\beta)$, which is defined by an asymptotic boundary condition with a Euclidean thermal circle of length $\beta=\tau+\tau'$. So, it is sufficient to define the theory via the amplitudes with asymptotic circular boundaries $Z(\beta)$. While this definition of the theory involves Euclidean quantities, with the answer in hand, we can nonetheless apply it to real-time evolution and Lorentzian questions.

As sketched in section \ref{ssec:introHilb}, there are two possible (closely related) perspectives. The first is to directly take the definition and interpretation of JT gravity given in \cite{Saad:2019lba}. With that definition, the theory computes amplitudes $Z_n(\beta_1,\ldots,\beta_n)$ with $n$ disconnected circular boundaries (order-by-order in a genus expansion). $Z_n$ is then interpreted as an $n$th moment $\overline{Z(\beta_1)\cdots Z(\beta_n)}$ of random variables $Z(\beta)$, with $\overline{\phantom{I}\cdot\phantom{I}}$ denoting the average with respect to some probability distribution. In this distribution, $Z(\beta)$ is the partition function $\Tr(e^{-\beta H})$ of a random Hamiltonian $H$, with discrete (random) spectrum $\{E_i\}_{i=0}^\infty$. We can describe this as an ensemble of quantum mechanical theories dual to pure JT. Directly importing this to the geodesic states, we will end up calculating statistics $\overline{\langle \ell_1'|\ell_1\rangle \cdots \langle \ell_n'|\ell_n\rangle}$, where $\langle\ell'|\ell\rangle$ is interpreted a random inner product.

For the second perspective, we can ask what the inner product $\langle\ell'|\ell\rangle$ looks like if we select a specific member of the ensemble. In practical terms, we can sample a single random matrix $H$  from the probability distribution, and use the resulting specific spectrum $\{E_i\}_{i=0}^\infty$ to define the theory. Then, an $n$ boundary amplitude factorizes into a product of $n$ partition functions $Z(\beta)=\sum_i e^{-\beta E_i}$ (so we need only ever consider a single asymptotic boundary of spacetime). It is reasonable to expect the result for a typical spectrum to be indicative of the situation in a theory with a conventional (non-ensemble) holographic dual. From the bulk perspective, this means that we are choosing a specific $\alpha$-state of closed `baby' universes \cite{Coleman:1988cy,Giddings:1988cx,Marolf:2020xie}. We can arrange this by modifying the boundary conditions, perhaps by adding a carefully arranged set of additional asymptotic boundaries \cite{Marolf:2020xie}, or similarly by additional `eigenbrane' boundary conditions on which spacetime can terminate \cite{Blommaert:2019wfy}, though we will not need to be very specific about the details. In any case, with this perspective, we take a single boundary dual quantum mechanics and compute the associated single matrix of inner products $\langle\ell'|\ell\rangle$.

We will mostly write with this latter perspective in mind, though they are essentially equivalent since the ensemble can always be recovered by thinking of $\{E_i\}$ as random variables rather than specific numbers. 

We should comment on how the two-sided Hilbert space of JT is related to the Hilbert space $\hilb_{\partial}$ of this boundary dual quantum mechanics. Since we have two asymptotic spatial boundaries, we expect to have two copies of the dual system $\hilb_R\simeq \hilb_\partial$ and $\hilb_L\simeq\hilb_\partial^*$ (the complex conjugate Hilbert space) living on the right and left boundaries respectively, and thus a total Hilbert space $\hilb \simeq \hilb_L \otimes \hilb_R$ with a basis $|i_L,i_R\rangle$ of eigenstates of left and right Hamiltonians $H_L,H_R$. While this is indeed the case when we consider JT gravity coupled to matter, it fails in pure JT since all states have $H_L=H_R$. This is clear from the definition of the bulk theory through the amplitudes as described above, since the TFD states $|\tau\rangle$ (which satisfy $H_L=H_R$ automatically) span all possible boundary conditions creating states of $\hilb$. Thus, $\hilb$ consists only of the `diagonal' subspace spanned by states $|i,i\rangle$ with $i_L=i_R=i$ (which we will simply write as $|i\rangle$).

Finally, we note that there is a third perspective, which is natural from the gravitational point of view, but gives an inequivalent construction. Once we include fluctuations of topology, it is no longer clear that we can describe a two-boundary bulk Hilbert space by only a single geodesic length, because there are processes which produce closed `baby' universes. If we keep track of these closed components, we arrive at an inherently different Hilbert space, which will be explored in future work (previewed in section \ref{ssec:BUdisc}). The ensemble is recovered by `tracing out' the closed universe sector, while the single dual emerges by placing the closed universes in an $\alpha$ state.%

\subsection{Relating geodesic and asymptotic states}

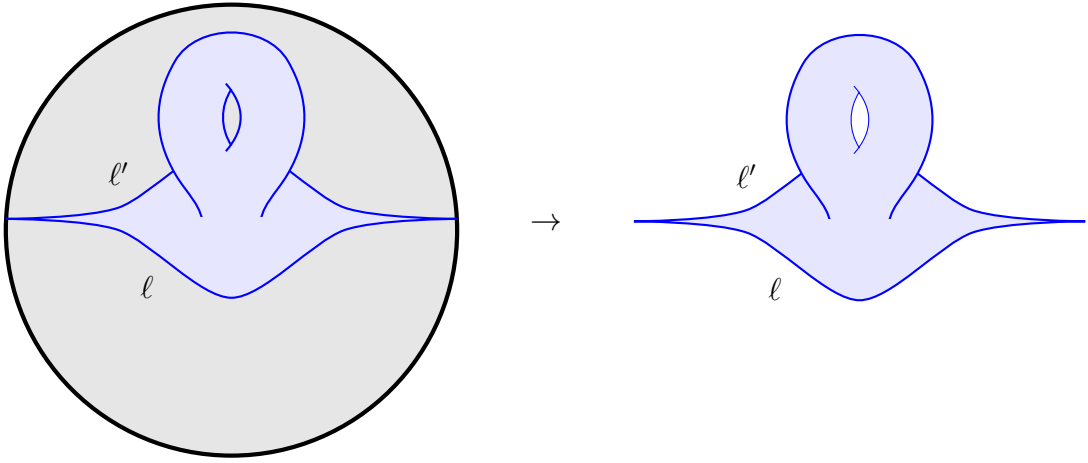
\begin{figure}
\[\begin{tikzpicture}[rotate=90,scale=0.75,baseline={([yshift=-0cm]current bounding box.center)}]
\draw[ultra thick, fill = gray!20] (2,0) circle (4);
\node at (1,1.5) {$\ell$};
\node at (3,2) {$\ell'$};
\draw [thick, blue, fill = llgray,rotate=90,yshift=-2.2cm,scale=2] plot [smooth] coordinates { (-2,0)(-1,0.1)(0,0.7)(1,0.1)(2,0)} -- plot [smooth] coordinates {(2,0)(1,-0.1)(0,-0.7)(-1,-0.1)(-2,0)} ;
\clip (2.25,-1.51) rectangle (5.6,1.51);
\draw[smooth, thick, blue, fill = llgray] (0,0.5) to[out=0,in=0] (2,0.5) to[out=0,in=210] (3,1) to[out=30,in=150] (5,1) to[out=-30,in=30] (5,-1) to[out=210,in=-30] (3,-1) to[out=150,in=0] (2,-0.5) to[out=0,in=0] (0,-0.5) to[out=150,in=-150] (0,0.5);
\fill[smooth, gray!20] (3.52,0) .. controls (3.8,-0.2) and (4.2,-0.2) .. (4.48,0);
\draw[smooth, thick, blue, fill = gray!20] (3.5,0) .. controls (3.8,0.2) and (4.2,0.2) .. (4.5,0);
\draw[smooth, thick, blue] (3.4,0.1) .. controls (3.8,-0.25) and (4.2,-0.25) .. (4.6,0.1);
\end{tikzpicture}  %
\qquad \to \qquad
\begin{tikzpicture}[rotate=90,scale=0.75,baseline={([yshift=-0.8cm]current bounding box.center)}]
\node at (1,1.5) {$\ell$};
\node at (3,2) {$\ell'$};
\draw [thick, blue, fill = llgray,rotate=90,yshift=-2.2cm,scale=2] plot [smooth] coordinates { (-2,0)(-1,0.1)(0,0.7)(1,0.1)(2,0)} -- plot [smooth] coordinates {(2,0)(1,-0.1)(0,-0.7)(-1,-0.1)(-2,0)} ;
\clip (2.25,-1.51) rectangle (5.6,1.51);
\draw[smooth, thick, blue, fill = llgray] (0,0.5) to[out=0,in=0] (2,0.5) to[out=0,in=210] (3,1) to[out=30,in=150] (5,1) to[out=-30,in=30] (5,-1) to[out=210,in=-30] (3,-1) to[out=150,in=0] (2,-0.5) to[out=0,in=0] (0,-0.5) to[out=150,in=-150] (0,0.5);
\fill[smooth, white] (3.52,0) .. controls (3.8,-0.2) and (4.2,-0.2) .. (4.48,0);
\draw[smooth, blue, fill = white] (3.5,0) .. controls (3.8,0.2) and (4.2,0.2) .. (4.5,0);
\draw[smooth, blue] (3.4,0.1) .. controls (3.8,-0.25) and (4.2,-0.25) .. (4.6,0.1);
\end{tikzpicture} \] %
\caption{On the left, we show a sample contribution to the disk partition function. It includes a handle disk. By stripping off the half-disk wavefunctions, we extract a contribution to the inner product $\braket{\ell}{\ell'}$. \label{fig:diskExtract} }
\end{figure}

With the preliminaries out the way, we can now proceed to compute the full inner product $\langle\ell'|\ell\rangle$, defined as the path integral over all Euclidean spacetimes bounded by two geodesics of lengths $\ell,\ell'$. To do this, we first consider computing the inner product between thermofield double states $\langle \tau'|\tau\rangle$, computed by the path integral $Z(\tau+\tau')$ with boundary conditions of an asymptotic thermal circle. We observe that for pure JT gravity, in every spacetime contributing to this path integral, there is a unique geodesic homotopic to each segment of the asymptotic boundary. Furthermore, the region between the boundary and this geodesic is a simple half-disk, namely the geometry pictured in \eqref{eq:half-disc}. We discuss the generality of this result and the underlying assumptions in section \ref{ssec:conditions}. Now, we are free to cut along these geodesics and divide the path integral into three parts: two half-discs and an integral over spacetimes bounded by two geodesics, which defines the length-basis inner product $\langle\ell'|\ell\rangle$:
\begin{equation}\label{eq:ZllIP}
        Z(\tau+\tau') =  e^{S_0}\int d\ell d\ell' \,Z_\HD(\tau;\ell)Z_\HD(\tau';\ell') \langle\ell'|\ell\rangle.
\end{equation}
This formula will be enough to compute $\langle\ell'|\ell\rangle$, see Figure~\ref{fig:diskExtract}.

We are defining $Z(\beta)$ non-perturbatively as the partition function of a dual quantum mechanics, so we can rewrite it as a sum over the spectrum $\{E_n\}$ of the Hamiltonian. Using this as well as \eqref{eq:ZHD} for the half-disk path integral, we can write \eqref{eq:ZllIP} as
\begin{equation}
      \sum_{i=0}^\infty e^{-(\tau+\tau')E_i}  = e^{S_0}\int d\ell d\ell' dE \rho_0(E) dE' \rho_0(E') e^{-(\tau E +\tau' E')} \phi_E(\ell)\phi_{E'}(\ell')\langle\ell'|\ell\rangle.
\end{equation}
Now, for this to hold for all $\tau,\tau'$ we we must have
\begin{equation}\label{eq:llIPintermediate}
      \sum_{i=0}^\infty \delta(E-E_i)\delta(E'-E_i)  = e^{S_0} \rho_0(E)  \rho_0(E') \int d\ell d\ell' \phi_E(\ell)\phi_{E'}(\ell')\langle\ell'|\ell\rangle.
\end{equation}
Finally, we can invert this using $\int dE \rho_0(E) \phi_E(\ell)\phi_E(\ell') = \delta(\ell-\ell')$ (the completeness relation for the states $|E\rangle$ in $\hilb_0$)\footnote{The fact that this integral runs only over $E>0$ might raise suspicions that something goes wrong when the spectrum contains negative energies. In fact, while the orthogonality of $\phi_E(\ell)$ states only makes sense for $E>0$, the integral $\int d\ell \phi_E(\ell) Z_\HD(\tau,\ell)$ still converges to $e^{-\tau E}$ for $E<0$. So while the intermediate step \eqref{eq:llIPintermediate} does not make sense because the order of integration cannot be exchanged, the final result is still correct even when $E_i<0$ for some states.} to obtain our final result for the non-perturbative geodesic inner product:
\begin{equation}\label{eq:llIP}
    \langle \ell'|\ell\rangle = e^{-S_0} \sum_{i=0}^\infty \phi_{E_i}(\ell)\phi_{E_i}(\ell').
\end{equation}
Note that if we approximate the sum $\sum_i$ by an integral $\int dE \, e^{S_0} \rho(E)$ with the disk density of states, we recover $\langle \ell'|\ell\rangle_0=\delta(\ell-\ell')$. This means that the inner product of wavefunctions with broad, coherent energy resolution will be very close to the perturbative result. Nonetheless, for some purposes \eqref{eq:llIP} is a dramatic departure from $\langle\ell'|\ell\rangle_0$.

We can express this result more directly in terms of the states themselves: we have learned that the relation \eqref{eq:ZHD} expressing the thermofield double state in terms of the geodesic states $|\tau\rangle = \int d\ell\, Z_\HD(\tau;\ell) |\ell\rangle$ continues to hold exactly non-perturbatively! We verify this by computing the overlap of $|\tau\rangle$ with any state (for which it's sufficient to take another thermofield double, since these span the Hilbert space), and cutting the resulting path integral along the geodesic homotopic to the boundary. In this form, it's also straightforward to write the geodesic states directly in terms of an orthonormal basis $|i\rangle$ of energy eigenstates ($\langle j|i\rangle=\delta_{i,j}$), using the decomposition $|\tau\rangle = \sum_n e^{-\tau E_i} |i\rangle$ of the thermofield double:
\begin{equation}\label{eq:ilIP}
    \langle i|\tau\rangle = e^{-\tau E_i} \implies \langle i|\ell\rangle = e^{-S_0/2} \phi_{E_i}(\ell) \qquad (i=0,1,2,\ldots).
\end{equation}

\subsection{An alternative ensemble perspective}\label{ssec:ensemble}

To reiterate the argument and to make the non-perturbative path integral somewhat more concrete, it is illuminating to restate things in the context of the JT ensemble \cite{Saad:2018bqo}, rather than a single specific dual spectrum as above. For this, consider the connected contribution to the product of $n$ inner products, defined by a path integral over connected geometries with $n$ pairs of geodesic boundaries. Such a product has a genus expansion:
\begin{equation}
\label{eq:average-l-def}
    \overline{\langle \ell_1'|\ell_1\rangle \cdots \langle \ell_n'|\ell_n\rangle}_\text{conn.} =  e^{(1-2g-n) S_0}\sum_{g=0}^\infty \eta_{n,g}(\ell_1,\ldots,\ell_n;\ell_1',\ldots,\ell_n'),
\end{equation}
where $ \eta_{n,g}(\ell_1,\ldots,\ell_n;\ell_1',\ldots,\ell_n')$ is computed by summing over all genus $g$ geometries with $n$ boundaries, each consisting of a pair of geodesics of renormalized lengths $\ell,\,\ell'$. The $n=1,g=0$ contribution gives the perturbative result $\delta(\ell-\ell')$, while other terms give non-perturbative corrections suppressed by powers of $e^{-S_0}$. For example, the following geometries can be used to compute $\eta_{1,1}$ and $\eta_{2,0}$ respectively which give the leading non-perturbative corrections to $\overline{\braket{\ell'}{\ell}}$ and $\overline{\braket{\ell'_1}{\ell_1} \braket{\ell'_2}{\ell_2}}_\text{conn.}$:\footnote{Here, we are assuming the sum over geometries solely includes orientable surfaces. A similar $e^{-S_0}$ expansion would apply to the unorientable case. }
\begin{align}
\overline{\braket{\ell'}{\ell}} \quad  = \quad \delta(\ell-\ell') +  \quad 
\begin{tikzpicture}[rotate=90,scale=0.75,baseline={([yshift=-0.8cm]current bounding box.center)}]
\node at (1,1.5) {$\ell$};
\node at (3,2) {$\ell'$};
\draw [thick, blue, fill = llgray,rotate=90,yshift=-2.2cm,scale=2] plot [smooth] coordinates { (-2,0)(-1,0.1)(0,0.7)(1,0.1)(2,0)} -- plot [smooth] coordinates {(2,0)(1,-0.1)(0,-0.7)(-1,-0.1)(-2,0)} ;
\clip (2.25,-1.51) rectangle (5.6,1.51);
\draw[smooth, thick, blue, fill = llgray] (0,0.5) to[out=0,in=0] (2,0.5) to[out=0,in=210] (3,1) to[out=30,in=150] (5,1) to[out=-30,in=30] (5,-1) to[out=210,in=-30] (3,-1) to[out=150,in=0] (2,-0.5) to[out=0,in=0] (0,-0.5) to[out=150,in=-150] (0,0.5);
\fill[smooth, white] (3.52,0) .. controls (3.8,-0.2) and (4.2,-0.2) .. (4.48,0);
\draw[smooth, blue, fill = white] (3.5,0) .. controls (3.8,0.2) and (4.2,0.2) .. (4.5,0);
\draw[smooth, blue] (3.4,0.1) .. controls (3.8,-0.25) and (4.2,-0.25) .. (4.6,0.1);
\end{tikzpicture}
\quad + \quad O(e^{-4S_0})
\end{align}
\begin{align}
\overline{\braket{\ell'_1}{\ell_1} \braket{\ell'_2}{\ell_2}}_\text{conn.} \quad = \quad
\begin{tikzpicture}[rotate=90,scale=0.75,baseline={([yshift=0cm]current bounding box.center)}]
\node at (1,1.5) {$\ell_2$};
\node at (3,2) {$\ell'_2$};
\node at (4,1.5) {$\ell_1$};
\node at (6,2) {$\ell'_1$};
\draw [thick, blue, fill = llgray,rotate=90,yshift=-2.2cm,scale=2] plot [smooth] coordinates { (-2,0)(-1,0.1)(0,0.7)(1,0.1)(2,0)} -- plot [smooth] coordinates {(2,0)(1,-0.1)(0,-0.7)(-1,-0.1)(-2,0)} ;
\draw [thick, blue, fill = llgray,rotate=90,yshift=-5.2cm,scale=2] plot [smooth] coordinates { (-2,0)(-1,0.1)(0,0.7)(1,0.1)(2,0)} -- plot [smooth] coordinates {(2,0)(1,-0.1)(0,-0.7)(-1,-0.1)(-2,0)} ;
\draw[smooth, line width=0.03cm, blue, fill = llgray, xshift=2.2cm, scale =1.5] (0,0.5) to[out=-15,in=195] (2,0.5) to[out=-30,in=30] (2,-0.5)  to[out=165,in=15] (0,-0.5) to[out=150,in=-150] (0,0.5);
\draw [thick, dashed, blue,rotate=90,yshift=-5.2cm,scale=2] plot [smooth] coordinates { (-2,0)(-1,0.1)(0,0.7)(1,0.1)(2,0)} -- plot [smooth] coordinates {(2,0)(1,-0.1)(0,-0.7)(-1,-0.1)(-2,0)} ;
\draw [thick, dashed, blue, rotate=90,yshift=-2.2cm,scale=2] plot [smooth] coordinates { (-2,0)(-1,0.1)(0,0.7)(1,0.1)(2,0)} -- plot [smooth] coordinates {(2,0)(1,-0.1)(0,-0.7)(-1,-0.1)(-2,0)} ;
\end{tikzpicture} \quad + \quad O(e^{-2S_0})\,.
\label{cylinder_fig}
\end{align}

Now, just as in the argument above, we can relate the quantities $\eta_{n,g}$ to the usual $n$ boundary genus $g$ amplitudes  $Z_{n,g}(\beta_1,\ldots,\beta_n)$ which compute the terms in the genus expansion of the cumulants of the partition function, by slicing along $2n$ geodesics. Explicitly,
\begin{equation}
    Z_{n,g}(\tau_1+\tau_1',\ldots,\tau_n+\tau_n') = \int \left(\prod_{k=1}^n d\ell_k d\ell_k' Z_\HD(\tau_k;\ell_k)Z_\HD(\tau'_k;\ell'_k)\right) \eta_{n,g}(\ell_1,\ldots,\ell_n;\ell_1',\ldots,\ell_n').
    \label{eq:Z-from-eta}
\end{equation}
 Here, we have the advantage that we can be very concrete about the geometries involved and the geodesic slicing since the path integral at fixed $n,g$ involves only a finite-dimensional integral over a moduli space of hyperbolic surfaces. For each such surface, one can easily verify the crucial fact that there is a unique geodesic homotopic to a given segment of boundary (in particular, the analysis reduces to a `trumpet' region lying between each asymptotic boundary and a closed geodesic of length $b$).%

Using \eqref{eq:Z-from-eta}, $ \eta_{n,g}(\ell_1,\ldots,\ell_n;\ell_1',\ldots,\ell_n')$ can be explicitly computed. Just like $ Z_{n,g}(\tau_1+\tau_1',\ldots,\tau_n+\tau_n')$ can be computed by using trumpet geometries glued to hyperbolic surfaces with arbitrary genus and geodesic boundaries, so too can the same algorithm be applied to $\eta_{n,g}(\ell_1,\ldots,\ell_n;\ell_1',\ldots,\ell_n')$. The only new ingredient we require is a modified trumpet,
\begin{align}
\psi(\ell, \ell';b)\equiv  \begin{tikzpicture}[baseline={([yshift=-0.1cm]current bounding box.center)}]
    \draw [thick, black, fill = gray, fill opacity=0.3] plot [smooth] coordinates { (-2,0)(-1,0.1)(0,0.7)(1,0.1)(2,0)} -- plot [smooth] coordinates {(2,0)(1,-0.1)(0,-0.7)(-1,-0.1)(-2,0)} ;
    \draw [thick, blue, fill = white] (0,0) circle  (0.4);
    \node at (0,-1) {$\ell_1$};
    \node at (0,1) {$\ell_2$};
    \node [blue] at (0,0) {$b$};
    \end{tikzpicture} &= \int dE \rho_\text{Trumpet}(E,b) \phi_E(\ell_1) \phi_E(\ell_2) \nonumber\\ 
    &= 4 K_0\left(4 \sqrt{e^{-\ell_2}+e^{-\ell_1}+2 e^{-\left(\ell_2+\ell_1\right) / 2} \cosh \frac{\color{blue} b}{2}}\right)\,,
\end{align}
where the trumpet density of states is given by $ \rho_\text{Trumpet}(E,b) = \frac{\cos(b\sqrt{E})}{\pi \sqrt{E}}$.
From this, we can obtain any connected contribution to $\eta_{n,g}(\ell_1,\ldots,\ell_n;\ell_1',\ldots,\ell_n')$ by gluing such trumpets to hyperbolic surfaces with $n$ geodesic boundaries,
\be \eta_{n,g}(\ell_1,\ldots,\ell_n;\ell_1',\ldots,\ell_n')_\text{conn.} =  \int \prod_{i=1}^n b_i\,db_i\, \psi(\ell_1, \ell_1';b_1) \cdots  \psi(\ell_{n}, \ell_{n}';b_n) V_{g, n}(b_1, \cdots, b_n)\,,
\ee
where $V_{g, n}$ is the volume of the moduli space of genus $g$ hyperbolic surfaces with geodesic boundaries of specified lengths.

As a concrete example, let us compute the leading order connected correction to $\overline{\braket{\ell'_1}{\ell_1} \braket{\ell'_2}{\ell_2}}$ shown in \eqref{cylinder_fig}. Using $V_{0, 2}(b_1,b_2) = \delta(b_1-b_2)/b_2$, the geometry appearing in \eqref{cylinder_fig} gives %
{\footnotesize 
\begin{align}
&\overline{\braket{\ell'_1}{\ell_1} \braket{\ell'_2}{\ell_2}}_\text{conn.} = \nonumber \\
& 16 \int {\color{blue} b} \,  \mathrm{d} {\color{blue} b} \, { K_0\left(4 \sqrt{e^{-\ell_1'}+e^{-\ell_1}+2 e^{-\left(\ell_1'+\ell_1\right) / 2} \cosh \frac{\color{blue} b}{2}}\right) K_0\left(4 \sqrt{e^{-\ell_2}+e^{-\ell_2'}+2 e^{-\left(\ell_2+\ell_2'\right) / 2} \cosh \frac{\color{blue} b}{2}}\right)}.
\end{align}
}
 By summing over $g$ and performing the integrals over the geodesic lengths $b_i$, the average product of inner-products can be simply expressed in terms of the spectral density correlator $\overline{\rho(E_1)\dots \rho(E_n)}$ in the JT ensemble, which is well known in the literature. From this, \eqref{eq:average-l-def} can be conveniently rewritten as
\be 
 \nonumber &\overline{\langle \ell_1'|\ell_1\rangle \cdots \langle \ell_n'|\ell_n\rangle} =\\ &\int \mathrm{d}E_1 \cdots \mathrm{d}E_n\, \overline{\rho(E_1)\cdots \rho(E_n)} \phi_{E_1}(\ell_1) \phi_{E_1}(\ell_1') \cdots  \phi_{E_n}(\ell_{n}) \phi_{E_n}(\ell_{n}')\,.
 \label{eq:average-prod-of-inner-prods}
 \ee

Since this holds for any $n$, from \eqref{eq:average-prod-of-inner-prods} we can deduce what the individual inner products are in each member of the JT ensemble with energies $\{E_i\}$. Namely, \eqref{eq:average-prod-of-inner-prods} implies
\begin{equation}
    \langle \ell'|\ell\rangle =e^{-S_0}\int dE \rho(E) \phi_{E}(\ell)\phi_{E}(\ell'),
\end{equation}
where $\langle \ell'|\ell\rangle$ and $\rho(E)$ are interpreted as random variables. More precisely, our argument shows that this equality holds for all moments to all orders in the genus expansion. Rewriting the density $\rho(E) = \sum_i \delta(E-E_i)$ in terms of a discrete spectrum, this is the same as \eqref{eq:llIP}.

 We note that it is tricky to write a direct formula for geodesic states in terms of energy eigenstates analogous to \eqref{eq:ilIP} in the ensemble language (unless we simply interpret the energies $E_i$ in \eqref{eq:ilIP} as random variables). This is due to a conflict between continuum and discrete normalized energy eigenfunctions. For similar reasons, it is subtle to attempt to argue directly for the relationship between energy eigenstates and geodesic states, which is why we choose to go via the thermofield double states where such issues do not arise.

\subsection{The Hamiltonian is unmodified}

From the decomposition of $|\ell\rangle$ into energy eigenstates, it is immediate that the Schr\"odinger operator \eqref{eq:Hpure} continues to generate time translations on superpositions of geodesic states non-perturbatively. More precisely, we mean that
\begin{equation}
   |\psi\rangle = \int d\ell \, \psi(\ell) |\ell\rangle \implies H|\psi\rangle = \int d\ell \left(-\tfrac{1}{2}\psi''(\ell)+2e^{-\ell}\psi(\ell)\right)|\ell\rangle \label{ham:LV}
\end{equation}
for sufficiently nice wavefunctions $\psi$ (in particular, decaying at large $\ell$). To verify this, take the inner product of each side with an energy eigenstate $\langle n|$, and use the fact that $\langle n|\ell\rangle$ is an eigenstate of the differential operator. Some care is needed with this and similar statements because the wavefunction $\psi(\ell)$ is not equal to $\langle \ell|\psi\rangle$ non-perturbatively, and (as we will explain in the next section) it is not even uniqely determined from the state $|\psi\rangle$.

This argument is a little abstract, but we can understand it intuitively with a more direct derivation from the path integral. To do this, we can generalize the above calculation of the inner product to compute a matrix element $\langle \ell'|e^{-\tau H_L}|\ell\rangle$, which means including insertion of a piece of Euclidean boundary on the left side (the right side would give the same answer in the end since $H_L=H_R$). Now, for a given geometry that appears in the path integral, consider continuously deforming the past boundary geodesic (defining $|\ell\rangle$) by moving its right endpoint forward in Euclidean time while keeping it as a geodesic. Under the same conditions as above, this process will sweep out a `wedge' bounded by geodesics of length $\ell,\ell''$ and the asymptotic boundary segment of length $\tau$. The path integral of this wedge computes the matrix elements of the time-evolution operator in the perturbative theory $\langle \ell''|e^{-\tau H_L}|\ell\rangle_0$. The non-perturbative matrix element is recovered by multiplying by $\langle \ell'|\ell''\rangle$ and integrating over $\ell''$. But the time dependence all appears in the piece which is independent of the non-perturbative corrections (analogous to $Z_\HD$ above), which gives us the above (unmodified perturbative) representation for the Hamiltonian. It is essential for this result that in every possible spacetime and for every possible initial geodesic, by moving an endpoint, we can obtain a continuous family of geodesics that never encounter interesting topology or other non-perturbative features:\footnote{See discussion in section \ref{sec:discussion}.}

\begin{align}
    \langle\ell'|e^{-H_L {\color{dgreen} \tau} }|{\color{red} \ell}\rangle = \int d\ell'' \,\langle\ell'|\ell''\rangle \, \langle \ell''|e^{-H_L{\color{dgreen} \tau} }|{\color{red} \ell}\rangle_0\  = 
    \begin{tikzpicture}[baseline={([yshift=0cm]current bounding box.center)}]
        \draw[thick,fill=lightgray, fill opacity=0.3] (0,0) circle (2);
        \draw [thick, blue, fill = blue, fill opacity=0.3] plot [smooth] coordinates { (-2,0)(-1,0.1)(0,0.35)(1,0.1)(2,0)} -- plot [smooth] coordinates {(2,0)(1,-0.1)(0,-0.35)(-1,-0.1)(-2,0)} ;
        \node[blue] at (0,.6) {$\ell'$};
        \node[blue] at (-1,-.4) {$\ell''$};
        \node[red] at(-0.1,-1) {$\ell$};
       \draw[thick, red] (2,0) arc (90:140:4.25);
       \draw[ultra thick,dgreen] (-2,0) arc (179:232:1.95);
        \crosscap at (0,0);
        \node[dgreen] at (-1.7,-.4) {$\tau$};
    \end{tikzpicture}\,.  \label{disk_crosscap}
\end{align}
In (\ref{disk_crosscap}), the cross symbol denotes any possible topology that could contribute to $Z(\beta)$.  Above, we have \textit{ not} inserted the resolution of the identity since the middle term contains $\langle \ell''|e^{-{\color{dgreen} \tau} H_L}|{\color{red} \ell}\rangle_0$ which is computed in the perturbative theory -- in the figure, this is illustrated through the fact that the region between $\ell$ and $\ell''$ does not contain any topology change.

\section{Null states: redundancies in the bulk Hilbert space}\label{sec:null}

In this section, we highlight a property of our inner product \eqref{eq:llIP} of central importance: it is highly degenerate. Our formulas allow us to be extremely explicit about this feature and the associated phenomenon of null states.

Consider a general superposition of geodesic states with wavefunction $\psi(\ell)$,
\begin{equation}
    |\psi\rangle = \int d\ell \, \psi(\ell) |\ell\rangle.
\end{equation}
The properties of the inner product of such states become manifest if we rewrite $\psi(\ell)$  as a superposition of scattering wavefunctions $\phi_E(\ell)$, defining $\hat{\psi}(E)$ by a sort of `Fourier transform' of $\psi$ adapted to the Hamiltonian:
\begin{equation}
    \hat{\psi}(E) := \int d\ell \,\phi_E(\ell) \psi(\ell) \iff \psi(\ell) = \int dE \,\rho_0(E) \phi_E(\ell) \hat{\psi}(E).
\end{equation}
Then, the norm of the state $|\psi\rangle$ is given by
\begin{equation}
    \langle\psi|\psi\rangle = e^{-S_0}\sum_{E_i}|\hat{\psi}(E_i)|^2.
\end{equation}
This is manifestly non-negative. However, it depends on the wavefunction only through $\hat{\psi}$ evaluated at the discrete values $E_i$.

A consequence is that there are many `null states' $|\chi\rangle$ having zero inner product with every state, whose wavefunction vanishes at all these discrete energies but can otherwise be nonzero:
\begin{equation}
    \hat{\chi}(E_i) =0 \quad \forall i \iff \langle\psi|\chi\rangle = 0 \quad \forall|\psi\rangle\in\hilb.
\end{equation}
Such a state must be regarded as equivalent to the zero state in the physical Hilbert space $\hilb$. This means that the representation of a given state $|\psi\rangle$ by a wavefunction $\psi(\ell)$ is not unique. We can add any null wavefunction $\chi(\ell)$, or equivalently modify $\hat{\psi}(E)$ in any way we please as long as its values on the discrete set $E_i$ remain unchanged, and arrive at the same physical state.

Since the energy eigenvalues $E_i$ have exponentially small typical spacing $\Delta E \sim e^{-S}$, null state energy wavefunctions $ \hat{\chi}(E)$ must be very narrowly supported or highly oscillatory on this scale. Using the intuition relating fine-scale features of a function to the decay rate of its Fourier transform, we can expect that null states will typically involve states $\chi(\ell)$ supported at exponentially long lengths $\ell \sim e^{S}$. This relates to previous discussions about the breakdown of semi-classical physics when the Einstein-Rosen bridge grows exponentially long \cite{Susskind:2015toa,Stanford:2022fdt}. We will see this intuition borne out by more precise calculations in section \ref{sec:a-length-operator}.

Another way to express the existence of null states is by the overcompleteness of the $|\ell\rangle$ basis. The modified inner product means that $\int d\ell \,|\ell\rangle\langle \ell|$ is no longer equal to the identity, but will instead be badly divergent (we give a geometric interpretation in the next section). Explicitly, the matrix elements of this operator between two energy eigenstates are formally given by $\frac{\delta(E_i-E_j)}{e^{S_0}\rho_0(E_i)}$ which gives an infinite answer for $E_i=E_j$. The correct resolution of the identity $\sum_i|i\rangle\langle i|$ does not have a unique representation in the $\ell$ basis. Similarly, one cannot simply take a trace in the $\ell$-basis. For example,  $\mathrm{Tr}_{\mathcal{H}_\text{bulk}} e^{-\beta_L H_{L} - \beta_R H_R} \neq \int d\ell\, \bra{\ell} e^{-\beta_L H_{L} - \beta_R H_R} \ket{\ell}$. The quantity on the RHS results in a rectangle with two geodesic boundaries of length $\ell$ that should be naively glued to each other to obtain a cylinder geometry. However, this does not correctly account for the action of the mapping class group, and the integral over $\ell$ produces a divergence instead of correctly reproducing the contribution of the cylinder geometry. Thus, from this perspective, the overcompleteness of the $\ell$-basis can be linked to not implementing the quotient by the mapping class group correctly at the path integral level. We will further explain how to correctly interpret the trace over states in this overcomplete basis in appendix \ref{sec:Hilbert-space-dimensions}.

\section{Defining non-perturbative bulk operators}
\label{sec:defining-non-perturbative-bulk-ops}

\subsection{Challenges and ambiguities}

A non-perturbative description of the bulk Hilbert space in geometric language promises to be extremely valuable in studying bulk physics. Indeed, since we can describe states as wavefunctions of bulk quantities, we might expect that operators defined perturbatively as linear operators on functions $\psi(\ell)$ continue to be useful as observables non-perturbatively. However, the modifications to the inner product mean that things are not quite so simple, and the challenges are particularly severe in the presence of null states.

To illustrate this, consider the construction of a simple length operator measuring the geodesic distance between the two boundaries. Perturbatively this just acts by multiplication, $\psi(\ell)\mapsto \ell \psi(\ell)$, meaning that the geodesic states $|\ell\rangle$ are eigenstates. We might be tempted to define an operator like this non-perturbatively, by
\begin{equation}
\label{eq:definition-of-length-operator}
    \hat{\ell} |\ell\rangle \stackrel{?}{=} \ell |\ell\rangle.
\end{equation}
However, this does not give a well-defined operator! The reason is that it does not map null states to null states. Two different wavefunctions describing the same physical state will typically be mapped to different physical states; if $\chi(\ell)$ is a null wavefunction with zero norm, then $\ell\chi(\ell)$ will typically have positive norm. And even without the issue of null states, such an operator would not be Hermitian with respect to the physical inner product and so would not be a candidate for an observable. 

\subsection{A class of length operators}

Since \eqref{eq:definition-of-length-operator} does not define an operator, we instead consider a large class of well-defined Hermitian operators, which are designed to reproduce the disk-level length operator in the absence of non-perturbative corrections.

For a real-valued function $\cF(\ell)$ we can define
\begin{equation}
    \widehat{\mathcal{F}(\ell)} = \int d\ell\, \cF(\ell) \ketbra{\ell}. 
\end{equation}
This avoids the above challenges since it is manifestly Hermitian and automatically annihilates null states. In particular, we can write the matrix elements in the energy basis using \eqref{eq:ilIP} as
\begin{equation}\label{eq:opMatrixElements}
    \langle j|\widehat{\mathcal{F}(\ell)}|i\rangle = e^{-S_0} \int d\ell\, \cF(\ell) \phi_{E_i}(\ell)\phi_{E_j}(\ell)
\end{equation}
if this integral converges for all $E_{i,j}$. This equation is valid only for discrete choices of $E_i, E_j$ that are in the list of energy eigenvalues $\{E_i\}$ for a given instance of the ensemble; it has zero matrix elements for any state $\ket{E}$ where $E \notin \{E_i\}$. Hence, the operator annihilates a null state, which has vanishing wavefunction on the particular $\{E_i\}$ of the ensemble draw.

If $\cF(\ell)$ is invertible, we can then define a candidate length operator as\footnote{There is one subtlety. For sufficiently nice functions $\cF$, our definition for $\widehat{\mathcal{F}}$ gives a symmetric operator defined on the dense subspace of finite linear combinations of energy eigenstates $|i\rangle$. But we can only unambiguously define $\cF^{-1}(\widehat{\mathcal{F}})$ if this operator is essentially self-adjoint, which could fail. Nonetheless, $\widehat{\mathcal{F}}$ certainly has some self-adjoint extension because it has real matrix elements. Furthermore, if $\cF(\ell)$ is a positive function (like $e^{-\Delta \ell}$), then $\widehat{\mathcal{F}}$ is a positive operator, so the Friedrichs extension gives us a canonical definition.}
\be
\hat \ell_\cF := \cF^{-1}(\widehat{\mathcal{F}(\ell)}) =\cF^{-1} \left(\int d\ell\, \cF(\ell) \ket{\ell}\bra{\ell}\right).
\label{eq:general-operator-definition}
\ee
At the disk level, since the states $\ket{\ell}$ are orthonormal, all such operators simply become $(\hat \ell_\cF)_\text{Disk} = \int d\ell \, \ell \, \ket{\ell}\bra{\ell}$, the correct length operator at the disk level. However, non-perturbatively, the operator will depend on the choice of $\cF$, so we have an ambiguity in the definition of the length.

From a geometric standpoint, this is not surprising. Once higher topology geometries are included in the path integral, the length of the black hole interior is also ambiguous. On any surface with genus $g>0$, there are an infinite number of geodesics going between any two boundary points (even if we require that they do not self-intersect). One strategy to define a length operator is to sum over all of these geodesic, with some relative weighting between them. As we shall emphasize shortly, the ambiguity in choosing the function $\cF$ in \eqref{eq:general-operator-definition}  is intimately related to the geometric ambiguity in weighing the different geodesics that exist between any two boundary points on a higher genus surface.

We consider two specific examples. First, take the simplest possibility $\cF(\ell)=\ell$:
\be 
\label{eq:naive-el}
\hat \ell  = \int_{-\infty}^{\infty} d\ell\, \ell\, \ket{\ell} \bra{\ell} \,.
\ee
This does not give a convergent result acting on any state as a result of the overcompleteness of the states $\bra{\ell}$. For this reason, one might think that the naive definition in \eqref{eq:naive-el} does not give a sensible operator. However, when we calculate the matrix elements on the energy eigenbasis \eqref{eq:opMatrixElements}, we can make sense of the off-diagonal elements (for example, by introducing a cutoff factor $e^{-\Delta \ell}$ in the integrand and taking $\Delta\to 0$), giving
\begin{equation}
    \langle j|\hat{\ell}|i\rangle = -e^{-S_0} \frac{(2\pi)^2}{(E_i- E_j)\sinh(\pi(\sqrt{2E_i}-\sqrt{2E_j}))\sinh(\pi(\sqrt{2E_i}+\sqrt{2E_j}))}, \qquad (i\neq j)
\end{equation}
while the divergence comes only from the diagonal ($i=j$). This means that while $\hat{\ell}$ itself is divergent, $[H,\hat{\ell}]$ (with vanishing diagonal matrix elements) is unambiguously defined, and hence so is the difference $\hat \ell(t) -\hat \ell(0)$ between different times.

In fact, this is precisely the length operator introduced in \cite{Iliesiu:2021ari}, and we will see further comparison when we consider the path integral interpretation below. They studied non-perturbative corrections to the matrix elements $\langle \tau_2|\hat{\ell}|\tau_1\rangle$ of this operator between thermofield double states with real-time evolution, taking $\tau_1 = \frac{\beta}2 + i t$ and $\tau_2 = \frac{\beta}2 - it$. While this is infinite, the $t$-dependence is insensitive to this divergence. The time-dependent finite contribution grows linearly for times $\beta\ll t \ll e^{S_0}$ and plateaus at very late times, $t\sim e^{S_0}$.  This implies that at such late times, the matrix elements of the velocity operator associated to this length vanish, which is consistent with an average non-zero probability for an infalling observer encountering a white hole instead of a black hole state. %

While $\hat{\ell}$ is an interesting operator to consider, it is preferable to have a definition without infinities arising. To get finite matrix elements requires us to suppress large values of $\ell$, and a natural choice is $\mathcal{F}(\ell) = e^{-\Delta \ell}$ for some $\Delta>0$:
\be
\label{eq:two-pt-function-op-definition}
\hat{e^{-\Delta \ell}} = 
 \int_{-\infty}^{\infty} d\ell\, e^{-\Delta \ell}\, \ket{\ell} \bra{\ell} \,, \qquad \hat \ell_{\Delta} \equiv -\frac{1}{\Delta} \log{\left(\hat{e^{-\Delta \ell}}\right) }.
\ee
This is similar  (up to a technical subtlety) to the length as determined by the two-point function of a probe field of dimension $\Delta$, which we comment on in the next subsection. We study this operator numerically in the next subsection.

\subsection{Path integral interpretation}

The definition \eqref{eq:general-operator-definition} of $\widehat{\mathcal{F}(\ell)}$ is somewhat abstract, but in fact, the expectation value of these operators has a very natural path integral interpretation. To see this, consider the matrix elements between thermofield double states  $\ket{\tau_1}$ and $\ket{\tau_2}$, taking the average over the JT ensemble:
\be 
\label{eq:naive-geodesic-length-operator}
\overline{\bra{\tau_1} \widehat{\mathcal{F}(\ell)} \ket{\tau_2}} = \int d\ell\, \mathcal{F}(\ell) \, \overline{\braket{\tau_1}{\ell} \braket{\ell}{\tau_2}}\,,
\ee
Evaluating the integrand amounts to performing the path integral over all geometries (connected and disconnected) with two boundaries, the first being a union between a geodesic with renormalized length $\ell$ and an asymptotic boundary segment with renormalized length $\tau_1$, and the second similarly consisting of a geodesic with renormalized length $\ell$ and an asymptotic boundary with renormalized length $\tau_2$. For example, all contributions from surfaces whose total genus is $2$ are schematically given by,\footnote{One can similarly study the geometries that contribute to the matrix element $\bra{\ell_1}\hat \ell \ket{\ell_2}$ by selecting similar geometries to those in \eqref{eq:naive-geodesic-length-operator-example-geometries} where the green boundary segments are replaced by homotopic geodesic segments.  From this, we can explicitly see that $\bra{\ell_1}\widehat{\mathcal{F}(\ell)} \ket{\ell_2} \ne \mathcal{F}(\ell_2) \bra{\ell_1} \ket{\ell_2}$ since the red geodesic need not coincide with the geodesic defining the ket, and therefore the $\ket{\ell}$ states are not eigenvectors of these operators.}
\be  
\label{eq:naive-geodesic-length-operator-example-geometries}
\overline{\bra{\tau_1} \widehat{\mathcal{F}(\ell)} \ket{\tau_2}} \supset \int \mathrm{d}{\color{red}{\ell}} \, \mathcal{F}({\color{red}{\ell}})\bigg(
\begin{tikzpicture}[baseline={([yshift=-.5ex]current bounding box.center)}, scale=0.5]
 \pgftext{\includegraphics[scale=0.25]{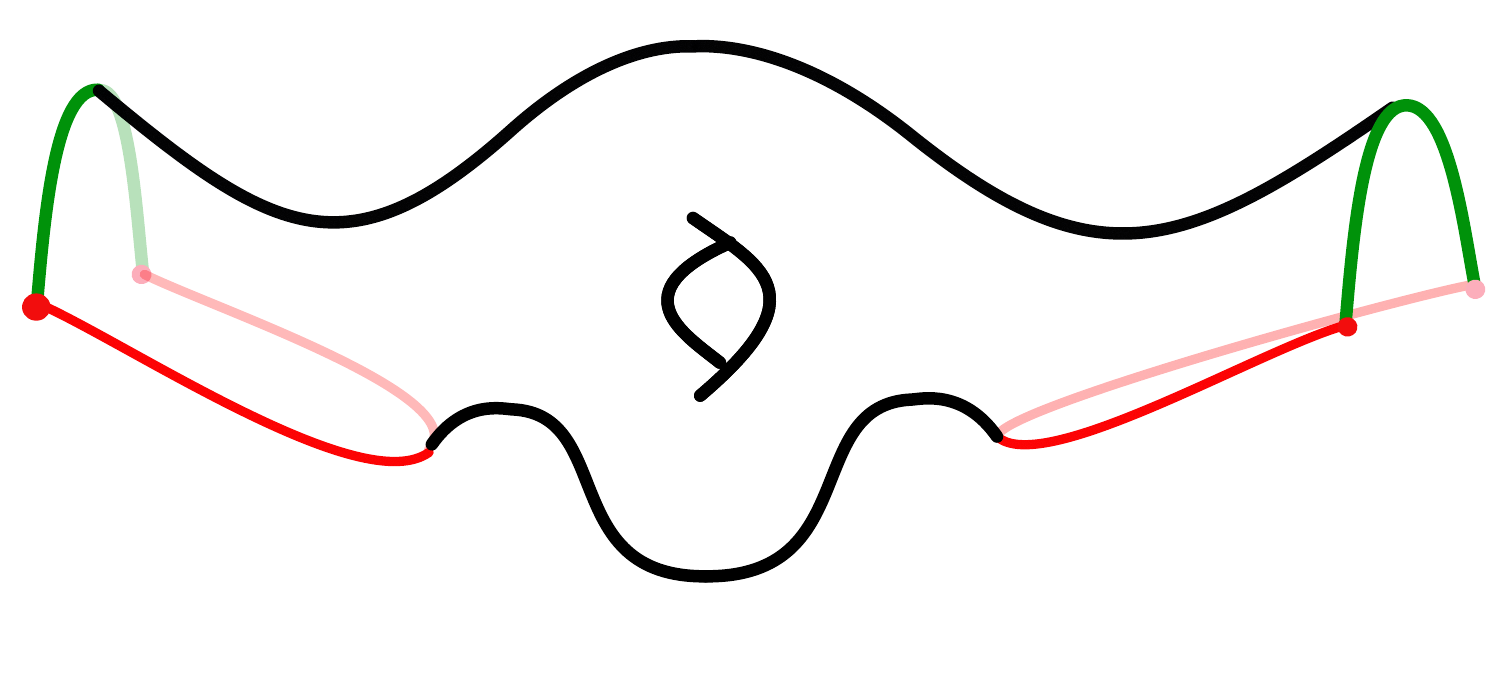}} at (0,0); 
  \end{tikzpicture}
  \hspace{0.1cm} 
  +
    \hspace{0.1cm} 
  \begin{tikzpicture}[baseline={([yshift=2ex]current bounding box.center)}, scale=0.5]
 \pgftext{\includegraphics[scale=0.25]{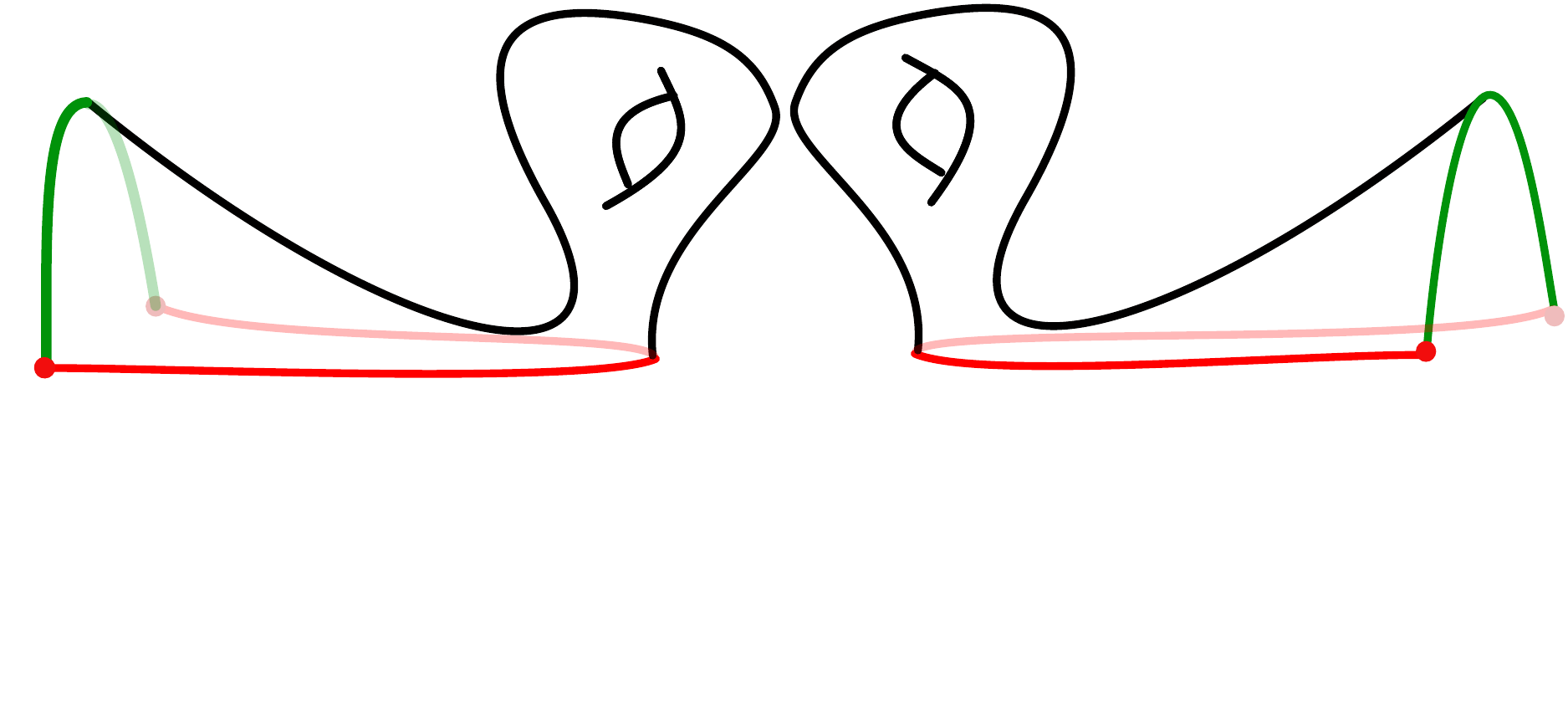}} at (0,0); 
  \end{tikzpicture}
    \hspace{0.1cm} 
  +
    \hspace{0.0cm} 
    \begin{tikzpicture}[baseline={([yshift=-.5ex]current bounding box.center)}, scale=0.5]
 \pgftext{\includegraphics[scale=0.25]{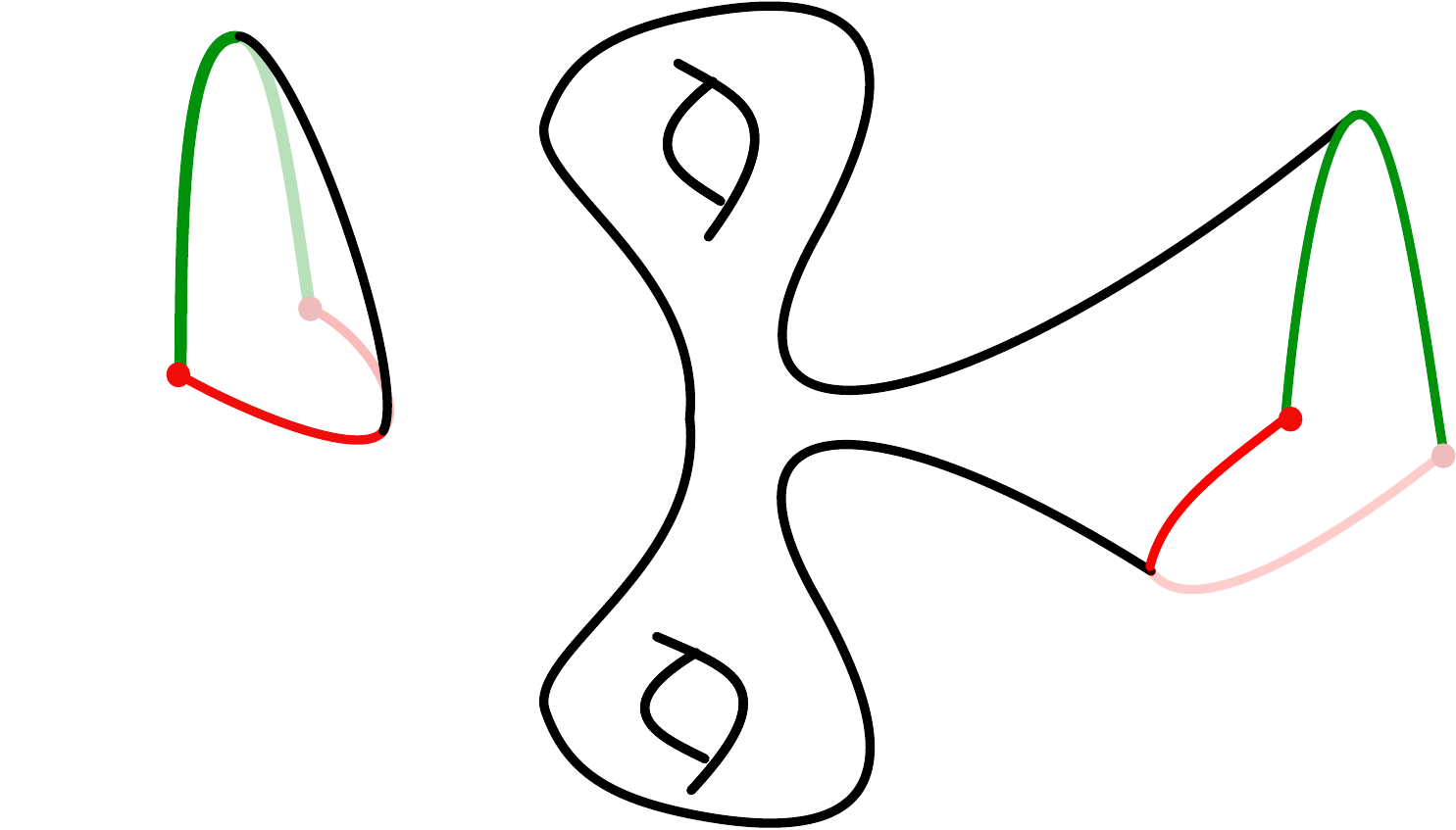}} at (0,0); 
  \end{tikzpicture}\bigg)
  \,,
\ee
where the {\color{dgreen} green curves} are asymptotic curves on which the dilaton is fixed, while the  {\color{red} red segments} are the geodesics of length $\ell$. These geometries are precisely those that contribute to the sum over all non-self-intersecting geodesics $\gamma$ going between two points on the boundary, separated by the curves of length $\tau_1$ and $\tau_2$, when properly taking into account the quotient by the mapping class group of the entire spacetime \cite{Saad:2019pqd, Blommaert:2020seb, Iliesiu:2021ari}. In other words, we can write \eqref{eq:naive-geodesic-length-operator} as a path integral
\be 
\overline{\bra{\tau_1} \widehat{\mathcal{F}(\ell)} \ket{\tau_2}} = \int%
\frac{Dg_{\mu\nu} D\Phi}{\text{Diffs}} \left(\sum_{\gamma} \mathcal{F}(\ell_\gamma)\right) e^{-I_\text{JT}}\,,
\label{eq:path-integral-expression-sum-over-geodesicsPI}
 \ee
where the integral over metrics includes a sum over all topologies, and the states determine the boundary conditions. This path integral definition of a length operator with $\mathcal{F}(\ell)=\ell$ was proposed in \cite{Iliesiu:2021ari}. 

In the case $\mathcal{F}(\ell) = e^{-\Delta \ell}$, it is tempting to use this path integral to identify $\hat{e^{-\Delta \ell}}$ with a bilocal  $O_L O_R$ of a dimension $\Delta$ boundary operator inserted on the left and right side. This is because the two-point function of a free field on a given local AdS$_2$ geometry can be computed by a sum $e^{-\Delta \ell_\gamma}$ over all geodesics $\gamma$ between the insertions. However, this is not quite the same as the insertion in \eqref{eq:path-integral-expression-sum-over-geodesicsPI}, because the sum there runs only over non-self-intersecting geodesics that split the geometry into two parts, while for the two-point function, the sum runs over all geodesics. This identification only exactly holds on the disk topology. Nevertheless, we might reasonably expect a close relation between $\hat{e^{-\Delta \ell}}$ and the boundary observable $O_L O_R$, particularly if $\Delta$ is large.

An important subtlety arises in this path integral interpretation when we consider not just expectation values but multiple insertions of these operators (or more general non-linear functions). For example, while an expectation value of $\hat{\mathcal{F}(\ell)}$ is computed by inserting a sum over non-self-intersecting geodesics $\sum_\gamma \mathcal{F}(\ell_\gamma)$ into the path integral, the expectation value of $\hat{\mathcal{F}(\ell)}^2$ is \emph{not} given by inserting $\left(\sum_\gamma \mathcal{F}(\ell_\gamma)\right)^2$. Instead, it is obtained by inserting a sum over pairs of non-self-intersecting geodesics $\gamma_{1,2}$, where $\gamma_2$ is constrained to lie completely to the future of (or to coincide with) $\gamma_1$. Similarly, inserting an $n$-th power corresponds to a path integral insertion of a sum over time-ordered $n$-tuples of geodesics. Heuristically, this happens because the path integral always computes products of operators in Euclidean time order. This fact makes the geometric interpretation of operators $\hat{\mathcal{F}(\ell)}$ complicated, and more so for non-linear functions like $\hat{\ell}_\mathcal{F}$. For example, one might naively think that $\frac{1}{\Delta} \log \left(\int d\ell e^{-\Delta \ell} \ket{\ell}\bra{\ell} \right)$ captures the minimal length geodesic in the limit $\Delta \to \infty$. However, this is not the case.\footnote{We thank D.~Stanford and Z.~Yang for discussions on this point.} %

\section{Case-study I: a length operator}
\label{sec:a-length-operator}

\def\i{\mathrm{i}}
In this section, we study various aspects of the operator $\hat{\ell}_\Delta$ numerically. First we consider the JT spectrum, truncated to some maximum energy $E$. We obtain this by generating a gaussian Hermitian matrix (drawn from the Gaussian unitary ensemble) and then rescaling the eigenvalues to obtain the desired density of states, see Appendix \ref{app:Numerics} for details.
\begin{figure}[h!]
    \centering
    \includegraphics[width=0.6\columnwidth]{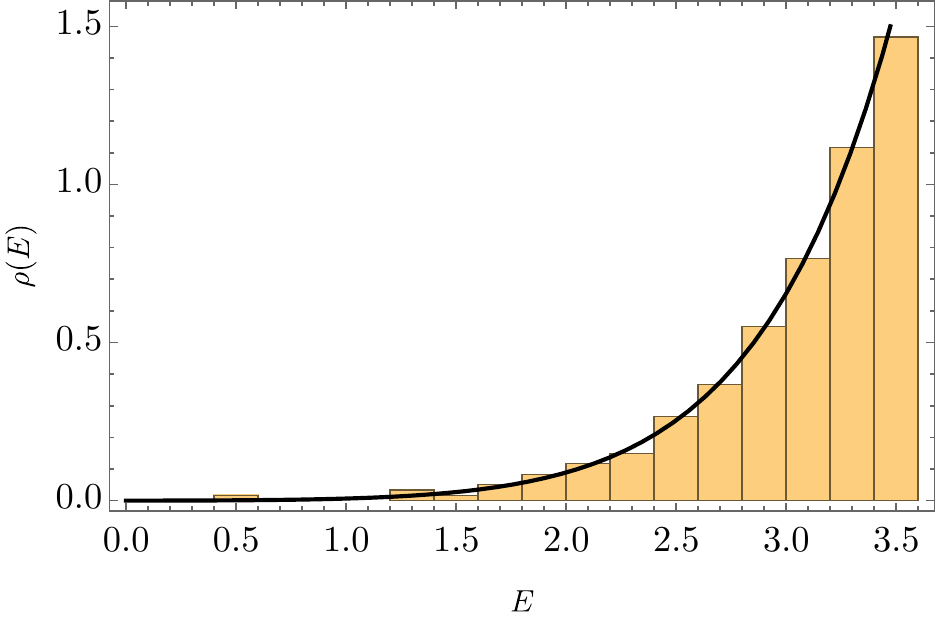}
    \label{fig:density_states}
        \caption{Spectrum of the JT Hamiltonian for one draw of the ensemble. The black curve is $\rho \propto \sinh (2 \pi \sqrt{2E})$, with the proportionality constant determined by requiring that there are 300 eigenvalues in the window.}
\end{figure}
We define the non-perturbative exponential operator in the obvious way and then define the length by computing the matrix logarithm:
\begin{align}
    \bra{i} \widehat{e^{-\Delta \ell}} \ket{j} = \frac{ \Gamma(\Delta \pm \i \sqrt{2E_i} \pm \i \sqrt{2E_j})}{4^{\Delta-1}\Gamma(2\Delta)}, \quad \widehat{\ell}_\Delta = - \frac{1}{\Delta} \log (\widehat{e^{-\Delta \ell}})_{E < E_\text{cutoff}}\,, \label{numerical_ldelta} 
\end{align} %
where the $\pm$ mean we take a product over four $\Gamma$ functions with all sign combinations.
Note that this numerical implementation of $\hat{\ell}_\Delta$ depends on the maximum energy cutoff of the JT spectrum.

\subsection{The Hamiltonian in the length eigenvector basis}
\begin{figure}[h]
    \centering
    \includegraphics[width=0.7\columnwidth]{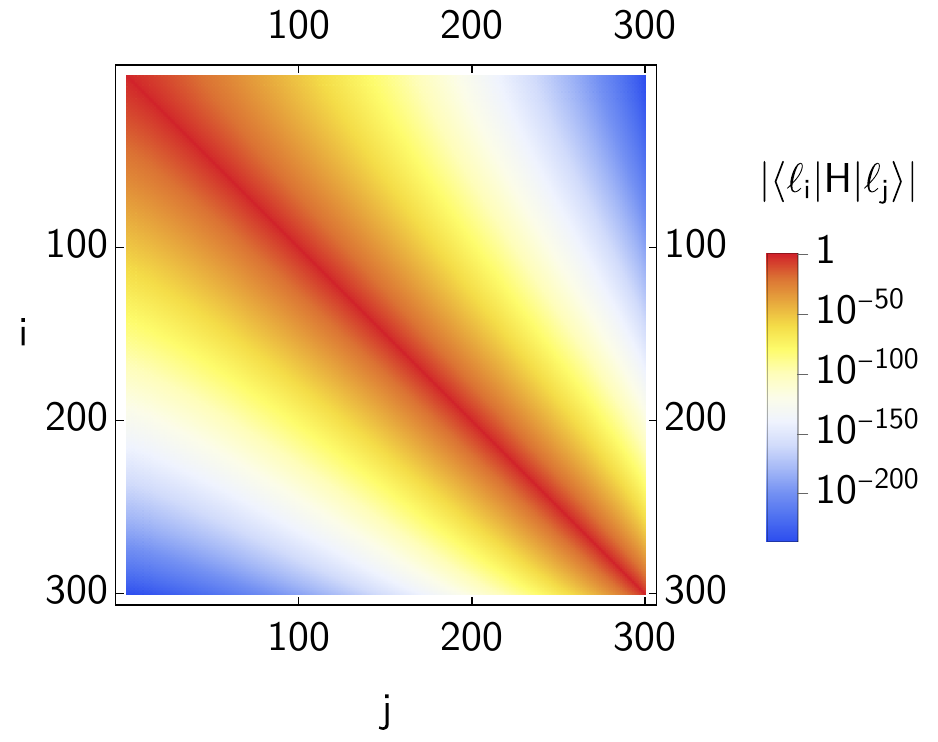}
    \caption{Matrix elements $|\mel{\ell_i}{H}{\ell_j}|$ of the Hamiltonian in the eigenbasis of the non-perturbative length operator $\hat\ell_\Delta$. Note that the colors are on a log scale. This plot was generated for a single draw of the ensemble.}
    \label{fig:hij}
\end{figure}

One can also consider the Hamiltonian in the non-perturbative length eigenvector basis 
\be
_\Delta\!\bra{\ell_a} H \ket{\ell_b}_\Delta =  \sum_i E_i \,\,_\Delta\!\bra{\ell_a }\ket{i}\bra{i}\ket{\ell_b}_\Delta, \quad \hat{\ell}_\Delta \ket{\ell_c}_\Delta = \ell_c \ket{\ell_c}_\Delta .
\ee
In the semi-classical approximation, there is no difference between $\ket{\ell}_\Delta$ and $\ket{\ell}$; therefore, in the semi-classical approximation the matrix $H$ only has overlaps between infinitesimally close $\ell_1$ and $\ell_2$ (given by the kinetic term $H \sim - \tfrac{1}{2} \partial_\ell^2$, see \eqref{eq:Hpure}).
In Figure~\ref{fig:hij}, we plot the absolute value of the matrix elements.\footnote{ Note that the matrix elements $e^{-\Delta \ell}$ are real and symmetric, see \ref{numerical_ldelta} and hence the eigenvectors $\bra{E}\ket{\ell_j}_\Delta$ are real. We find that while the absolute values of the matrix elements are close to self-averaging, each matrix element has a random sign. }

We see that the Hamiltonian now contains non-zero off-diagonal elements, although the numerical values of these matrix elements rapidly decay away from the diagonal. 
In the non-perturbative description, we see that the Hamiltonian now becomes non-local in the length eigenspace. We can say that we have traded the wormhole effects for non-locality in ``superspace,'' i.e., in the space of possible spatial metrics. These small off-diagonal elements can be interpreted as enhancing the probability that the wormhole dramatically changes its length, even after a very small amount of Hamiltonian time evolution. It would be interesting to study this further (in particular, to study the statistics of this matrix numerically and analytically) and to see whether this Hamiltonian gives an effective description of time evolution where we have ``integrated out'' effects like the emission of baby universes, etc., \cite{Stanford:2022fdt}.

Note that here, we are choosing to write the non-perturbative Hamiltonian on a basis where the inner product is trivial. By conservation of evil, this makes the matrix elements complicated. Alternatively, we could choose to write the Hamiltonian in the original $\ket{\ell}$ basis, where it retains its simple Liouville form. The disadvantage is then that the inner product is complicated in this basis.

\subsection{Evolution of the wormhole length }
Using our numerical implentation of the length, we can compute its evolution in a particular state. We choose a ``microcanonical'' version of the thermofield double, e.g.,
\begin{align}
   e^{\i H_R t} \ket{\mathrm{TFD},\bar{E},\sigma} = \frac{1}{Z} \sum_E e^{\i E t} e^{-\frac{(E-\bar{E})^2}{2\sigma^2}} \ket{E}_L\ket{E}_R.
\end{align}
We expect that the results should have a well-defined $\tau$-scaling limit \cite{Saad:2022kfe}, defined in this context as 
\begin{align}
    t \to \infty, \quad e^{S_0} \to \infty, \quad \tau = t e^{-S_0} = \text{fixed}
\end{align}
We expect that the length should plateau at a value given by $\sim e^{S_0}$, so it is useful to work with $\ell_\Delta / N$, where $N \propto e^{S_0}$ is the dimension of the Hilbert space once we truncate the energy spectrum.
We plot this for various different values of $N$ below. Our results are consistent with $\tau$-scaling. More precisely, we expect a plateau length that scales with $e^{S(\bar{E})}$. Consistent with this expectation, in Appendix \ref{app:Numerics}, we show that lowering $\bar{E}$ gives a plateau where the wormhole length is shorter.
\begin{figure}[H]
    \centering
    \includegraphics[width=0.9\columnwidth, trim = 0 0cm 0cm 0, clip=true]{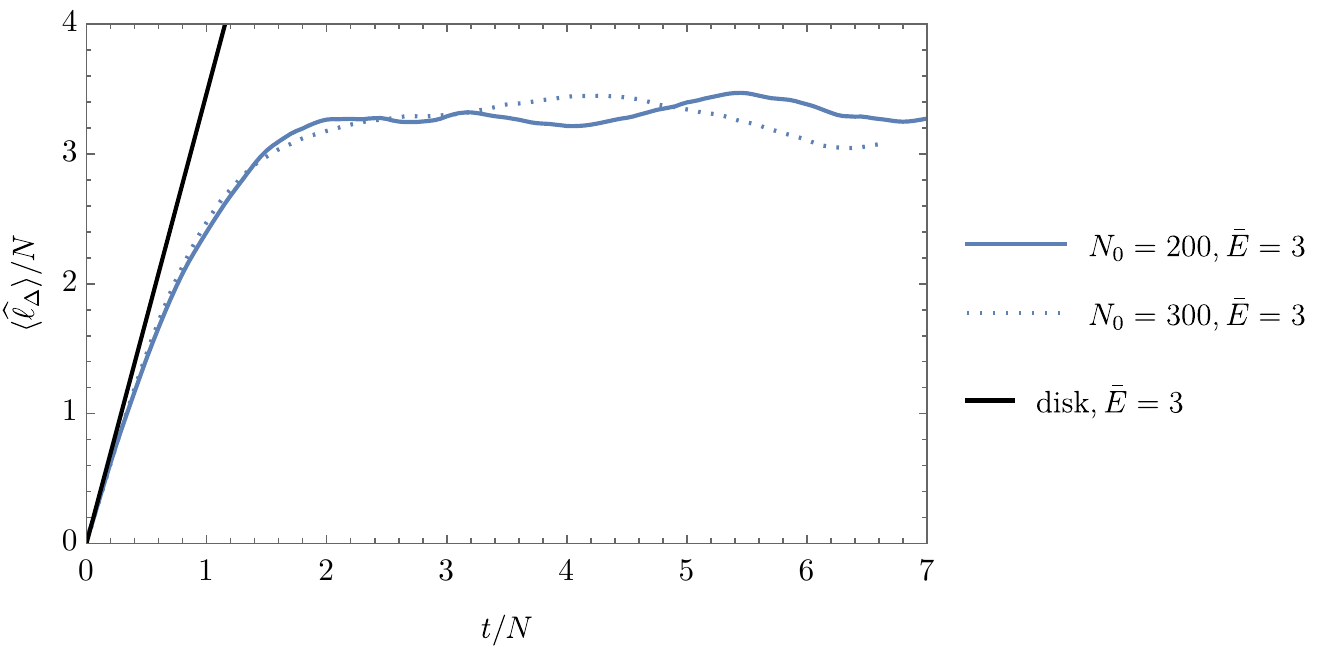}
        \caption{Numerical demonstration of the $\tau$ scaling limit. We plot the average length as a function of $t/N \sim \tau = e^{-S_0} t$. We show two draws of the ensemble with different values of $e^{S_0}$. After an appropriate rescaling, the two curves are quite similar. We also show the slope predicted by the disk.
        In Appendix \ref{app:Numerics}, we show a version of this diagram where we vary more parameters, see Figure~\ref{fig:length_expectation_tau_complete}. \label{fig:length_expectation_tau}}
\end{figure}
Of course, besides the average value of the length, one may also compute the wavefunction in the length basis. This is displayed in Appendix \ref{app:Numerics}, see Figure~\ref{fig:probDistributionVelocity}.

\subsection{The wormhole velocity }

\begin{figure}[H]
    \centering
    \includegraphics[width=0.6\columnwidth]{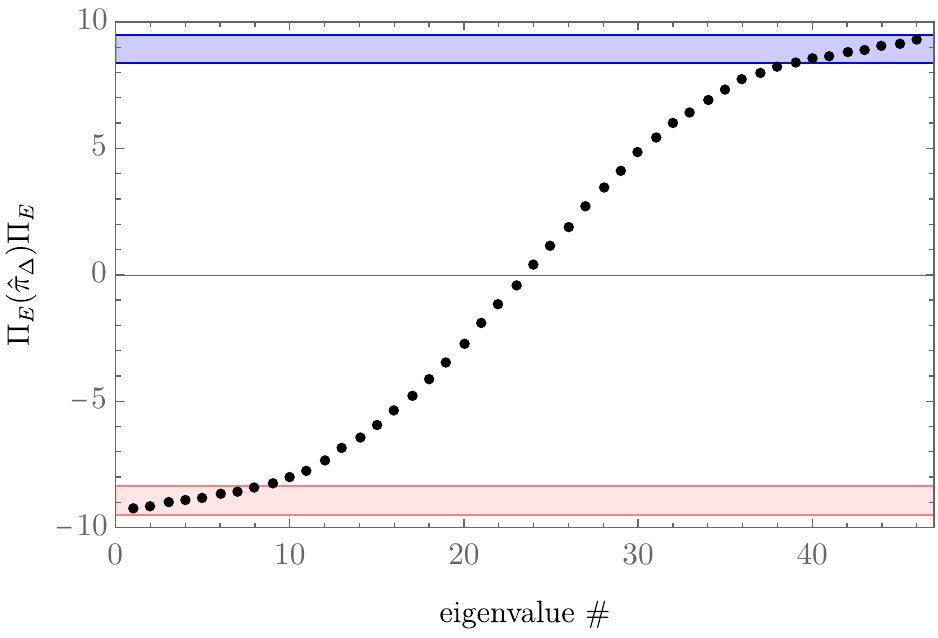}
    \label{fig:speed_spectrumproj}
        \caption{The spectrum of the velocity operator projected into the energy window around $\bar E$. The blue/red horizontal lines represent the semi-classical velocity of the wormhole for the black hole/white hole $v^2 \propto E$. \label{velocity_spec}}
\end{figure}

We may also consider the velocity operator, which should directly tell us whether the wormhole is growing or shrinking:
\begin{align}
    \hat\pi_\Delta = \i [H, \widehat{\ell}_\Delta ].
\end{align}
We can compute the eigenvalues of the velocity operator. Semi-classically $\pi$ is a continuous operator, but non-perturbatively $\pi$ has discrete eigenvalues.
The distribution of eigenvalues will be sensitive to the cutoff $E_\text{max}$. A quantity with a clearer interpretation is the non-zero eigenvalues of the operator
\begin{align}
    \Pi_{E} \hat\pi_\Delta  \Pi_{E},
\end{align}
where $\Pi_E$ is a projector onto an energy window $E \in  (\bar{E}-\epsilon, \bar{E}+\epsilon)$. These projections should remove very large eigenvalues that would otherwise correspond to large energies. We plot the eigenvalue spectrum for a single draw of the ensemble in Figure~\ref{velocity_spec}.

\begin{figure}[t!]
    \centering
    \includegraphics[width=0.46\columnwidth]{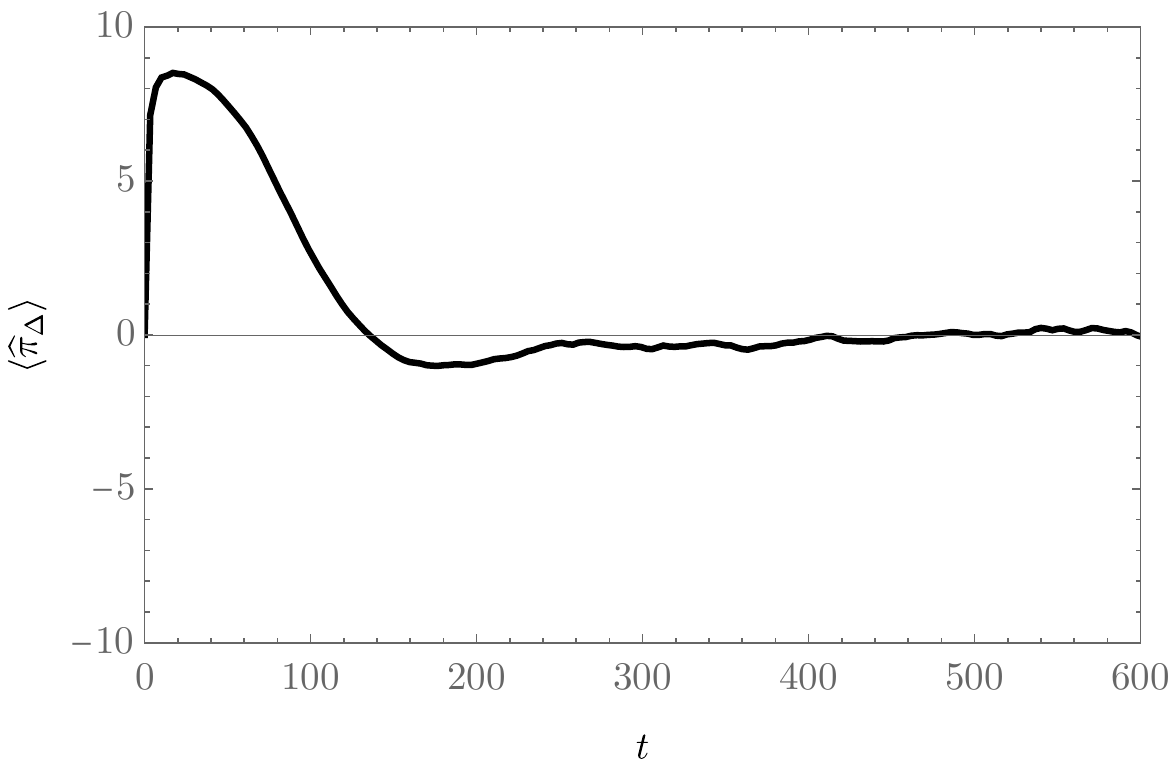}
    \includegraphics[trim={0 0 0 0},clip,width=0.43\columnwidth]{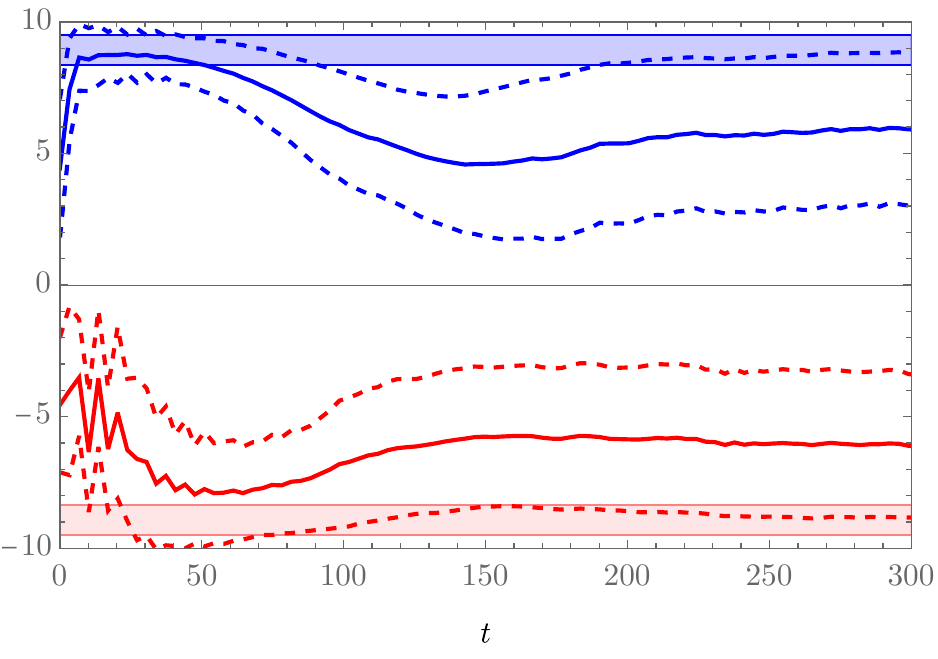}
            \caption{  \label{fig:velocity}Expectation value of $\hat{\pi}_\Delta$ (black). We also show in {\color{red} red} ({\color{blue} blue}) the expectation values when we project into {\color{red} positive} ({\color{blue} negative}) velocity sub-sectors, as well as their quantum standard deviations through the dotted curves. The shaded regions indicate the disk predictions for a black hole/firewall. Note that the noisy estimate of the standard deviation of the red curve at early times reflects the very small ``firewall'' probability at early times which is better reflected in figure~\ref{fig:pvelocity2}.  } %
\end{figure}

The energy projected velocity has eigenvalues that are clustered around the semi-classical black hole/white hole velocities, e.g.,
    $v^2/2 = (\bar{E} \pm \epsilon)$.
We can also consider the evolution of the $\pi_\Delta$ expectation value. This is done in figure~\ref{fig:velocity}, where we consider the expectation value of $\pi_\Delta$ as well as the expectation value of $\pi_\Delta$ after a projection to its positive and negative eigenvalues. To reduce the noise, we disorder average over a number of draws of the ensemble $n_\mathrm{trials}$ but find that at late times, all such expectation values are self-averaging. Moreover, they are insensitive to the exact choice of $\Delta$ and thus appear to be universal.

\begin{figure}[h!]
    \centering
    \includegraphics[width=0.32\textwidth]{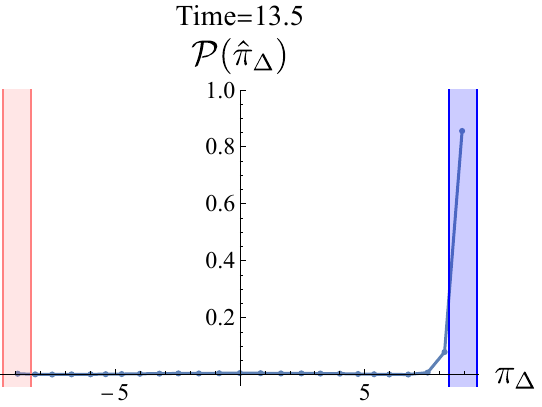}
      \includegraphics[width=0.32\textwidth]{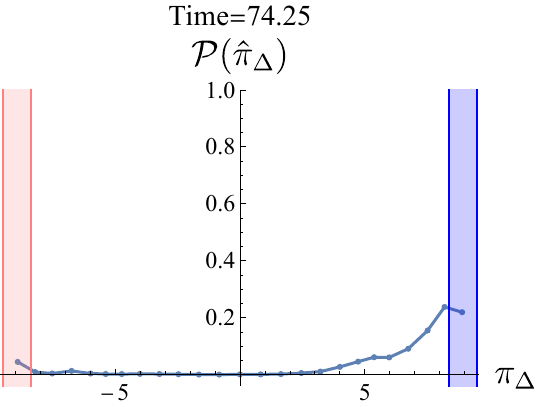}
        \includegraphics[width=0.32\textwidth]{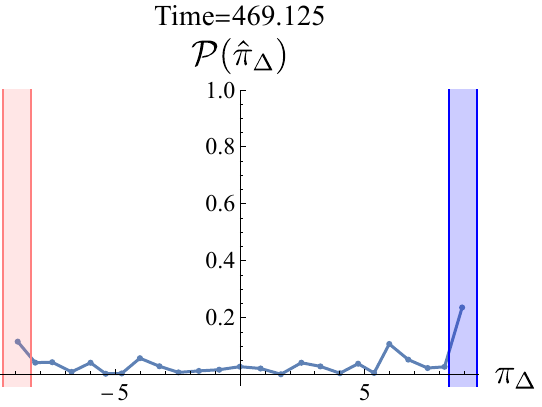}
            \caption{The probability of detecting a velocity within a given velocity window (centered around each point)  at different times. The blue and red vertical bands indicate the semi-classical velocities for black holes and white holes within the energy band that we study. The \textit{left plot} shows the probability at early times (with $t \ll e^{S_0}$) where the wavefunction almost entirely has support on black hole states whose velocity is $+\sqrt{2\overline{E}}$. The \textit{middle plot} shows the probability at intermediate times (with $t \lesssim e^{S_0}$) where the wavefunction now starts having small support on white hole states whose velocity is $- \sqrt{2\overline{E}}$.  The \textit{right plot} indicates the probability at very late times (with $t > e^{S_0}$) where the wavefunction now has support not only on black hole and white hole states but also on gray hole states whose velocity is far from $\pm \sqrt{2\overline{E}}$. A detailed video of this probability distribution as a function of time can be found at \cite{video1}.\label{fig:probDistributionVelocity}}
\end{figure}

\begin{figure}[h!]
    \centering
    \includegraphics[width=0.65\textwidth]{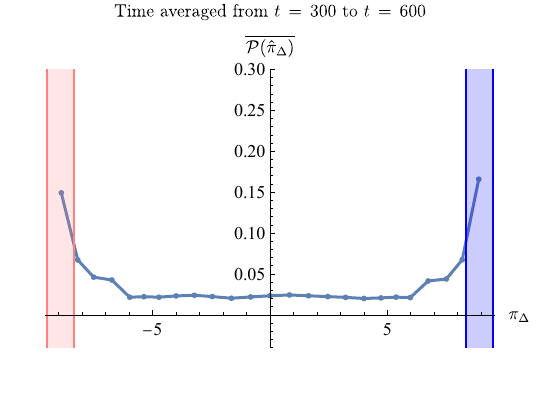}
            \caption{The probability of detecting a velocity within a given velocity window averaged over a large time window with all times $t > e^{S_0}$. The blue and red vertical bands once again indicate the semi-classical velocities for black holes and white holes within the energy band that we study. While the probability distribution is equally peaked at the black hole and white hole velocity values, but it also has non-negligible support for gray hole velocities.   \label{fig:probDistributionVelocityAvg} }
\end{figure}

We find that at late times, the expectation value of $\pi_\Delta$ goes to zero, consistent with the plateau in the length expectation value seen in Figure~\ref{fig:length_expectation_tau}. Because of this, the probability of detecting a negative or positive velocity eigenvalue becomes 1/2 at late times. However, we can be more detailed about the properties of the wavefunction in the velocity basis. In figure~\ref{fig:probDistributionVelocity}, we show the probability distribution over velocities at different times. An additional video of this probability distribution over time can be found at \cite{video1}. Since the probability distribution that we find at late times has rapidly fluctuating noise, we also plot a time average of this probability distribution over an interval in the plateau region. The result is figure~\ref{fig:probDistributionVelocityAvg}, which shows that the probability distribution is largely bimodal, with peaks around $\pm \sqrt{2 \bar E}$ and a small, though non-negligible, probability for velocities between $\pm \sqrt{2 \bar E}$. Unsurprisingly, since we have averaged over time, this probability distribution is also self-averaging with respect to the choice of ensemble member.

\begin{figure}[t!]
    \centering
    \includegraphics[width=0.6\columnwidth]{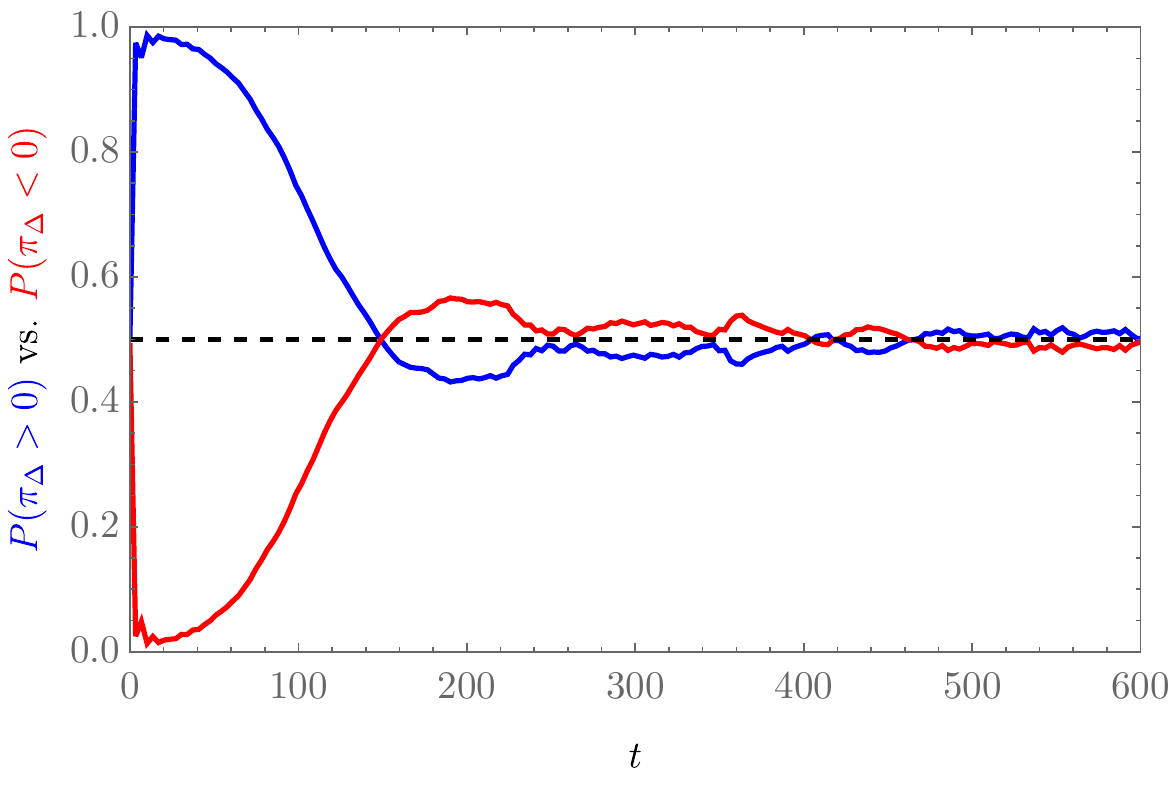}
            \caption{Probability that $\hat{\pi}_\Delta > 0$ (or $<0$) as a function of time averaged over 86 trials. A possible interpretation is that $P(\pi_\Delta < 0)$ is the probability of a ``firewall.'' At late times, the probability seems to converge to 50/50. \label{fig:pvelocity2}}
\end{figure}

Let us now briefly interpret these results. In this section, we have in mind that the velocity operator $\hat{\pi}_\Delta$ is at least a crude proxy for firewalls, e.g., that a state with $\ev{\hat{\pi}} \approx - \sqrt{2 \bar E}$ is a dangerous wormhole to jump into. As we shall further discuss in section \ref{sec:discussion}, this interpretation should be scrutinized further. Nevertheless, our main emphasis is that there are definitions of the wormhole velocity that are non-perturbatively well-defined and are, in principle, computable, even for individual members of the ensemble.

\section{Generalization to JT with matter}
\label{sec:JT-with-matter}

In this section, we discuss the generalization of the preceding discussion to JT coupled to an arbitrary matter CFT. We are able to follow essentially the same strategy, using the condition that a segment of asymptotic Euclidean boundary (preparing a TFD-like state) is homotopic to a unique geodesic, and these bound a hyperbolic half-disk. The geodesic states are labeled by a state of the matter as well as the length $\ell$, and correspondingly the asymptotic states include matter sources. The main novelty is that the energy on left and right boundaries need not be equal, so we have two different commuting Hamiltonians $H_L$ and $H_R$.

\subsection{Perturbative JT with matter}

As for pure JT, we must first understand the $S_0\to\infty$ Hilbert space on the disk topology. The first part of this subsection is mostly a review of previous work \cite{Maldacena:2016upp, Lin:2019qwu,Penington:2023dql,Kolchmeyer:2023gwa}, emphasising the results we will later require. In the remainder, we give a concrete description of the matter Hilbert space in terms of the symmetries of QFT on AdS$_2$ and then describe the calculation of a half-disk amplitude generalized to include matter. %
However, we will not directly use these results in the later parts, so the reader eager to get to the non-perturbative theory can skip from the end of section \ref{sssec:matterstates} to section \ref{ssec:matterNPdef}.

\subsubsection{Geodesic and asymptotic states}\label{sssec:matterstates}

It is relatively simple to describe the Hilbert space of JT coupled to an arbitrary matter theory since the matter sector does not couple directly to the dilaton. Excepting for its influence on the boundary cutoff (which will appear in the Hamiltonians later), the matter behaves as a QFT on a fixed AdS$_2$ background, with Hilbert space $\hilbM$. This means that the perturbative (disk-level) Hilbert space is simply a tensor product of length wavefunctions and QFT states,
\begin{equation}\label{eq:geoHS}
    \hilb_0 = L^2(\mR) \otimes \hilbM \,.
\end{equation}
We will not need to know very much about the matter theory, only requiring knowledge of how the isometries of AdS$_2$ act on it. There should be a unitary representation of the $\widetilde{SL}(2,\mR)$ symmetry group acting on the matter states, and we can decompose $\hilbM$ into irreducible representations. The algebra $\mathfrak{sl}(2,\mR)$ has a basis of Hermitian operators $\mathsf{H}$, $\mathsf{P}$ and $\mathsf{B}$, which act as global time translation (not to be confused with asymptotic time translations generated by $H_{L,R}$), spatial translation along the geodesic, and a boost (or rotation of the disk in Euclidean signature). Assuming that the global energy $\mathsf{H}$  is bounded from below,  the only possible non-trivial representations are the discrete series $\mathcal{D}_{\Delta}^+$ labeled by a lowest weight (minimal $\mathsf{H}$ eigenvalue) $\Delta>0$; \cite{Kitaev:2017sl2} reviews of the relevant representation theory. In the next subsection we give a concrete description of these representations with a direct physical interpretation in terms of QFT on AdS$_2$.  The upshot is the decomposition
\begin{equation}\label{eq:Hmirreps}
    \hilbM =   \mC \oplus \bigoplus_{\Delta} \mathcal{D}_{\Delta}^+,
\end{equation}
where the first term is the trivial representation corresponding to an $\mathfrak{sl}(2,\mR)$ invariant vacuum state (supposing that this exists and is unique).\footnote{We use $\Delta$ as an index for the representations, though this is a slight abuse of notation since there may be degeneracies in $\Delta$ (such as in the example of a free theory).} Thus, a basis of geodesic states is labeled by a length $\ell$, a representation $\Delta$, and a label $m$ for an orthonormal basis of $\mathcal{D}_{\Delta}^+$, and this basis is orthonormal on the disk:
 \begin{equation}
    \text{Geodesic states:}\quad |\Delta;\ell,m\rangle, \qquad \langle\Delta',\ell',m'|\Delta;\ell,m\rangle_0 = \delta_{\Delta\Delta'}\delta_{mm'}\delta(\ell-\ell').
 \end{equation}

The coupling of the matter theory to the gravitational sector appears through the Hamiltonians $H_{L,R}$ generating asymptotic time translation. These are differential operators in $\ell$ generalising \eqref{eq:Hpure}, but also include generators of the $\mathfrak{sl}(2,\mR)$ symmetry acting on the matter %
\cite{Lin:2019qwu, Harlow:2021dfp, Penington:2023dql, Kolchmeyer:2023gwa}:
\begin{align}
   H_L &=  \tfrac{1}{2}(p+\tfrac{1}{2}\mathsf{P})^2 +(\mathsf{H}-\mathsf{B})e^{-\ell/2}+2e^{-\ell} \label{eq:HL} \, , \\
   H_R &= \tfrac{1}{2}(p-\tfrac{1}{2}\mathsf{P})^2 +(\mathsf{H}+\mathsf{B})e^{-\ell/2}+2e^{-\ell} \, .\label{eq:HR} 
\end{align}
Here, we have introduced $p$, which is conjugate to the length, e.g., $\i [p, \ell] = 1$.
 These two Hamiltonians are no longer equal (except for the trivial representation when we recover \eqref{eq:Hpure}), but they commute, $[H_L,H_R]=0$. Since they depend only on the symmetry generators, they act diagonally on the decomposition \eqref{eq:Hmirreps} into irreducible representations. In particular, the quadratic Casimir 
\begin{align}
    \mathsf{C} = \mathsf{H}^2 - \mathsf{B}^2-\mathsf{P}^2 =\Delta(\Delta-1) \label{eq:casimir_def}
\end{align}
is conserved \cite{Lin:2019qwu}. We give more concrete formulas for \eqref{eq:HL}, \eqref{eq:HR} in the next subsection by writing generators in the representation $\mathcal{D}_{\Delta}^+$ as differential operators acting on a function of the AdS$_2$ spatial coordinate $u$, so $H_{R,L}$ act on wavefunctions of two variables $\ell,u$.

Not only do $H_L,H_R$ commute; in fact, they form a complete set of commuting operators acting on each representation space $L^2(\mR) \otimes \mathcal{D}_{\Delta}^+$ with a spectrum consisting of all positive energies $E_{L,R}>0$. This means that there is a basis of energy eigenstates $|\Delta;E_L,E_R\rangle$ for this space, with a change of basis matrix $\phi^\Delta_{E_L E_R}(\ell,m)$ analogous to the scattering wavefunctions $\phi_E(\ell)$ above:
\begin{equation}
    \begin{gathered}
    L^2(\mR) \otimes \mathcal{D}_{\Delta}^+ \simeq L^2(\mR_+\times \mR_+)\nonumber, \\
    |\Delta;E_L,E_R\rangle =  \sum_m \int d\ell \, \phi^\Delta_{E_L E_R}(\ell,m)|\Delta;\ell,m\rangle.
\end{gathered}
\end{equation}
We prove this directly by giving explicit expressions for $\phi^\Delta_{E_L E_R}(\ell,m)$ below; here, we note only that they can be chosen to be real. The Hermiticity and spectrum of $H_{L,R}$ imply an  orthogonality relation
\begin{equation}
    \begin{aligned}
        \langle \Delta,E_L',E_R'|\Delta;E_L,E_R\rangle_0 &= \sum_m \int d\ell\, \phi^\Delta_{E_L E_R}(\ell,m)\phi^\Delta_{E_L' E_R'}(\ell,m) \\
        &= \frac{\delta(E_L-E_L')\delta(E_R-E_R')}{\rho_0(E_L)\rho_0(E_R)}\Gamma^\Delta_{E_LE_R},
    \end{aligned}
\end{equation}
where $\Gamma^\Delta_{E_LE_R}$ sets the normalisation of $|\Delta;E_L,E_R\rangle$, with a convenient choice made below. %

To follow the argument we used in section \ref{sec:pureJT}, we would like a set of asymptotically-defined states that can be used to prepare any geodesic state, now with matter excitations. For this, we generalize the TFD state to include insertion of a local `primary' boundary operator $\mathcal{O}^\Delta$, with Euclidean boundary segments of length $\tau_{L,R}$ to the left and right of the operator.  $\mathcal{O}^\Delta$ is defined to prepare the ground state $\mathsf{H}=\Delta$ of the given representation on a geodesic very close to the boundary (i.e, for $\tau_{L}=\tau_R=\epsilon\to 0$); see the following subsection for more details. Analogous to \eqref{eq:TFD}, we can write these TFD-like  states as
\begin{equation}\label{eq:matterAsympStates}
    |\Delta;\tau_L,\tau_R\rangle = e^{-(\tau_L H_L+\tau_R H_R)} |\mathcal{O}^\Delta\rangle = e^{S_0/2} \int dE_L \rho_0(E_L) dE_R\rho_0(E_R) e^{-(\tau_L E_L+\tau_R E_R)} |\Delta;E_L,E_R\rangle,
\end{equation}
where $|\mathcal{O}^\Delta\rangle= |\Delta;\tau_L=0,\tau_R=0\rangle$ is an `infinite temperature' state produced by $\mathcal{O}^\Delta$ with no Euclidean evolution. This state is non-normalizable with respect to the perturbative inner product, becoming normalizable after any Euclidean evolution $\tau_{L,R}>0$.

To recover the interpretation as a local primary operator insertion, we will have to make the correct choice of $\Gamma^\Delta_{E_LE_R}$. For this, we consider the matrix elements
\begin{equation}\label{eq:JT2pt}
    \langle \mathcal{O}^\Delta|e^{-(\tau_L H_L+\tau_R H_R)} |\mathcal{O}^\Delta\rangle_0 = \delta_{\Delta \Delta'} e^{S_0} \int  dE_L \rho_0(E_L) dE_R\rho_0(E_R) e^{-(\tau_L E_L+\tau_R E_R)}\Gamma^\Delta_{E_LE_R},
\end{equation}
which should be computed by a path integral on a boundary circle of total length $\tau_L+\tau_R$ with two local primary operator insertions. This must reproduce the perturbative two-point function in JT (or equivalently, the Schwarzian theory), with boundary quantum mechanics interpretation $\Tr(e^{-\tau_L H}\mathcal{O}^\Delta e^{-\tau_R H}\mathcal{O}^\Delta)$. From the results of \cite{Mertens:2017mtv, Yang:2018gdb}, we should choose
\begin{equation}\label{eq:Gamma}
    \Gamma^\Delta_{E_LE_R} = \frac{\Gamma(\Delta\pm i \sqrt{2E_L}\pm i \sqrt{2E_R})}{4^{\Delta-1} \Gamma(2\Delta)}.
\end{equation}
 Not coincidentally, this is equal to the matrix element $\langle E_R|e^{-\Delta \ell}|E_L\rangle = \int d\ell \phi_{E_L}(\ell)\phi_{E_R}(\ell)e^{-\Delta \ell}$  in the perturbative Hilbert space of pure JT gravity, appearing also in \eqref{numerical_ldelta}. We have normalized our definitions based on the short-time singularity of the resulting two-point function \eqref{eq:JT2pt}: taking $\tau_L\to \beta$ and $\tau_R = \tau\to 0$, we have $\langle \Delta|e^{-(\tau_L H_L+\tau_R H_R)} |\Delta\rangle_0 \sim Z_0(\beta)\tau^{-2\Delta}$.

 We emphasize that this is entirely independent of the details of the bulk theory and holds for any matter state, being determined only by the AdS$_2$ symmetry of the matter sector (it does not require a free QFT, for example). %

Combining these definitions, we can express the asymptotic TFD-with-operator states in terms of geodesic states:
\begin{gather}\label{eq:HDmatter}
    |\Delta;\tau_L,\tau_R\rangle  = e^{S_0/2} \sum_m \int d\ell \,Z_{\HD}^\Delta(\tau_L,\tau_R;\ell,m)|\Delta;\ell,m\rangle, \\
    Z_{\HD}^\Delta(\tau_L,\tau_R;\ell,m)= \int dE_L \rho_0(E_L) dE_R\rho_0(E_R) e^{-(\tau_L E_L+\tau_R E_R)}  \phi^\Delta_{E_L E_R}(\ell,m).
\end{gather}
Just like in the pure JT case \eqref{eq:ZHD}, the wavefunction $Z_{\HD}$ is the path integral on the half-disc bounded by an asymptotic boundary with primary operator insertion and a geodesic of length $\ell$ with matter state $m$ in the representation $\Delta$. We will explicitly compute this in section \ref{ssec:HDmatter}, and in doing so, obtain an explicit formula for $\phi_{E_L,E_R}^\Delta(\ell,m)$.

\subsubsection{QFT in AdS$_2$ and $\mathfrak{sl}(2)$ representations}

Here we give a concrete description of the $\mathfrak{sl}(2)$ representations $\mathcal{D}_{\Delta}^+$ in terms of QFT on AdS$_2$. We describe this first for the single-particle states of a free real scalar $\chi$ of mass $m^2=\Delta(\Delta-1)$, afterwards explaining why the result holds for all states in a general QFT\footnote{See \cite{Bak:2023zkk} for some discussion about the $m^2=0$ case.}.

To describe states on a geodesic Cauchy surface of AdS$_2$, it is convenient to use coordinates which make manifest the translation symmetry $\mathsf{P}$ fixing the geodesic. To do this, we can write the Euclidean hyperbolic metric as
\begin{equation}
    ds^2 = \frac{dv^2+du^2}{\sin^2 v}, \qquad 0<v<\pi,
\end{equation}
\begin{figure}
    \centering
    \includegraphics[width=0.25\columnwidth]{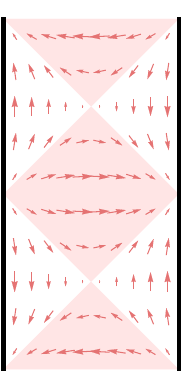}
    \caption{Illustration of the (Lorentzian signature) coordinates $ds^2 = \frac{-dv^2+du^2}{\sin^2 v}$. The pink arrows are $\partial_u$ (corresponding to the generator $\mathsf{P}$).  This figure is adapted from \cite{Lin:2019qwu}. }
    \label{fig:enter-label}
\end{figure}
with the geodesic in question at $v=0$, and asymptotic Euclidean conformal boundaries at $v=\pm\frac{\pi}{2}$. Wick rotation of $v$ gives Lorentzian AdS$_2$ in the `Wheeler-DeWitt patch', the domain of dependence of the geodesic. To relate this to half-plane coordinates $(x,y)$ with $ds^2 = \frac{dx^2+dy^2}{2}$ we can use $x+i y = \tanh\left(\frac{u+iv}{2}\right)$, so the `past' $v=0$ boundary corresponds to $-1<x<1$ and $x=\tanh\frac{u}{2}$, while the future $v=\pi$ boundary is $|x|>1$ with $x=\coth\frac{u}{2}$. 

Now, single-particle states are created by acting on the vacuum with an operator inserted in the Euclidean past $0<v<\frac{\pi}{2}$. In fact, to span all states, it's sufficient to insert operators only close to the asymptotic boundary $v=0$. For this, we can define boundary operators $\mathcal{O}^\Delta(u)$ as a $v\to 0$ limit of bulk operators, rescaled to get a finite limit for correlation functions, and acting with these on the vacuum $|\Omega\rangle$ creates a basis of single-particle states:
\begin{equation}
    \hilb_\mathrm{single-particle}\simeq \mathcal{D}_{\Delta}^+ \ni \int_{-\infty}^\infty du\, f(u)\mathcal{O}^\Delta(u) |\Omega\rangle  , \qquad \mathcal{O}^\Delta(u) = \lim_{v\to 0} v^{-\Delta} \chi(u,v).
\end{equation}
This means that we can describe elements of $\mathcal{D}_{\Delta}^+$ as functions $f(u)$ which are normalizable in the relevant norm (introduced in a moment). The isometries are generated by Killing vectors $\partial_u$, $\cosh u \cos v \partial_u +  \sinh u \sin v \partial_v$, $\sinh u \cos v \partial_u +  \cosh u \sin v \partial_v$, which correspond to the generators $ i \mathsf{P}$, $\mathsf{H}$ and $\mathsf{B}$ respectively (with factors of $i$ depending on the Wick rotations to real vectors in Lorentzian signature). Since $\chi$ is a scalar, we find that $\mathcal{O}^\Delta$ transforms like a conformal primary of weight $\Delta$, and integrating by parts gives us the action on the function $f$:
\begin{gather}
    \mathsf{P}  = -i\partial_u, \quad \mathsf{H} = (\Delta-1)\cosh u- \sinh u \, \partial_u , \quad \mathsf{B} = (\Delta-1)\sinh u- \cosh u \partial_u \,.
\end{gather}
These indeed obey the $\mathfrak{sl}(2)$ commutation relations and have Casimir $\mathsf{C}=\Delta(\Delta-1)$.

The inner product, which makes the representation unitary, is not diagonal in this $u$ basis. We can determine it from the QFT inner product, where hermitian conjugation acts on operators by reflecting them in Euclidean time, so $(\mathcal{O}^\Delta(u))^\dag$ is a boundary operator on the future $v=\pi$ boundary. Hence, we find the inner product by integrating against the boundary limit of the two-point function of $\chi$, which (choosing an appropriate normalization) gives 
\begin{equation}\label{eq:inner-prod-u-basis}
    \langle g|f \rangle = \int du du' \frac{f(u)\bar{g}(u')}{\left(2\cosh\left(\tfrac{u-u'}{2}\right)\right)^{2\Delta}} \,.
\end{equation}
This can be understood as a conformal correlation function, which in upper half-plane coordinates would be $\frac{1}{(x-x')^{2\Delta}}$, with the change of variables $x=\tanh \frac{u}{2}$, $x'=\coth \frac{u}{2}$ and Jacobian factors to account for the cutoff at constant $v$ instead of constant $y$.

With this representation of the $\mathfrak{sl}(2)$ generators, we can write the left and right Hamiltonians  (which act on functions of $u,\ell$) as
\begin{align}
    H_L &= -\tfrac{1}{2}(\partial_\ell+\tfrac{1}{2}\partial_u)^2 +e^{-u-\ell/2}(\Delta -1 + \partial_u)+2e^{-\ell} \, ,\label{eq:HLlu}\\
    H_R &=  -\tfrac{1}{2}(\partial_\ell-\tfrac{1}{2}\partial_u)^2 +e^{u-\ell/2}(\Delta -1 - \partial_u)+2e^{-\ell}  \,.\label{eq:HRlu}
\end{align}
This is precisely the same expression for the Hamiltonians that appears naturally from the double-scaled SYK chord Hilbert space \cite{Lin:2022rbf} (though note that the definition of $\ell$ in that work differs by a shift). From our explicit expression, we see that $u$ naturally labels the \emph{boundary} location of a particle, not a location on the geodesic (where a local insertion on the boundary will have spread to a wavefunction of width $\Delta^{-1/2}$ in AdS units).

For an arbitrary representation in a general QFT, we will not have an analog of the local operator $\chi$. Nonetheless, we can define $\mathcal{O}^\Delta$ by preparing a primary state on a small geodesic between nearby points on the $v=0$ boundary, taking a limit including a renormalizing factor analogous to the boundary limit of $\chi$. All the same results follow.

We should explain how this relates to the orthonormal discrete $m$ basis used above, which diagonalizes $\mathsf{H}$ with eigenvalues $\Delta+m$ ($m=0,1,2,\ldots$). First, the ground state $m=0$ corresponds simply to the primary operator $\mathcal{O}^\Delta$ inserted at $u=0$, so $f(u)\propto \delta(u)$. To get higher $m$ states, first form the combinations $\mathsf{L}_\pm = \mathsf{B}\pm i \mathsf{P}$ (and $\mathsf{L}_0=\mathsf{H}$) (this matches the usual conventions for the `global' subalgebra $n=-1,0,1$ of the Virasoro algebra, with $[\mathsf{L}_m,\mathsf{L}_n]=(m-n)\mathsf{L}_{m+n}$). Then $\mathsf{L}_+$ and  $\mathsf{L}_-$ are lowering and raising operators for $\mathsf{L}_0$, so the state $|m\rangle$ is obtained by acting $m$ times with $\mathsf{L}_-$ on $\delta(u)$, giving a linear combination of derivatives of $\delta$-functions. In the coordinate $x=\tanh\frac{u}{2}$ we simply have $\mathsf{L_-}=-\partial_x$, so by changing coordinates back we can write $f_m(u) \propto \left(2\cosh^2 \frac{u}{2}\right)^{1-\Delta}\delta^{(m)}(\tanh\frac{u}{2})$.

\subsubsection{The half-disk and Hamiltonian eigenstates}\label{ssec:HDmatter}

We now compute the half-disk path integral $Z_\HD^\Delta(\tau_L,\tau_R;\ell,m)$ with matter representation $\Delta$. By comparing the result to the formula \eqref{eq:HDmatter}, we will extract the eigenfunctions $\phi^\Delta_{E_L,E_R}(\ell,m)$ which diagonalise the Hamiltonians \eqref{eq:HR}, \eqref{eq:HL}. The argument that follows can be made more concrete using the boundary particle formalism for JT described in \cite{Yang:2018gdb}; in appendix \ref{app:JTparticle}, we extend this formalism to describe the geodesic Hilbert space and explain how to use it to compute the half-disk amplitude. We will use the $u$ basis for the matter representation $\mathcal{D}^+_\Delta$ described above, getting expressions for $Z_\HD^\Delta(\tau_L,\tau_R;\ell,u)$ and $\phi^\Delta_{E_L,E_R}(\ell,u)$, explaining afterwards how to change to the $m$ basis.

\begin{figure}
    \centering
    \begin{tikzpicture}[baseline={([yshift=0cm]current bounding box.center)}, scale=1.25]
    \draw[thick, fill=lightgray, fill opacity=0.3] (0,0) circle (2.0);
    \draw[thick, red] (-2,0) arc (90:45:4.83) node[midway, below] {$\ell_L$};
    \draw[thick, dgreen] (2,0) arc (90:225:0.829); %
    \draw[thick,dashed] ({2*cos(45)},{-2*sin(45)}) arc (-135:-225:2);
    \node[dgreen] at (1.5,-0.6) {$ \ell_R$};
    \node[red] at (-0.4,-2.2) {$\tau_L$};
    \node[dgreen] at (2.2,-0.8) {$\tau_R$};
    \node[blue] at (0,0.25) {$\frac{1}{2}\ell+u$} ;
    \node[blue] at (1.4,0.25) {$\frac{1}{2}\ell-u$} ;
    \draw[thick, blue] (-2,0)--(2,0) ;%
    \fill ({2*cos(45)},{-2*sin(45)}) circle (0.1);
    \fill ({2*cos(45)},{2*sin(45)}) circle (0.1);
    \fill[gray] (-2,0) circle (0.1);
    \fill[gray] (2,0) circle (0.1);
    \fill[blue] (.82,0) circle (0.1);
    \node[black] at (1.7,-1.7) {$\mathcal{O}^\Delta$} ;
    \end{tikzpicture}
    \caption{A diagram indicating the geometric meaning of the label $u$ for the wavefunctions $\tilde{\phi}^\Delta(\ell_L,\ell_R;\ell,u)$. }
    \label{fig:halfdiskmatter}
\end{figure}
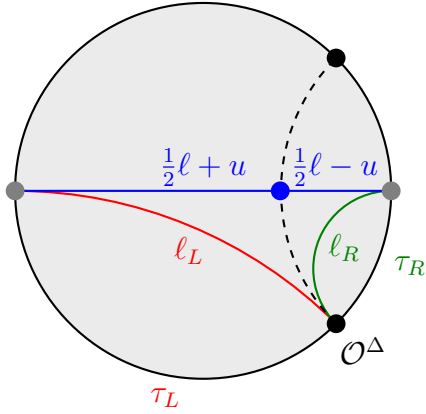

The main idea behind this calculation is illustrated in figure~\ref{fig:halfdiskmatter}. We cut the half-disk along geodesics which join the location of the $\mathcal{O}^\Delta$ insertion to the left and right boundary points, writing their renormalized lengths as $\ell_L,\ell_R$. This splits the half-disk into three parts: two half-disks between asymptotic boundaries of length $\tau_{L,R}$ and geodesics of length $\ell_{L,R}$, and a hyperbolic triangle with (renormalized) side length $\ell_L,\ell_R,\ell$. Accordingly, the amplitude splits as
\begin{equation}\label{eq:HDcomp1}
    Z_\HD^\Delta(\tau_L,\tau_R;\ell,u) = \int d\ell_L d\ell_R Z_\HD(\tau_L;\ell_L)Z_\HD(\tau_R;\ell_R) \tilde{\phi}^\Delta(\ell_L,\ell_R;\ell,u)
\end{equation}
for some $\tilde{\phi}^\Delta(\ell_L,\ell_R;\ell,u)$, where $Z_\HD$  the pure JT half-disk given in \eqref{eq:ZHD}.

More carefully, we can think of this perturbative JT calculation as taking place on a fixed hyperbolic disk but with a `wiggly cutoff' \cite{Yang:2018gdb,Kitaev:2018wpr}. This means that there is not a direct relation between the proper time along the cutoff and the points of the fixed disk; instead, we integrate over deomorphisms that relate the proper boundary time to the coordinate time. In particular, the location of the operator insertion (which is determined by $\ell_L,\ell_R$) fluctuates, even though the proper times $\tau_{L,R}$ to the boundaries along the cutoff are fixed. This is accounted for by integrating over $\ell_L,\ell_R$, while the fluctuations of the cutoff on either side of the operator are computed by the vacuum half-disk factors.

The remaining part of the computation $\tilde{\phi}^\Delta(\ell_L,\ell_R;\ell,u)$ at fixed $\ell_{L,R}$ involves the hyperbolic triangle with renormalized side lengths $\ell_L,\ell_R$ and $\ell$ (so the actual lengths are $2\log\epsilon^{-1}$ larger, and we take $\epsilon\to 0$), with the operator $\mathcal{O}^\Delta$ inserted at one corner. For pure JT gravity (ignoring the operator for now), the contribution of this triangle is $I(\ell,\ell_L,\ell_R)$, which can be determined from the requirement that we recover $Z_\HD(\ell;\tau_L+\tau_R)$:
\begin{align}
Z_\HD(\tau_L+\tau_R;\ell) &= \int d\ell_L d\ell_R Z_\HD(\tau_L;\ell_L)Z_\HD(\tau_R;\ell_R) I(\ell,\ell_L,\ell_R) \\
\implies
    I(\ell,\ell_L,\ell_R) &= \int dE \rho_0(E) \phi_E(\ell)\phi_E(\ell_L)\phi_E(\ell_R)  \label{eq:Iformula} \\
    &= \exp\left[-2 e^{-(\ell+\ell_L+\ell_R)/2}(e^{\ell}+e^{\ell_L}+e^{\ell_R})\right].\nonumber
\end{align}
The exponent in $I$ has the geometric interpretation in terms of the area of the triangle (it is proportional to $\pi$ minus the area when taking the cutoff $\epsilon\to 0$).

It remains only to understand the effect of the operator insertion, which determines the matter state. For this, it's helpful to use the $(u,v)$ coordinates introduced above, with the original geodesic boundary at $v=\frac{\pi}{2}$. We fix the residual translation symmetry by choosing $u=0$ to be the midpoint of the geodesic. Now, the length of the triangle side lengths is sufficient to fix the coordinates of the corner where we insert the operator $\mathcal{O}_\Delta$, with $u \sim \frac{1}{2}(\ell_L-\ell_R)$ and $v\sim \epsilon \, e^{-\frac{1}{2}(\ell_L+\ell_R-\ell)}$ as we take $\epsilon\to 0$. The $v$ coordinate gives us a renormalisation of the operator, contributing a factor $e^{-\frac{\Delta}{2}(\ell_L+\ell_R-\ell)}$. The $u$ coordinate simply tells us that we get a $\delta$-function matter wavefunction in the $u$ basis for fixed $\ell_{L,R}$.

Putting all these ingredients together, we obtain our final result (we check the normalisation in appendix \ref{app:JTparticle}):
\begin{gather}
    \label{eq:HDcomp2}
    \tilde{\phi}^\Delta(\ell_L,\ell_R;\ell,u) =  e^{\frac{\Delta}{2}(\ell-\ell_L-\ell_R)} I(\ell,\ell_L,\ell_R) \delta(u-\tfrac{\ell_L-\ell_R}{2}).
\end{gather}

To recover the original eigenfunctions, their relation \eqref{eq:HDmatter} to the half-disk gives us
\begin{equation}\label{eq:phiELERlu}
    \phi_{E_L,E_R}^\Delta(\ell,u) = \int d\ell_L d\ell_R \phi_{E_L}(\ell_L)\phi_{E_R}(\ell_R) \tilde{\phi}^\Delta(\ell_L,\ell_R;\ell,u).
\end{equation}
The condition that $\phi_{E_L,E_R}^\Delta(\ell,u)$ is an eigenfunction for $H_{L,R}$ is equivalent to an `intertwining' property for $\tilde{\phi}^\Delta(\ell_L,\ell_R;\ell,u)$. Namely, we require that acting with $H_L$ on the $\ell,u$ variables gives the same result as acting with the pure JT Hamiltonian \eqref{eq:Hpure} on $\ell_L$, and similarly for the right. It is straightforward to check that our result \eqref{eq:HDcomp2} indeed has this property.

To use a different basis for the representation, we can write the delta function $\delta(u-u_0)$ more abstractly by $e^{-i \mathsf{P} u_0}|0\rangle$, where $|0\rangle$ is the lowest weight state of the representation. In particular, for the $m$ basis, we find
\begin{equation}
\begin{gathered}
   \tilde{\phi}^\Delta(\ell_L,\ell_R;\ell,m) = \sqrt{\frac{\Gamma(2\Delta+m)}{m! \Gamma(2\Delta)}} I(\ell,\ell_L,\ell_R) \left(\frac{e^{\frac{\ell-\ell_L-\ell_R}{2}}}{\cosh^2(\tfrac{\ell_L-\ell_R}{4})}\right)^\Delta \left(\tanh(\tfrac{\ell_L-\ell_R}{4})\right)^m \,. %
\end{gathered}
\end{equation}
One way to get this is to write both $e^{-iP\tfrac{\ell_L-\ell_R}{2}}|0\rangle$ and $|m\rangle$ states in the $x$ representation (a $\delta$-function supported at $x=\tanh(\tfrac{\ell_L-\ell_R}{4})$ and $\delta^{(m)}(x)$, up to normalisations), and take the inner product.

\subsection{The non-perturbative definition of the theory}\label{ssec:matterNPdef}

With matter, we continue to follow the same philosophy to give a non-perturbative definition of the theory in terms of a boundary dual, though with a few novelties.

First, since $H_L$ and $H_R$ are no longer equal, we have a more conventional relationship between the bulk and boundary dual Hilbert spaces, namely $\hilb\simeq \hilb_\partial \otimes \hilb_\partial^*$ where the tensor factors correspond to left and right boundaries respectively.

To populate states with matter we need to include definitions for asymptotic boundaries with operator insertions, creating the asymptotic states $|\Delta;\tau_L,\tau_R\rangle = e^{-(\tau_L H_L+\tau_R H_R)}|\mathcal{O}^\Delta\rangle$ defined in \eqref{eq:matterAsympStates}. This means that the boundary dual is defined not only by its spectrum, but also by the wavefunctions of the states $|\mathcal{O}^\Delta\rangle$. We can give this data in terms of the overlaps $\langle i,j|\mathcal{O}^\Delta\rangle$ with an orthonormal basis $|i,j\rangle$ of simultaneous eigenstates of left and right Hamiltonians, with $H_L|i,j\rangle = E_i|i,j\rangle$,  $H_R|i,j\rangle = E_j|i,j\rangle$ and  $\langle i',j'|i,j\rangle = \delta_{i i'}\delta_{j j'}$. In terms of the boundary dual Hilbert space $\hilb_\partial$, our bulk two-sided Hilbert space $\hilb\simeq \hilb_\partial \otimes \hilb_\partial^*$ is equivalent to the space of operators on $\hilb_\partial$, and the wavefunctions on $\hilb$ can be interpreted as matrix elements $\mathcal{O}^\Delta_{ij}$ of local primary operators in $\hilb_\partial$:
\begin{equation}
    \langle i,j|\mathcal{O}^\Delta\rangle = \mathcal{O}^\Delta_{ij} := \langle i|\mathcal{O}^\Delta|j\rangle_\partial.
\end{equation}
To summarise, our input will be the spectrum $\{E_i\}$ as before, along with matrix elements $\mathcal{O}^\Delta_{ij}$ of a collection of primary operators which are in one-to-one correspondence with bulk $\mathfrak{sl}(2)$ matter representations. We will be able to use this data to define the bulk geodesic Hilbert space.

Note that we need the data for primary operators $\mathcal{O}_\Delta$ dual to \emph{every} matter representation, not just some collection of `fundamental fields'. For example, in a free scalar theory we have not only the boundary limit $\mathcal{O}$ of the bulk field $\chi$ with dimension $\Delta_\mathcal{O}$, but also `double-trace' operators $[\mathcal{O}\mathcal{O}]_n$ with dimension $\Delta=2\Delta_\mathcal{O}+n$ for $n=0,1,2,...$ (built from a product of operators dressed with $n$ time derivatives), and similarly higher multi-trace operators. However, the two-point functions of these will be sufficient. We could instead keep only a finite number of `fundamental' operators but also ask for the data of higher-point functions, considering boundary states with several operator insertions like $|e^{-\tau_L H}\mathcal{O}e^{-\tau_M H}\mathcal{O}e^{-\tau_R H}\rangle$. To relate this to our approach, products $\mathcal{O}e^{-\tau_M H}\mathcal{O}$ can then be decomposed into a sum of primary states using a `Schwarzian-dressed OPE', studied in \cite{Jafferis:2022wez, Kolchmeyer:2023gwa}, also discussed in section \ref{sec:CM-collision-energy}. This depends on the bulk theory only through the three-point functions of boundary operators, which are determined up to a constant by symmetry (just like for CFTs).

The final novelty is that one encounters difficulties in defining the genus expansion for a putative ensemble dual to JT with matter. The reason is that matter loops give a large contribution to spacetimes with small cycles, and so the integral over moduli diverges and must be regulated in some way. There is, therefore, no unambiguous order-by-order calculation of moments like we have for pure JT (though there is nonetheless a rich formal duality \cite{Jafferis:2022wez}). Given this challenge, our approach of defining a single theory by the data ($\{E_i\}$, $\mathcal{O}^\Delta_{ij}$) is particularly beneficial, because it allows us to study certain aspects of non-perturbative bulk physics without directly confronting this issue. The drawback is that we do not have an obvious distinguished ensemble from which to select a representative.

\subsection{The non-perturbative inner product}

To construct the non-perturbative bulk geodesic Hilbert space $\hilb$, the argument we used in section \ref{sec:pureJT} goes through essentially unchanged. Once again, for any amplitude involving asymptotic states $|\Delta;\tau_L,\tau_R\rangle$ we cut the path integral along a geodesic homotopic to the boundary bounding a half-disk, which is assumed to exist for every spacetime in the path integral. The simplest way to express the outcome is that the relation \eqref{eq:HDmatter} (writing the $|\Delta;\tau_L,\tau_R\rangle$ in terms of geodesic states $|\Delta;\ell,m\rangle$, integrated against the half-disk amplitude $Z_{\HD}^\Delta(\tau_L,\tau_R;\ell,m)$) continues to be true non-perturbatively.

We can use this relation to directly write the bulk states $|\Delta;\ell,m\rangle$ in the energy eigenbasis. Taking the inner product of \eqref{eq:HDmatter} with an energy eigenstate  $|i,j\rangle$ and demanding that the result hold for all $\tau_{L,R}$, we find that
\begin{equation}
    \langle i,j|\mathcal{O}^\Delta\rangle \frac{\delta(E_L-E_i)}{\rho_0(E_L)}\frac{\delta(E_R-E_j)}{\rho_0(E_R)}  = e^{S_0/2}   \sum_m \int d\ell\,   \phi^\Delta_{E_L E_R}(\ell,m) \langle i,j|\Delta;\ell,m\rangle\,.
\end{equation}
We can invert this equation using the completeness relation for $\phi^\Delta_{E_{L} E_{R}}(\ell,m)$ to get
\begin{equation}\label{eq:ijDeltalm}
    \langle i,j|\Delta;\ell,m\rangle = e^{-S_0/2} \,  \frac{ \phi^\Delta_{E_i E_j}(\ell,m)}{\Gamma^\Delta_{E_i E_j}} \mathcal{O}^\Delta_{ij} \,,
\end{equation}
where we use the notation writing $ \langle i,j|\mathcal{O}^\Delta\rangle$ as the matrix elements of a boundary operator. The only exception to this is the trivial representation (with matter in the vacuum state) or taking $\mathcal{O}^\Delta$ to be the identity operator, for which we have the result of section \ref{sec:pureJT} restricted to `diagonal' states $i=j$:
\begin{equation}
    \langle i,j|\ell\rangle = \delta_{ij} e^{-S_0/2}  \phi_{E_i}(\ell).
\end{equation}

We can use these results to give expressions for the non-perturbative inner products, by inserting a complete set of energy eigenstates:
\begin{equation}
    \langle \Delta';\ell',m'|\Delta;\ell,m\rangle = e^{-S_0}   \sum_{i,j}\frac{\phi^{\Delta'}_{E_i E_j}(\ell',m')\phi^\Delta_{E_i E_j}(\ell,m)}{\Gamma^\Delta_{E_i E_j}\Gamma^{\Delta'}_{E_i E_j} }(\mathcal{O}^{\Delta'}_{ij})^*\mathcal{O}^\Delta_{ij}.
\end{equation}
Note that if we choose $\Delta=\Delta'$,  replace the squared matrix elements $|\mathcal{O}^\Delta_{ij}|^2$ by the `disk' value $\Gamma^{\Delta}_{E_i E_j}$, and approximate the sum over $i,j$ by an integral $\int dE_L \rho_0(E_L)dE_R \rho_0(E_R)$ with the disk density of states, then we recover the completeness relation for $\phi^\Delta_{E_L,E_R}$ giving the orthonormal result $\delta_{m m'}\delta(\ell-\ell')$.
An interesting special case is the overlap between the trivial representation and a non-trivial matter state, which gives us
\begin{equation}
     \langle \ell'|\Delta;\ell,m\rangle = e^{-S_0}   \sum_{i}\frac{\phi_{E_i}(\ell')\phi^\Delta_{E_i E_i}(\ell,m)}{\Gamma^{\Delta}_{E_i E_i} }\mathcal{O}^\Delta_{ii}.
\end{equation}

\subsection{Null states \& redundancies}

As for pure JT gravity, an outcome is that the bulk Hilbert space described as wavefunctions of $\ell$ and matter states is highly redundant, with many null states. But the inclusion of matter makes this much more striking because a given state can be written as a wavefunction with essentially \emph{any} matter configuration!

 We can write a general geodesic state in the bulk Hilbert space as a wavefunction of length $\ell$ along with a matter state. Decomposing into representations, this can be expressed as a sequence of wavefunctions $\psi_\Delta(\ell,m)$ for each $\Delta$ (along with $\psi_0(\ell)$ for the matter vacuum state):
\begin{equation}\label{eq:bulkstatematter}
    |\psi\rangle = \int d\ell \,\psi_0(\ell) |\ell\rangle + \sum_\Delta \sum_m \int d\ell \, \psi_\Delta(\ell,m) |\Delta;\ell,m\rangle.
\end{equation}
As in section \ref{sec:null}, to understand this state non-perturbatively it is most transparent to first `Fourier transform' the wavefunctions $\psi_\Delta(\ell,m)$ from the $(\ell,m)$ basis to $\hat{\psi}_\Delta(E_L,E_R)$ in an $(E_L,E_R)$ basis, using the kernel $\phi^\Delta_{E_L,E_R}(\ell,m)$ in each fixed $\Delta$ sector:
\begin{equation}
    \begin{aligned}
        \hat{\psi}_\Delta(E_L,E_R) &= \sum_m \int d\ell \,\phi^\Delta_{E_L,E_R}(\ell,m) \psi_\Delta(\ell,m), \\
        \psi_\Delta(\ell,m) &= \int dE_L \rho_0(E_L) dE_R \rho_0(E_R) \frac{\phi^\Delta_{E_L,E_R}(\ell,m)}{\Gamma^\Delta_{E_L,E_R}}\hat{\psi}_\Delta(E_L,E_R),
    \end{aligned}
\end{equation}
where the inverse in the second line follows from the completeness relation for $\phi^\Delta_{E_L,E_R}(\ell,m)$. Using the result \eqref{eq:ijDeltalm} for the overlap of geodesic states with energy eigenstates $|i,j\rangle$, we find that the state \eqref{eq:bulkstatematter} can be written as
\begin{equation}\label{eq:psiij}
    |\psi\rangle = e^{-S_0/2} \sum_{i,j}\left[\delta_{i,j} \hat{\psi}_0(E_i)  + \sum_\Delta \frac{\mathcal{O}^\Delta_{ij}}{\Gamma^\Delta_{E_i,E_j}} \hat{\psi}_\Delta(E_i,E_j)\right]|i,j\rangle.
\end{equation}

From this result, we find that null states appear in two qualitatively different ways. First, as for the case of pure JT, the physical state depends only on the value of $\hat{\psi}_\Delta(E_L,E_R)$ evaluated at energies $E_{L,R}$ belonging to the discrete spectrum $\{E_i\}_{i=0}^\infty$, so null states can be constructed from any function that vanishes at those points. The second is novel: we can have a null state even if $\hat{\psi}_\Delta(E_i,E_j)$ is nonzero for a given $\Delta$, as long as the sum in the square brackets of \eqref{eq:psiij} vanishes.

Generically, we expect all matrix elements $\mathcal{O}^\Delta_{ij}$ to be nonzero (unless there are more symmetries in the boundary theory). This has the striking result that every state can be represented by a bulk state with any matter content we choose!  States $|\Delta;\ell,m\rangle$ span the Hilbert space (and are, in fact, highly overcomplete) even for a single fixed non-trivial representation $\Delta$; this observation was also made in  \cite{Hsin:2020mfa, Balasubramanian:2022gmo, Boruch:2023trc}. Specifically, $|i,j\rangle$ can be written as a matter state in representation $\Delta$ by choosing $\hat{\psi}^\Delta(E_i,E_j) =  \frac{\Gamma^\Delta_{E_i,E_j}}{\mathcal{O}^\Delta_{ij}}$, and $\hat{\psi}^\Delta(E_L,E_R)=0$ for all other pairs of values in the spectrum. For example, a very complicated state containing many particles can be rewritten as a superposition of single-particle states, with some highly-tuned wavefunction $\psi_\Delta(\ell,m)$ for the length and location of the particle.

A specific consequence is that the Casimir $\mathsf{C} = \Delta(\Delta-1)$ which labels representations \cite{Lin:2019qwu} is no longer conserved non-perturbatively: in fact, it is not even a well-defined operator! We will discuss this more in the next section.

\section{Case study II: a center-of-mass collision energy operator}
\label{sec:CM-collision-energy}
In this section, we discuss our second operator case study. We would like a probe of what an infalling observer measures as they fall behind the horizon of an old black hole. We imagine a situation, as studied by \cite{Stanford:2022fdt}, where the observer jumps into an old black hole in the thermo-field double state from the right side. At some time, much earlier, some matter fell into the black hole from the left. We would like to know whether the observer gets hit by this matter or not. Ideally, we would like to understand the observer's experience by computing the center-of-mass energy of any collision that the observer participates in.

\subsection{Semi-classical approximation}
\label{sec:semi-classical-approx}

It will be helpful to understand what the observer sees in the classical limit, where quantum gravitational effects are turned off. In what follows, we will restrict to the case where there is a single operator, $V$, inserted on the left side of the eternal black hole, with the observer represented by an operator $W$ inserted on the right. By the boost symmetry of the thermofield double, we can insert the operators at some time $t_L = t_R = T/2$.  The observer will experience a collision with the excitation generated by $V$. 

The center of mass energy, $\mathfrak{c}$, of this collision is given by the product of their momenta as
\begin{align}
    \mathsf{c} \sim -p_+^{W}q_-^{V} - p_-^{W}q_+^{V}.
\end{align}
As we increase the time at which these two operators are inserted, the momenta will become exponentially boosted\footnote{A quick argument for this is the following. In the setup we are considering, the Casimir is dominated by the global energy $\mathsf{C} \sim \mathsf{H}^2$. At large times, global energy grows exponentially with time, essentially because of the $\mathfrak{sl}_2$ algebra (e.g. by conjugating $\mathsf{H}$ by the boost.)  } so that for times $T>\beta$ we have
\begin{align}
    \mathsf{c}(T) \sim \mathsf{c}_0 \ e^{2\pi T/\beta}.
\end{align}
The exact form of the coefficient in front will be unimportant for us except to say that it depends linearly on both masses of the operators $\mathsf{c}_0 \sim \Delta_W \Delta_V$. Furthermore, the collision energy is enhanced when the operators start out near the boundary. How far out the operators are inserted can be adjusted by moving the operators $V$ and $W$ off into the Euclidean section by a small amount in order to regulate the state. In the end, using that the momenta have units of inverse length, we expect
\begin{align}
    \mathsf{c}_0 \sim \Delta_V \Delta_W/\varepsilon^2,
\end{align}
up to order one coefficients. 

Note that, as illustrated in the figure above, $V$ and $W$ will participate in a violent future collision if they jump in very early or very late. This fact is special to the situation where the geometry is that of a rigid AdS$_2$ as in JT gravity. In more realistic, higher dimensional situations, $V$ and $W$ are expected to collide with high energies only if they jump in very early since a spacelike singularity will prevent them from running into each other if they jump in too late. This means that our center of mass collision energy will actually be symmetric in time, reflecting the fact that the observer sees no difference between late and early times in AdS$_2$.

\subsection{Relation to the Casimir operator}

 Given this semi-classical expectation, we turn now to an exact quantum calculation of an observable that measures the collision energy between $V$ and $W$. Consider the state $\ket{VW}$, where $V$ is inserted on the left at Rindler time $t$ and $W$ on the right at time $0$. As discussed briefly in Sec. \ref{sec:JT-with-matter}, we can expand the resulting state into SL(2,$\mathbb{R}$) representations, labeled by the value of their Casimir

For the scenario where there is a single matter operator inserted on the left at some early time, this $\mathfrak{sl}(2)$ Casimir operator computed in the joint matter state actually measures precisely this center of mass collision energy $\mathfrak{c}$. When we expand the joint matter state into $SL(2,\mathbb{R})$ blocks, the Casimir value associated with these blocks will be given in terms of a discrete index because matter states are labeled by discrete series representations of $SL(2,\mathbb{R})$. We also need to couple these conformal blocks to the boundary particle, Schwarzian mode. This leads to a ``gravitationally dressed" block, discussed at length in \cite{Jafferis:2022wez}.

Since the boundary particle states are associated with principal series representations, the blocks, $P_n^{\Delta_V,\Delta_W}(s_2; s_1,s_3)$, are labeled by six Casimir values, three discrete series representations and three principal series representations. 
They tell us the amplitude for a state with two matter particles, of Casimir given by $\Delta_V, \Delta_W$, and two boundary particles of energies $s_1^2/2, s_3^2/2$ to fuse into a state with a single matter ``particle" of fixed Casimir value labeled by the index $n$, as discussed in Sec. \ref{sec:JT-with-matter}. Here $s_2^2/2$ labels the internal energy for the boundary particle in the region between the matter particle insertions. This can be represented in the equation \cite{Jafferis:2022wez}
\def\red{\color{red}}
\def\blue{\color{blue}}
\begin{align}
   P^{\Delta_V, \Delta_W}_n(s_2;s_1,s_3) \quad = \quad   \begin{tikzpicture}[scale=0.7, baseline={([yshift=-0.1cm]current bounding box.center)}]
    \node at (-1.1,0) {$s_1$} ;
    \node at (1.1,0) {$s_3$} ;
    \node at (0,1.5) {$s_2$} ;
 	\node at (-1.4,1.8) {$\blue \Delta_V$} ;
 	\node at (1.4,1.8) {$\red \Delta_W$} ;
 	\draw[thick, blue] (-1,1.5) -- (0,0.5); 
 	\draw[thick, red] (1,1.5) -- (0,0.5);
 	\draw[thick] (0,0.5) -- (0,-0.72) node[below] {$\Delta_V + \Delta_W + n$};
\end{tikzpicture}\,.\label{P}
\end{align}
One can then expand a four-point function of boundary operator insertions into these gravitationally dressed blocks. Schematically, this decomposition is given by \cite{Jafferis:2022wez}
\begin{align}
\begin{tikzpicture}[scale=0.7, baseline={([yshift=-0.1cm]current bounding box.center)}]
    \node at (-1.8,1.8) {$\blue \Delta_V$} ;
    \node at (1.8,1.8) {$\red \Delta_W$} ;
    \node at (-1.8,-1.8) {$\blue \Delta_V$} ;
    \node at (1.8,-1.8) {$\red \Delta_W$} ;
    \node at (-1.2,0) {$s_1$} ;
    \node at (1.2,0) {$s_3$} ;
    \node at (0,1.2) {$s_2$} ;
    \node at (0,-1.2) {$s_4$} ;
 	\draw[thick, blue] (-1.4,1.4) -- (0,0); 
 	\draw[thick, blue] (-1.4,-1.4) -- (0,0); 
 	\draw[thick, red] (0,0) -- (1.4,-1.4);
 	\draw[thick, red] (0,0) -- (1.4,1.4);
\end{tikzpicture}  \quad 
=  \quad \sum_{n=0}^\infty (-1)^n\,
\begin{tikzpicture}[scale=0.7, baseline={([yshift=-0.1cm]current bounding box.center)}]
    \node at (-1.2,0) {$s_1$} ;
    \node at (1.2,0) {$s_3$} ;
    \node at (0,1.5) {$s_2$} ;
    \node at (0,-1.5) {$s_4$} ;
 	\node at (-1.4,1.8) {$\blue \Delta_V$} ;
 	\node at (1.4,1.8) {$\red \Delta_W$} ;
 	\node at (-1.4,-1.6) {$\blue \Delta_V$} ;
 	\node at (1.4,-1.6) {$\red \Delta_W$} ;
 	\draw[thick, blue] (-1,1.5) -- (0,0.5); 
 	\draw[thick, red] (1,1.5) -- (0,0.5);
 	\draw[thick] (0,0.5) -- (0,-0.5) node[midway, right] {$n$};
 	\draw[thick, blue] (0,-0.5) -- (-1,-1.5);
 	\draw[thick, red] (0,-0.5) -- (1,-1.5);
\end{tikzpicture}\,.\label{fourpointfunction}
\end{align}
The upshot is that these blocks allow us to simply define a wavefunction for the state $\ket{VW}$ in the basis of matter states, $\ket{[VW]_n}$, with definite Casimir value given by
\begin{align}\label{eqn:Cn}
    \mathsf{C}_n = \left(\Delta_V + \Delta_W + n-1\right)\left(\Delta_V + \Delta_W + n\right).
\end{align}
Using the formulae of Appendix C in \cite{Jafferis:2022wez}, we see that the full Casimir probability distribution for the state $\ket{VW}$ is given by
\begin{align}\label{eqn:fulldistributionmain}
   & p(n) = \frac{4}{Z(\beta)\braket{VW}} \times \nonumber \\
&\int_0^{\infty} \prod_{j=1}^4 \left( ds_j \rho(s_j) e^{-\tau_j s_j^2}\right) \left(\Gamma_{12}^{\Delta_V}\Gamma_{23}^{\Delta_W}\Gamma_{34}^{\Delta_W}\Gamma_{41}^{\Delta_V}\right)^{1/2} P_n^{\Delta_V \Delta_W}(s_4;s_1, s_3) P_n^{\Delta_V \Delta_W}(s_2;s_1,s_3),
\end{align}
where here 
\begin{align}
    \Gamma^{\Delta}_{ij} = \frac{\Gamma(\Delta \pm \i \sqrt{2E_i} \pm \i \sqrt{2E_j})}{4^{\Delta -1} \Gamma(2\Delta)}
\end{align}
is the necessary vertex factor for a matter particle to join with a boundary particle. Using the completeness identity 
\begin{align}
    \frac{\delta(s_2-s_4)}{\rho_0(s_2)}=4\sum^{\infty}_{n=0} P_n^{\Delta_V, \Delta_W}(s_2;s_1, s_3) P_n^{\Delta_V, \Delta_W}(s_4;s_1,s_3),
    \label{schannelIdentity}
\end{align}
one can check that this distribution is indeed normalized. This completeness relation can be summarized in the following schematic \cite{Jafferis:2022wez}
\begin{align}
\begin{tikzpicture}[scale=0.7, baseline={([yshift=-0.1cm]current bounding box.center)}]
    \node at (-2.2,1.8) {$\blue \Delta_V$} ;
    \node at (2.2,1.8) {$\red \Delta_W$} ;
    \node at (-2.2,-1.6) {$\blue \Delta_V$} ;
    \node at (2.2,-1.6) {$\red \Delta_W$} ;
    \node at (-2.2,0) {$s_1$} ;
    \node at (2.2,0) {$s_3$} ;
    \node at (0,1.5) {$s_2$} ;
    \node at (0,-1.5) {$s_4$} ;
 	\draw[thick, blue] (-2,1.3) -- (-1,0); 
 	\draw[thick, blue] (-2,-1.3) -- (-1,0); 
 	\draw[thick, dashed] (-1,0) -- (1,0) node[midway,above] {$\mathbf{1}$} ;
 	\draw[thick, red] (1,0) -- (2,-1.3);
 	\draw[thick, red] (1,0) -- (2,1.3);
\end{tikzpicture}  \quad 
=  \quad \sum_{n=0}^\infty 
\begin{tikzpicture}[scale=0.7, baseline={([yshift=-0.1cm]current bounding box.center)}]
    \node at (-1.2,0) {$s_1$} ;
    \node at (1.2,0) {$s_3$} ;
    \node at (0,1.5) {$s_2$} ;
    \node at (0,-1.5) {$s_4$} ;
 	\node at (-1.4,1.8) {$\blue \Delta_V$} ;
 	\node at (1.4,1.8) {$\red \Delta_W$} ;
 	\node at (-1.4,-1.6) {$\blue \Delta_V$} ;
 	\node at (1.4,-1.6) {$\red \Delta_W$} ;
 	\draw[thick, blue] (-1,1.5) -- (0,0.5); 
 	\draw[thick, red] (1,1.5) -- (0,0.5);
 	\draw[thick] (0,0.5) -- (0,-0.5) node[midway, right] {$n$};
 	\draw[thick, blue] (0,-0.5) -- (-1,-1.5);
 	\draw[thick, red] (0,-0.5) -- (1,-1.5);
\end{tikzpicture}\label{identityBlock}
\end{align}

Note that in the distribution  \eqref{eqn:fulldistributionmain}, we integrate over the energies $E_i = s_i^2/2$ to move us from fixed boundary energies between the operator insertions to fixed time separation. We would like to understand this Casimir distribution for states with a large, Lorentzian time separation between the operators $V$ and $W$. This means we are interested in the kinematics
\begin{align}
\tau_4 = \varepsilon-\i 2\pi T/\beta, \ \tau_1 = -\pi+\varepsilon,\ \tau_2 = \pi- \varepsilon,\ \tau_3 = -\varepsilon-\i 2\pi T/\beta,
\end{align}
where the $\tau$'s are angular coordinates on the boundary thermal circle and are associated with the bulk AdS$_2$ metric $ds^2 = d\rho^2 + \sinh^2 \rho \,d\tau^2$. Here $\varepsilon$ is a small parameter that, roughly speaking, controls how far out near the boundary we insert the operators $V$ and $W$. Thus, we will be interested in the scaling limits $\varepsilon \ll 1 \ll T/\beta \lesssim \beta \log 1/G_N$.

It is informative to compute the expectation value of the Casimir in the semi-classical limit, where Schwarzian effects are suppressed in order to match with the behavior of the CM collision energy described in section \ref{sec:semi-classical-approx}. In this limit, we have an emergent $SL(2, \mathbb R)$ conformal symmetry, and so it is convenient to express everything in terms of the conformal cross ratio relevant to this correlation function. This ratio is
\begin{align}\label{eqn:crossratio}
\chi = \frac{\sin \frac{\tau_{13}}{2} \sin \frac{\tau_{42}}{2}}{\sin \frac{\tau_{14}}{2} \sin \frac{\tau_{32}}{2}} = \cos^2 \varepsilon + \sin^2 \varepsilon \tanh^2 \frac{\pi T}{\beta}
\end{align}
As $T$ increases, $\chi$ moevs along the real axis from a bit less than 1, approaching 1 as $T \to \infty$.

When the Schwarzian effects are weakly coupled, the authors of \cite{Jafferis:2022wez} worked out the form of these gravitational blocks. Not surprisingly, they are given in terms of $SL(2,\mathbb{R})$ conformal blocks. Plugging in $\mathsf{C}_n$ from \eqref{eqn:Cn} into the distribution \eqref{eqn:fulldistributionmain}, we get 
\begin{align}
&\ev{\mathsf{C}} = \sum_n p(n) \mathsf{C}_n = \sum_{n} \mathsf{C}_n \frac{\left(2\Delta_W\right)_n \left(2\Delta_V\right)_n}{n! (2\Delta_W + 2\Delta_V + n-1)_n}  \times \nonumber \\
&(1-\chi)^{2\Delta_W} \chi^{n} \,{}_2 F_1(n+ 2\Delta_W, n +2\Delta_W, 2\Delta_W + 2 \Delta_V + 2n,\chi).
\end{align}
Using equations (352) and (353) in \cite{Lin:2023trc}, we get the answer
\begin{align}\label{eqn:semiclassicalcasimir}
\ev{\mathsf{C}} = \frac{4\Delta_V \Delta_W \chi}{1-\chi} + (\Delta_V + \Delta_W)(\Delta_V + \Delta_W -1).
\end{align}
Using eq. \eqref{eqn:crossratio}, we see that $\ev{\mathsf{C}} \sim 4 \Delta_V\Delta_W \frac{\cosh^2\left(\pi T/\beta\right)}{\varepsilon^2}\sim  \Delta_V\Delta_W \frac{e^{2\pi T/\beta}}{\varepsilon^2}$ as $T \to \infty$. As expected from the semi-classical discussion, this grows exponentially in time, is proportional to the product of dimensions, and is enhanced at small $\varepsilon$.

\subsection{Perturbative probability of detecting a fixed Casimir}

We would now like to compute the probability distribution, $p(n)$, for Casimir given in eq. \eqref{eqn:fulldistributionmain} in more detail. Since these gravitationally dressed conformal blocks are quite difficult to work with analytically, we save a discussion of this distribution using the explicit form of these blocks for Appendix \ref{app:casimir}.  

In the rest of this section, we take a simpler route to computing $p(n)$ by arguing that at late times the Casimir operator is approximately proportional to the exponential of the length operator, which measures the geodesic length between the two boundary particles \cite{Harlow:2018tqv}. In other words, in order to find the Casimir distribution for the state $\ket{VW}$ at large times, we just need to know the length distribution in $\ket{\beta}$ at large times. 

To understand the claim that the Casimir is approximately $e^{\ell}$, we note that to find the effect of gravity on the Casimir distribution, we just need to couple the in-falling particles to the Schwarzian mode. This means that we can just compute the Casimir associated to a state with $V$ and $W$ inserted at some position in rigid AdS$_2$. Then, to couple to the Schwarzian mode, we just make these positions dynamical, weighted by the measure induced by the Schwarzian action.

Alternatively, we can compute the Casimir for two operators inserted at a definite position in rigid AdS$_2$, re-write this expression in terms of gauge-invariant observables (like the length), and then use known results for the wavefunctions for these observables in the Hartle-Hawking state. We will take this route.

Actually, we can relate the Casimir operator to the length operator more generally. In the previous section, we have already encountered the $\mathfrak{sl}(2)$ Casimir (see eq. \ref{eq:casimir_def}), which is defined for JT gravity $+$ quantum fields. We note that one can write an explicit expression for the Casimir in terms of the length operator by inverting equation \ref{eq:HR}  to solve for $\mathsf{H}, \mathsf{B}, \mathsf{P}$ in terms of the $\ell, H_L, H_R$. One arrives at %
\begin{align}
\mathsf{C} = -(p_L -p_R)^2 + \frac{1}{2} e^{\ell} \big\{  H_L-\frac{1}{2} p_L^2 - 2 e^{-\ell},  H_R-\frac{1}{2} p_R^2 -2 e^{-\ell} \big\} .
\end{align}
where we have introduced
two operators that are (at a perturbative level) conjugate\footnote{This follows from the JT gravitational algebra of \cite{Harlow:2021dfp}. For a proof of this algebra in quantum JT, see the appendix of \cite{Lin:2022rbf}.} to the length:
\begin{align} \label{eq:conjugateP}
    p_{L/R} = -\i [H_{L/R}, \ell], \quad \i [p_{L/R},\ell] = 1
\end{align} (As an aside, by promoting $\ell \to \hat{\ell}_\Delta$ one could define a non-perturbative Casimir operator.)

Assuming the $V$ and $W$ particles only interact via gravity, then with gravitational Schwarzian interactions tuned off, we can just use the conformal answer for the Casimir. This is given in \eqref{eqn:semiclassicalcasimir}. Plugging in the formula for the conformal cross ratio in \eqref{eqn:crossratio} and expanding at small $\varepsilon$, we have
\begin{align}
    \mathsf{C} = \frac{\beta^2 \Delta_V \Delta_W}{\pi^2 \varepsilon^2} \cosh^2 \frac{\pi T}{\beta},
\end{align}
where we also moved from Rindler to Schwarzschild time coordinates, rescaling $\varepsilon \to 2\pi \varepsilon/\beta$.

Using the formula (2.32) in \cite{Harlow:2018tqv}, we see that the time $T$ between the two insertions is, in fact, related to the renormalized geodesic length between the two points by 
\begin{align}
    \log \cosh ^2 \pi T/\beta = \ell + \log \frac{\pi^2}{\beta^2}.
\end{align}
Plugging this in above, we have that
\begin{align}
    e^{\ell} = \frac{\mathsf{C}}{4\mathsf{C}_0},\ \ \mathsf{C}_0 \equiv \frac{\Delta_V \Delta_W}{4\varepsilon^2}. 
\end{align}
Thus, we see that for late times and for small enough $\varepsilon$, the Casimir operator is essentially just the exponential of the length operator. In what follows, we will not be careful about tracking all possible perturbative corrections. Some of these corrections can shift the definition of $C_0$ by order one factors. Again, for a more careful presentation of the Casimir distribution, the interested reader should consult Appendix \ref{app:casimir}.

Using the wavefunctions for the length operator, with the normalization convention given in \eqref{eq:phiEl}, we have 
\begin{align}
   \mathsf{C} p(\mathsf{C}) = \frac{16}{Z(\beta)}\int_0^{\infty} \! ds_1 ds_2 \rho_0(s_1) \rho_0(s_2)  e^{-\frac{\beta}{4}(s_1^2 + s_2^2) +i\frac{T}{2}(s_1^2 -s_2^2)} K_{2is_1}\!\!\left(\! 8 \sqrt{\frac{\mathsf{C}_0}{\mathsf{C}}}\right) K_{2is_2}\!\!\left(\! 8 \sqrt{\frac{\mathsf{C}_0}{\mathsf{C}}}\right).
\end{align}
The factor of $\mathsf{C}$ on the left-hand side is from switching between the flat measure on $\ell$ to that on $\mathsf{C}$. Note that this distribution is normalized (since the $\ell$ distribution is normalized). 

Assuming the distribution is dominated at late times by increasingly large $C$, we can expand the Bessel functions at small argument. We get
\begin{align}\label{eqn:latetimedist}
    p(\mathsf{C}) &= \frac{4}{Z(\beta) \mathsf{C}}\int_{-\infty}^{\infty} ds_1 ds_2 \rho_0(s_1) \rho_0(s_2) \times e^{-\frac{\beta}{4}(s_1^2 + s_2^2) +i\frac{T}{2}(s_1^2 -s_2^2)} \nonumber \\
    & \times \Gamma(-2is_1) \Gamma(2is_2) \exp \left(i (s_1 - s_2) \log \frac{4 \mathsf{C}_0}{\mathsf{C}}\right) \nonumber \\
    &\equiv \frac{1}{Z(\beta) \mathsf{C}}f(\mathsf{C}/\mathsf{C}_0).
\end{align}
Note that the original Casimir distribution $p(n)$ was a function of a discrete index. Here, we have lost the discreteness. This is because we have taken the large Casimir limit and so have, in a sense, coarse-grained over the discreteness of $C$. In Appendix \ref{app:casimir}, we show more explicitly how the discrete observable becomes continuous by obtaining eq. \eqref{eqn:latetimedist} more directly from the gravitationally-dressed conformal blocks.

Given eq. \eqref{eqn:latetimedist}, we would like to evaluate this probability distribution more explicitly. To make our lives easier, we will consider a black hole that is semi-classical at $T =0$ by also taking the limit $1/\beta \gg 1$ so that Schwarzian effects are suppressed at early times. We will also be interested in the large time and small $\varepsilon$ limit, and so we take $\beta/\varepsilon$ and $T/\beta$ large. To perform the $s$ integrals in \eqref{eqn:latetimedist} in this limit, we can find saddle points in $s_1$ and $s_2$. One natural place to look is at large $s$. As we will see, such saddles always exist for arbitrarily large $T$. It will be convenient to switch to sum and difference coordinates on $s$-space given by 
\begin{align}
    s_{\pm} = \frac{s_1 \pm s_2}{2}.
\end{align}
Then we look for saddles in the limit where $s_+ \gg s_-$ and $s_+ \gg 1$. We expand out the Gamma function factors in the denominator of \eqref{eqn:latetimedist} to get 
\begin{align}
    f(x) = \frac{1}{8\pi^3} \int ds_+ds_-\times e^{-\frac{\beta}{2}( s_+^2 + s_-^2) +2iTs_+s_- + 2\pi |s_+| + \log |s_+| -4i s_- \log |s_+|} \exp \left(-2i s_- \log \frac{x}4\right).
\end{align}
Now, in the limit that $|Ts_+ s_-| \gg \beta s_-^2$ then, the expression in the exponential becomes linear in $s_-$. We thus get a $\delta$-function constraint on $s_+$, imposing the equation
\begin{align}\label{eqn:smonshell}
2T s_+ -4 \log s_+ - 2 \log \frac{x}4 = 0.
\end{align}
Solving this equation, we see that as long as we scale $\log x \sim T/\beta$ then $ s_+ \sim 1/\beta \gg 1$; thus, the delta function is consistent with our assumptions on the saddle point. Integrating over this $\delta$-function in $s_+$ and dividing by the partition function to normalize the distribution, we have
\begin{align}\label{eqn:growingdist}
    \mathsf{C} p(\mathsf{C}) = \frac{ \beta^{3/2}}{\sqrt{8\pi^3}T^2} \times \log \frac{\mathsf{C}}{4\mathsf{C}_0} \times \exp \left(  - \frac{\beta}{2T^2} \left(\log \frac{\mathsf{C}}{4\mathsf{C}_0} - \frac{2\pi |T|}{\beta}\right)^2 \right).
\end{align}
One can check that this distribution is normalized by integrating over $ \mathsf{C}$ from $0$ to $\infty$. 
\begin{figure}
    \centering
    \includegraphics[width=.75\columnwidth]{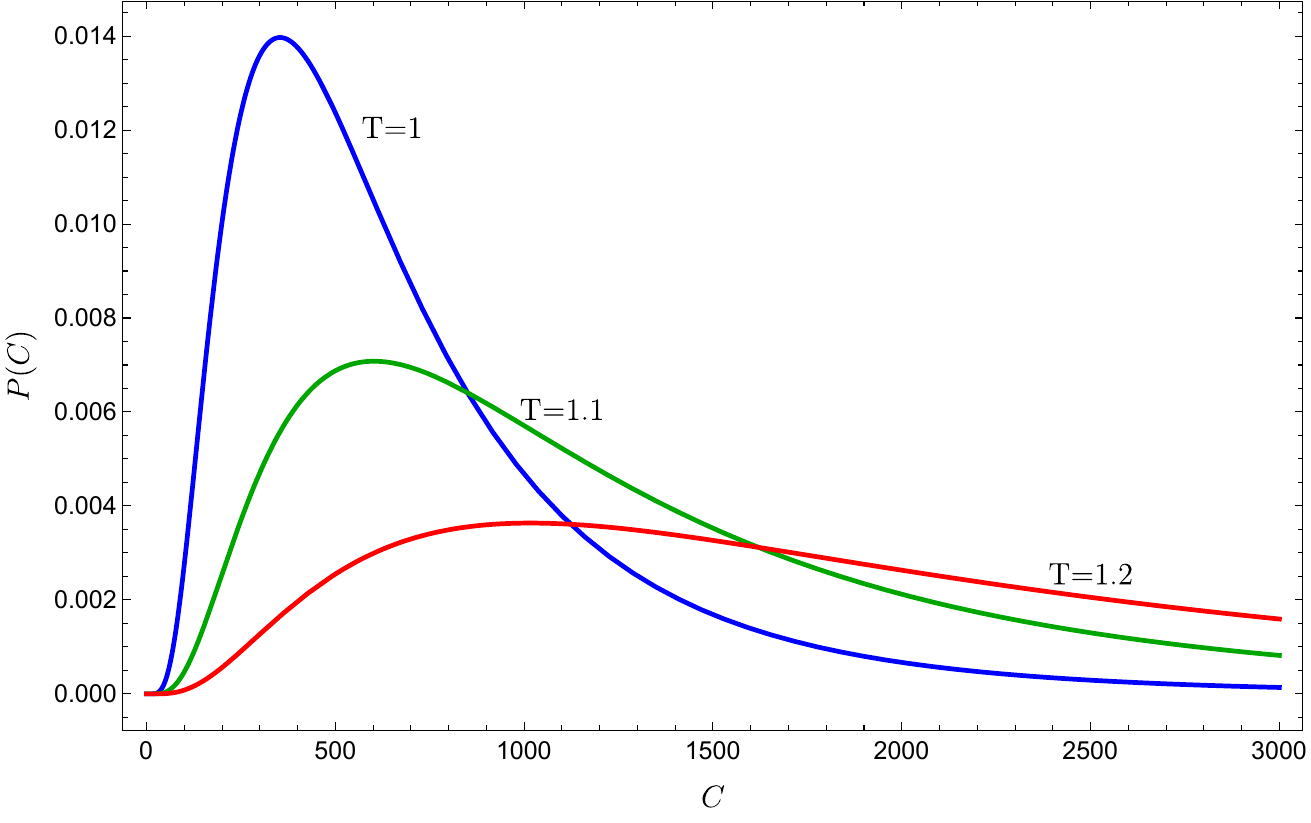}
    \caption{We illustrate the Casimir distribution for various times, showing how the mean and standard deviation broaden with increasing times. We have chosen the temperature $\beta =1$ and set $C_0=1$.}
    \label{fig:casdist}
\end{figure}

Note that this is a Gaussian in $\log \mathsf{C}$, and so the standard deviation of this distribution also grows exponentially in time. This distribution has a standard deviation of order $e^{S_0}$ at times only of order $S_0$. Naively, this suggests that perhaps non-perturbative corrections to the Casimir distribution kick in at much earlier times than naively thought. Furthermore, this distribution implies that for times $T>1$, the expectation value of the Casimir, in fact, grows super exponentially like an inverted Gaussian, $e^{T^2}$. The reason is that distributions over a variable $x$, which are Gaussian in $\log x$, have moments that are dominated by the tails of the distribution. When discussing what an observer may or may not measure for such a distribution, it is best to discuss probabilities instead of moments since moments do not characterize the distribution well.

\subsection{A violation of charge conservation for the Casimir operator}
Given the distribution in eq. \eqref{eqn:growingdist}, we can ask how many states with Casimir, $\mathsf{C}_n \sim n^2$, labeled by the index $n$ are within one standard deviation from the center of the Gaussian in \eqref{eqn:growingdist}. This is given by 
\be 
n_\text{max} - n_\text{min} = \sqrt{C_0}\left(e^{\frac{\pi T}{\beta}  \left(1 +\frac{1}{2\pi} \sqrt{\beta}\right)}  - e^{\frac{\pi T}{\beta}  \left(1 -\frac{1}{2\pi} \sqrt{\beta}\right)} \right) 
\ee
Thus, when $T \approx \frac{2S_0}{1 +\frac{1}{\pi} \sqrt{\beta}}$, the number of states that have an equal probability is $n_\text{max} - n_\text{min} 
\sim e^{2S_0}$. These states then have a probability $\mathcal{P}(n) \sim e^{-2S_0}$, for $n \in (n_\text{min},\,n_\text{max})$. Since the number of states is proportional to the dimension of the Hilbert space, it is at this time that we expect our expansion of $\ket{VW}$ into Casimir eigenstates, labeled by a discrete index $n$, to no longer have meaning.  %
 
To emphasize how the expansion into exact Casimir eigenstates stops having any meaning at times of order $S_0$, note that we have so far assumed that the Casimir is a Hermitian operator whose eigenvalues are set by the scaling dimension $\Delta_n = \Delta_V+ \Delta_W + n$ with the eigenvalue $\mathsf{C}_n = \Delta_n(\Delta_n-1)$. Assuming Hermiticity, the eigenvectors associated with different values of $n$ are, therefore, orthogonal to each other.

As is by now standard in the study of JT gravity, these states are in fact not orthogonal when including non-perturbative corrections. Consider the inner-product of two naive eigenstates of $ \hat{\mathsf{C}}$, $\braket{[VW]_n}{[VW]_m}$, for $n$, $m\gg 1$. If one considers the transition probability between two such states, this can be viewed as a fundamentally two-boundary observable. These two boundaries can be connected by a wormhole. This wormhole encodes how the states $\ket{[VW]_n}$ and $\ket{[VW]_m}$ are no longer orthogonal. This probability then receives a contribution of the form
\be 
|\braket{[VW]_n}{[VW]_m}|^2 &= \begin{tikzpicture}[baseline={([yshift=-.5ex]current bounding box.center)}, scale=0.35]
 \pgftext{\includegraphics[scale=1.8]{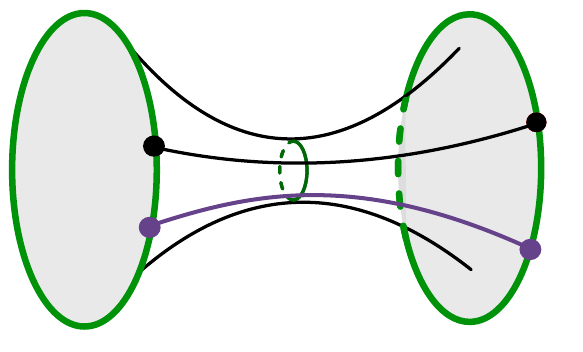}} at (0,0);
 \draw (-6.5,+1.05) node  {$[VW]_n$};
 \draw (-6.5,-1.05) node  {$[VW]_m$};
 \draw (9.5,+1.05) node  {$[VW]_n$};
 \draw (9.5,-1.4) node  {$[VW]_m$};
  \end{tikzpicture}\nonumber\\ &= 4^4 \rho(E_1) \rho(E_3) \int d\ell \, d\ell' e^{-\Delta_n \ell} e^{-\Delta_m \ell'} K_{2i s_1}(4 e^{-\frac{\ell}2}) \times \nonumber \\ &\qquad \qquad \times K_{2i s_1}(4 e^{-\frac{\ell'}2}) K_{2i s_3}(4 e^{-\frac{\ell}2}) K_{2i s_3}(4 e^{-\frac{\ell'}2})  \nonumber \\ 
&= \rho(E_1) \rho(E_3) \frac{\Gamma(\Delta_n\pm \i \sqrt{2E_1} \pm \i \sqrt{2E_3})}{2^{2\Delta_n - 2} \Gamma(2\Delta_n)}\, \frac{\Gamma(\Delta_m\pm \i \sqrt{2E_1} \pm \i \sqrt{2E_3})}{2^{2\Delta_m - 2} \Gamma(2\Delta_m)}\,,
\ee
where the \textcolor{OliveGreen}{green} boundary segments, in this case, have fixed ADM energies given by $E_1$ and $E_3$.
To obtain the above result, we used the fact that at large $\Delta_n$ and large $\Delta_m$, the double-trace operators $[VW]_n$ and $[VW]_m$ are only connected to identical operators on the opposite boundary. 
The states should be normalized by the norm of each state, 
\be 
\braket{[VW]_n}{[VW]_n} = e^{S_0}  \rho(E_1) \rho(E_3) \frac{\Gamma(\Delta_n\pm \i \sqrt{2E_1} \pm \i \sqrt{2E_3})}{2^{2\Delta_n - 2} \Gamma(2\Delta_n)}\,.
\ee
Thus, the transition probability between states with different Casimirs is given by 
\be 
\frac{|\braket{[VW]_n}{[VW]_m}|^2 }{\braket{[VW]_n}{[VW]_n} \braket{[VW]_m}{[VW]_m}} = e^{-\left(S(E_1) + S(E_3)\right)} = \frac{e^{-2S_0}}{\rho(E_1) \rho(E_3)}\,.
\ee 
For this reason, we can think of the conservation of the Casimir as being violated. This is exactly analogous to the violations of other global symmetries due to wormhole effects in \cite{Chen:2020ojn,Hsin:2020mfa}. Moreover, we can ever have a non-zero inner product between states containing $V$ and $W$ and states they are absent. Thus, in principle, the observer could even disappear due to non-perturbative corrections.  

Such corrections drastically affect the expectation value of the Casimir in the original state
\be 
\label{eq:Casimir-non-pert}
\langle \mathsf{C} \rangle = \sum_n \Delta_n(\Delta_n-1) {|\braket{VW}{[VW]_n}|^2}.
\ee
The inner-product ${|\braket{VW}{[VW]_n}|^2}$ can be computed from the gravitational path integral following the same procedure used for ${|\braket{[VW]_m}{[VW]_n}|^2}$. Specifically, the leading non-perturbative correction is given by the wormhole connecting the two inner products that appear in the overlaps, which for large values of $\Delta_V$ and $\Delta_W$ is  once again given by 
\be 
\label{eq:inner-prod-VW-VWn}
|\braket{V W}{[VW_n]}|^2_\text{conn.} = \frac{e^{-2S_0}}{\rho(E_1) \rho(E_3)}
\ee
Thus, these non-perturbative corrections dominate over the perturbative result for all $n$ and all times $T\gtrsim 2S_0$. 
These corrections also imply that the expectation value of the Casimir is divergent for all values of $T$ since \eqref{eq:inner-prod-VW-VWn} is completely $n$-independent and $\Delta_n(\Delta_n-1) \sim n^2$ at large $n$.

One important point is that the overlaps $|\braket{V W}{[VW_n]}|^2$ have the same scaling with $S_0$ as the average overlap $|\braket{V W}{X}|^2$, where $\ket{X}$ denotes a Haar random state, in which the observer $V$ and the particle $W$ might not be present. There is, however, a difference which makes the states $\ket{[VW]_n}$ atypical. While the probabilities $|\braket{V W}{[VW_n]}|^2$ and $|\braket{V W}{X}|^2$ are the same on average, the amplitudes obey
\begin{align}\label{eqn:nothaar}
\frac{\braket{VW}{[VW]_n}}{\sqrt{\braket{VW} \braket{[VW]_n}}} \sim e^{-\# S_0}\,, \ \text{while} \ \ \frac{\braket{VW}{X}}{\sqrt{\braket{VW} \braket{X}}} = 0\,.
\end{align}
In other words, the basis of states $\ket{[VW]_n}$ is close to being typical, but the phases of the overlaps with $\bra{VW}$ are not completely random.

The results above thus suggest that at times $t \sim S_0$, an observer can, in principle, detect large deviations for the eigenvalues and associated probabilities of the exact Casimir operator $\hat{\mathsf{C}}$. Whether an observer can actually measure properties of the exact Casimir, i.e.~measure the CM collision energy exactly, we do not know. One can imagine replacing the exact Casimir operator with a coarse-grained version. However, we were not yet able to find a definition for a coarse-grained Casimir for which such a breakdown is not visible when $t \sim S_0$. We hope to further analyze this in the near future.

\subsection{Non-perturbative definition of the Casimir}
Since non-perturbative corrections make different eigenfunctions of the perturbative Casimir non-orthogonal, we can no longer associate a probability to each term in the sum over $n$. Just as in the case of the length operator in pure JT gravity without matter, we are thus forced to come up with a definition of the Casimir operator that makes sense non-perturbatively. One such definition amounts to measuring the Casimir on every time slice (in between the two boundary points where we were previously measuring the Casimir in the perturbative calculation). In the $\ket{\Delta, \ell, m}$ basis introduced in Sec. \ref{sec:JT-with-matter}, this version of the Casimir operator takes the form
\begin{align}
    \hat{\mathsf{C}} = \sum_{\Delta} \sum_m \int d\ell\, \Delta (\Delta -1) \ket{\Delta, \ell, m} \bra{\Delta, \ell, m}.
\end{align}
Using the non-perturbative inner product, the states $\ket{\Delta, \ell, m}$ are no longer orthogonal, and so this operator is not diagonal anymore. For this reason, we can also define the operator 
\begin{align}
    \hat{\mathsf{C}}_{\mathcal{F}} = \mathcal{F}^{-1} \left( \sum_{\Delta} \sum_m \int d\ell\,  \mathcal{F} \left(\Delta (\Delta -1) \right) \ket{\Delta, \ell, m} \bra{\Delta, \ell, m}\right) ,
\end{align}
for any invertible function $\mathcal{F}$. For all such $\mathcal{F}$, this operator will agree with the perturbative Casimir up to non-perturbative corrections. Expanding into the energy eigenbasis, these operators will be diagonal up to non-perturbative effects, which will account for the non-conservation of the operator.

This same story also applies to the coarse-grained Casimir, for which there are an infinite number of non-perturbative completions, $\hat{\mathsf{C}}_{\mathrm{coarse-grained}}^{\mathcal{F}}$, also labeled by an invertible function $\mathcal{F}$. As mentioned in the previous sub-section, if we work with $\log \hat{\mathsf{C}}_{\mathrm{coarse-grained}}^{\mathcal{F}}$ instead of $\hat{\mathsf{C}}_{\mathrm{coarse-grained}}^{\mathcal{F}}$, we expect the long time behavior of this observable to match that of the operators $\hat{\ell}_{\mathcal{F}}$, discussed in Case I.

\section{Discussion}
\label{sec:discussion}

\subsection{Firewalls}

One of the main motivations for studying the bulk Hilbert space of quantum gravity is to understand the experience of an observer that is part of a spacetime where non-perturbative effects are important. To quantify this, we considered two bulk observables in JT gravity, both of which could serve as proxies for the experience of an observer falling into a two-sided black hole: a wormhole length operator with its associated velocity operator and the CM collision operator between the observer and a particle falling from the opposite side. In going between the perturbative and non-perturbative definitions of these observables, there is a large ambiguity -- there are an infinite number of operators specified by the function $\mathcal F(\ell)$, which all agree perturbatively. Out of all such operators, which one truly captures the experience of the infalling observer? One possibility is that the exact operator that describes the experience of the observer sensitively depends on the details of the experiment that they perform. In such a case, one might think that it is impossible to find a perfect proxy for the observer's experience without fully specifying both the details of the experiment and those of the UV completion of the theory. Nevertheless, the universality of the results shown in section \ref{sec:a-length-operator} makes us hopeful that one can gain a qualitative understanding without specifying all such details. Regardless of the exact definition of the length operator (e.g., regardless of our choice of $\Delta$ in the function $\mathcal F(\ell) = e^{-\Delta \ell}$), we found that the length plateaued at a time $t\sim e^{S_0}$. Consequently, the expectation value of the associated velocity operator vanished at those times, and because the spectrum of the velocity operator is symmetric around $\pi_\Delta=0$, the probability of detecting a negative velocity was $1/2$. Nevertheless, the late-time wavefunction in the velocity basis always had an $O(1)$ probability for eigenvalues that were not associated with a good semi-classical geometry. This suggests that while the firewall operator can be exactly defined at the perturbative level (to have eigenvalues +1 if a firewall is detected and $-1$ if not) once inner products are modified, the definition of such an operator is only approximate: an $O(1)$ fraction of its eigenvalues are non-perturbatively modified to now take values very different than $\pm 1$. The corresponding eigenstates would not have a good semi-classical description (i.e., far from $\pm \sqrt{2\bar E}$) and the probability of detecting such states at late times is $O(1)$.  Even if such states exist, we found that the remaining probability of detecting a state that has an approximate white hole description and thus has a firewall is $O(1)$, making the problem of studying the firewall operator non-perturbatively still sensible and physically interesting.    

Another issue that complicates our analysis is that the notion of an observer is also only approximate at late times. As seen in section \ref{sec:CM-collision-energy}, states, where the observer $V$ is inserted, have non-perturbatively small overlaps with states where no observers (or, equally confusingly, multiple observers) are present. Thus, if the velocity operator is a good proxy for firewalls, which ones of its eigenstates can actually be detected by an observer? Presumably, eigenstates whose eigenvalue are close to $\sqrt{2\bar E}$ contain the observer since they describe early-time semi-classical physics. Similarly, eigenstates whose eigenvalue are close to $-\sqrt{2\bar E}$ should also contain the observer since they also have a good semi-classical description: they describe the state of the black hole at $t<0$ with $|t|\ll e^{S_0}$. However, for gray hole states whose eigenvalues are far from $\pm \sqrt{2\bar E}$, the presence of an observer is unclear. We hope to clarify this question in the near future by studying whether any such states have an $O(1)$ probability for an observer to be present. 

One assumption that we've made in our above interpretation is that the experience of the infalling observer is correctly captured by a linear operator. Whether or not this is true is unclear. On the one hand, in our construction, we can, in principle, associate a probability for the overlap of any two gravitational states in which we specify the state of the infalling observer. On the other hand, it is unclear whether other geometric definitions of the length operator, such as the shortest geodesic length between two points or the related definition proposed in \cite{Stanford:2022fdt},  properly act on the bulk Hilbert space that we defined. One risk is that when acting on null states with such operators, the resulting states would no longer be null; thus, such operators would not respect the modified inner products that we derived. We summarize the status of a few past proposed definitions of the length operator in the table below. 

It would also be interesting to explore the proposal of \cite{Penington:2023dql}, who also define a bulk Hilbert space, in more detail. Unlike our proposal, their inner product is not diagonal in the $\ket{\alpha}$ basis \cite{Marolf:2020xie}. 
\begin{table}[h!]
\begin{tabular}{ |p{3cm}||p{3.5cm}|p{3cm}|p{3cm}|p{3cm}|  }
 \hline
 \multicolumn{5}{|c|}{{\bf Length definitions}} \\
 \hline
Authors & Reproduces length on the disk  & Non-perturbatively well-defined & Respects inner product & Relevant for infalling observer\\
 \hline
Iliesiu, Mezei, Sarosi \cite{Iliesiu:2021ari}  & yes, after a subtraction    &  ? $\to \checkmark$  & ? $\to \checkmark$ &   ??? \\
\hline
Stanford \& Yang \cite{Stanford:2022fdt}  &  $\checkmark$  & Only for a subclass of topologies ($g=0 \,\& \,1$)  & ? & ??? \\
\hline
present work & $\checkmark$ & $\checkmark$ & $\checkmark$ & ??? \\
 \hline
\end{tabular}
\caption{A summary of previous proposals for the non-perturbative definition of the length operator.}
\end{table}

\subsection{The conditions underlying our construction}\label{ssec:conditions}

Our construction of the inner products of geodesic states $|\ell\rangle$ rested crucially on the assertion that every asymptotic boundary segment is uniquely associated with a bulk geodesic, such that they bound a half-disk. Here we critically examine the circumstances under which this is true.

In the calculations with the 
JT ensemble in section \ref{ssec:ensemble}, for any spacetime appearing in the path integral (which localizes to constant curvature geometries) this is a fact of hyperbolic geometry. So, there is no doubt that our results make sense for all moments, to all orders in the genus expansion. Nonetheless, it is something of a leap to claim that it holds non-perturbatively at finite $S_0$. A definitive argument would require a complete bulk definition of the theory (beyond the topological expansion), which may not even exist. In particular, we would ideally like a bulk description of a single member of the ensemble, giving us a definite discrete spectrum $\{E_i\}$.

While this lack of a full bulk definition precludes a complete argument, we can gain some confidence by studying modifications of the JT path integral which partially fix the ensemble. There are two closely related approaches. One proposal is to add `eigenbranes', new boundary conditions on which spacetimes can end, with each brane designed to fix a single eigenvalue of the dual Hamiltonian \cite{Blommaert:2019wfy, Blommaert:2021gha}. Alternatively one can add additional asymptotic boundaries, which modify the probability distribution of Hamiltonians as described in \cite{Marolf:2020xie}. Alternatively, the addition of asymptotic boundaries that fix the spectrum to that of a single member of the ensemble can be viewed as turning on a non-local interaction in the gravitational theory \cite{Blommaert:2021fob, Blommaert:2022ucs}. In all such cases, the necessary property continues to hold with these modifications. This at least takes us closer to a single member of the ensemble with a geometric definition and without violating our key assumption. 

Nonetheless, the requirement that all amplitudes factorize on half-disks in this way is not completely innocuous. For a general theory of gravity, one might hope that there is an unambiguous way to identify a `nearest' geodesic (or locally minimal volume hypersurface in higher dimensions) to a given asymptotic boundary in any (Euclidean) geometry. If this held, then the region between an asymptotic boundary and the corresponding hypersurface would be analogous to our half-disk. Unfortunately, for a general geometry, no such slice exists.

To understand these issues better, a class of interesting models is given by deformations of JT gravity which include dynamical conical defects \cite{Maxfield:2020ale, Witten:2020wvy, Kruthoff:2024gxc}. In the presence of such defects some geometries do not split on half-disks in the way that pure JT does, so this provides a tractable set of models where we can study the consequences of this failure.

\subsection{A non-perturbative Wheeler-DeWitt equation}\label{ssec:NPWDW}

The non-perturbative bulk Hilbert space discussed above consists of arbitrary wavefunctions in the perturbative variables (general wavefunctions $\psi(\ell)$ for pure JT), but with a degenerate inner product and hence identifications between wavefunctions which differ by a null state. There is an alternative (mathematically equivalent) way of describing the Hilbert space, which has a non-degenerate inner product, instead restricting the allowed states. This restriction is a non-perturbative analogue of the Wheeler-DeWitt equation, which imposes invariance under infinitesimal diffeomorphisms.

Algebraically, we can explain this by thinking of our inner product on $\hilb$ as an operator $\eta$ in the perturbative Hilbert space $\hilb_0$:
\begin{equation}
    \langle \psi'|\psi\rangle = \langle \psi'|\eta |\psi\rangle_0 \,.
\end{equation}
Then, informally we can think of $\hilb$ as the quotient by null states $\hilb_0/\ker \eta$. But the cosets $\{|\psi\rangle + |\chi\rangle: \eta|\chi\rangle=0\}$ in this quotient are in one-to-one correspondence with their image $|\Psi\rangle = \eta|\psi\rangle$. So an equivalent way to describe $\hilb$ uses wavefunctions in this image, $\hilb \simeq \operatorname{im} \eta\subseteq \hilb_0$. In this way of doing things, we reduce the number of physical states using a constraint on allowed states rather than equivalence under null states. The inner product on such wavefunctions can then be thought of as a generalized inverse $\eta^g$ of $\eta$ (satisfying $ \eta \eta^g \eta = \eta$) so that $\langle \Psi'|\Psi\rangle = \langle\Psi|\eta^g|\Psi\rangle_0$. Since $\eta$ is degenerate, there are many such $\eta^g$, so there are many ways to write this form of the inner product in the length basis.

To explain this from the path integral (restricting to pure JT for simplicity), consider two possible ways to obtain a `length wavefunction' from an asymptotic state $|\tau\rangle$. In our construction, we would use the wavefunction $\psi_\tau(\ell)$ obtained from the half-disk path integral $Z_\HD(\tau;\ell)$, satisfying $|\tau\rangle = \int d\ell \,\psi_\tau(\ell) |\ell\rangle$. By varying the asymptotic boundary conditions, we can get any wavefunction this way, but the inner product on such length wavefunctions is degenerate so we must quotient-out the null states. An alternative wavefunction $\Psi_\tau(\ell)$ is computed by the path integral over \emph{all} spacetimes bounded by the asymptotic boundary segment and geodesic including all topologies, which we would write as $\Psi_\tau(\ell) = \langle \ell|\tau\rangle$. In this case, $\Psi$ can not be an arbitrary function of $\ell$; by varying boundary conditions, we can obtain only linear combinations of $\phi_{E_i}(\ell)$ for energies $E_i$ in the spectrum. But the set of all such allowed wavefunctions is now in one-to-one correspondence with the physical states, so we can use the alternative length wavefunction $\Psi(\ell)$ to describe the Hilbert space without null states.

These two alternative descriptions are analogous to two possible constructions of Hilbert spaces with constraints, such as the Hamiltonian constraint $\mathcal{H}=0$ of gravity. These are the `co-invariant' construction where physical states are cosets with identifications $|\psi\rangle\sim |\psi\rangle + \mathcal{H}|\chi\rangle$ for any $|\chi\rangle$, and the `invariant' construction where we restrict to states annihilated by constraints $\mathcal{H}|\Psi\rangle=0$ (the Wheeler-DeWitt equation). In this analogy, the restriction $|\Psi\rangle\in \operatorname{im} \eta$ is a non-perturbative version of the Wheeler-DeWitt equation. The path integral construction of $\Psi$ above fits this analogy, since such path integrals satisfy the Wheeler-DeWitt equation. One way to think about this paper is that we have already accounted for perturbative diffeomorphisms by `gauge-fixing' to a geodesic slice, but since non-perturbatively a spacetime can have several geodesic slices, there are residual large diffeomorphisms to deal with. Then $|\Psi\rangle\in\operatorname{im} \eta$ expresses invariance under these residual diffs. It would be interesting to make this more precise since it gives an explanation of the null states as arising from redundancy under a gauge symmetry.

\subsection{An enlarged bulk Hilbert space with baby universes}\label{ssec:BUdisc}

When we have topology change, we could have processes where an initial geodesic state evolves to a disconnected space by emitting a closed  `baby universe'. It is, therefore, natural to expect that the Hilbert space (with two spatial asymptotic boundaries as considered here) is described not only by a slice connecting the boundaries, but also allows for one or more closed universe components. It is possible to construct this in JT, and we arrive at a different Hilbert space from the one described here. This will be explored in future work, which we preview here.

Strictly, the full space of states even perturbatively (at infinite $S_0$) contains closed universes. We can write the full (two-sided) perturbative Hilbert space as $L^2(\bR) \otimes \bigoplus_{n=0}^\infty \operatorname{Sym}^n L^2(\bR_+)$, where $L^2(\bR)$ is $\hilb_0$ as a wavefunction of geodesic length $\ell$ considered above, while the second factor is a Fock space of closed universes ($n$ denoting the number of such universes). $L^2(\bR_+)$ is the single-universe Hilbert space, given by wavefunctions of a positive geodesic length $b$. The generalisation of our geodesic states $\psi(\ell)$ is a collection of wavefunctions $\psi_n(\ell;b_1,\ldots,b_n)$ for $n=0,1,2,\ldots$. Once we include topology change the components with different values of $n$ will mix, replacing the Fock space with an interacting many-universe Hilbert space with a non-diagonal inner product.

We recover our construction of $\hilb$ by `tracing out' the closed components. When we do this, the result will depend on the state of these closed pieces. One possibility is to put them in the `Hartle-Hawking' no-boundary state, in which case we recover the SSS ensemble average for the remaining piece as in section \ref{ssec:ensemble}, with the squared wavefunction for the closed universes (in the $\alpha$-state basis) reinterpreted as probabilities in the ensemble \cite{Marolf:2020xie}. Another is to choose an $\alpha$-state for the closed components, in which case we recover our construction with a definite spectrum. We can always do this if we consider only operators that depend on the $\ell$ piece but not the closed part of the wavefunction. But it is interesting to keep track of the closed components and to study operators which depend on the $b$ variables.

A particularly interesting outcome is that we can construct natural bulk operators on this Hilbert space that do not commute with boundary operator insertions (for example, operators $\hat{Z}(\beta)$ that add an asymptotic circle as described in  \cite{Marolf:2020xie}). An example is the total geodesic length $\ell+b_1+\cdots+b_n$ including all baby universe components, defined using the same ideas as we used in section \ref{sec:defining-non-perturbative-bulk-ops}. This failure to commute means that such operators are not `superselected' in the algebra of asymptotic operators, so they are not diagonal in the basis of $\alpha$ states for closed universes. These operators cannot be given an ensemble interpretation: they map one member of the ensemble to a superposition over all members. This should be contrasted with the class of operators discussed in this paper, $\hat{\ell}_{\Delta}$ and $\hat{\mathsf{C}}_{\mathcal F}$, which have a universal definition in each member of the ensemble and can hopefully be defined for actual holographic CFTs.%

\subsection{Relation to non-isometric codes}

It is instructive to consider our construction of a non-perturbative Hilbert space in the language of non-isometric codes \cite{Akers:2022qdl} (see also \cite{Penington:2019kki}). For us, a non-isometric map $V$ would be a linear map from the perturbative to the non-perturbative Hilbert space  $V:\hilb_0 \to \hilb$. In this way of doing things, the `null states' in $\hilb_0$ are defined as the kernel of $V$.

An obvious choice of $V$ maps the geodesic state $\int d\ell \psi(\ell)|\ell\rangle$ regarded as an element of the perturbative Hilbert space $\hilb_0$ to the same geodesic state in $\hilb$: roughly, this is a quotient map that takes a wavefunction $\psi$ to the coset (under adding an arbitrary null state) containing $\psi$. (We restrict to pure JT for simplicity, but similar comments apply with matter.) This gives
\begin{equation}\label{eq:Vdef}
    |\psi\rangle = \int d\ell \psi(\ell)|\ell\rangle \in \hilb_0 \implies V|\psi\rangle = e^{-S_0/2}\sum_i  \hat{\psi}(E_i) |i\rangle \in \hilb,
\end{equation}
where $\hat{\psi}(E)$ is the energy-transform of $\psi$ introduced in section \ref{sec:null}. More precisely, this $V$ is defined on a domain consisting of square integrable functions $\psi(\ell)$ such that the integral $\hat{\psi}(E_i) = \int d\ell \, \psi(\ell) \phi_{E_i}(\ell) $ exists for all $i$, and  the norm $e^{-S_0}\sum_i |\hat{\psi}(E_i)|^2$ of the result is finite. 

A natural condition that singles out \eqref{eq:Vdef} is compatibility with maps taking asymptotic boundary conditions to states. That is, if we define a perturbative state by some boundary condition and then apply $V$, we should get the same result as directly defining the non-perturbative state with the same boundary condition. Using TFD states, this means that \eqref{eq:Vdef} must hold for $\psi(\ell) = Z_\HD(\tau;\ell)$ for any $\tau>0$.%

An interesting subtlety here is that we could have negative energies $E_i<0$ in the physical spectrum. Since any square-integrable perturbative wavefunction $\psi(\ell)$ can be written as a superposition $\int_0^\infty dE \,\rho_0(E) \phi_E(\ell) \hat{\psi}(E)$ of positive-energy states, one might think that \eqref{eq:Vdef} would simply restrict to \emph{positive} energies in the non-perturbative spectrum, and the negative energy states would not be in the image of $V$. However, if we take our guiding principle for defining $V$ as compatibility with asymptotic states, this is not the case! Instead, there are slightly stronger restrictions on $\psi(\ell)$ in the domain of $V$ so that the integral defining $\hat{\psi}(E_i)$ always exists,\footnote{This will require $\psi(\ell)$ to decay faster than an exponential $e^{-\sqrt{-2E_i}\ell}$, in which case $\hat{\psi}(E)$ will have an analytic extension to the required negative energy. This is similar to the standard Fourier transform in the variable $k = \sqrt{2E}$, where exponential decay guarantees analyticity of the Fourier transform in a strip.} but the range of $V$ is nonetheless dense in $\hilb$, even covering negative energies.

As a word of warning, since the non-isometric map $V$ acts by evaluating functions of a continuous energy at specific discrete values, it is extremely badly behaved from a technical point of view. This leads to many counterintuitive properties and indicates that $V$ is probably not very physical.\footnote{One might think that negative energies make $V$ especially badly behaved in some technical way. For example, $V$ is not closable because there is a sequence of states $|\psi_n\rangle\in\hilb_0$ which tend to zero but such that $V|\psi_n\rangle$ has a non-zero limit (an example is $e^{E_i\tau_n}$ times the TFD state $|\tau_n\rangle$ where $E_i<0$ is a negative energy in the spectrum). But in fact, $V$ fails to be closable even if there are no negative energies (consider a sequence such that $\hat{\psi}(E)$ is unity at $E=E_i$ but has increasingly narrow support around this energy)! We do not know of any specific property of $V$ that's any worse when we have a spectrum containing negative energies.} 

Finally, note that when we include matter in the gravitational Hilbert space, one generally expects non-perturbative effects to kick in once the number of ``relevant'' bulk states is of order $e^S$. If one considers throwing in one of several different perturbations into the black hole at a constant rate, the number of states grows exponentially with time. Thus one might guess that already at a time of order $\sim S$ some firewall-like phenomena might occur (see section 5 of \cite{Akers:2022qdl}). We are investigating whether this might be related to the non-perturbative corrections to the expectation value of the Casimir, as discussed around (\ref{eq:inner-prod-VW-VWn}).

\subsection{Embedding into other microscopic models}
We have discussed the case of JT gravity (pure or with matter). We could imagine obtaining this as a limit of a more general bulk theory. For example, one can consider the SYK model in the appropriate large $N$, low temperature $\beta J \gg 1$ limit. In this context, we would expect further perturbative and non-perturbative effects that go beyond what we have already analyzed. For example, in the double scaling limit, there are new perturbative corrections to the inner product coming from the discreteness of the length. In the SUSY SYK context, there are also non-perturbative corrections to the length wavefunction, see \cite{Boruch:2023bte}. 

In the context of near-extremal black holes, one could imagine embedding our discussion in a variety of more conventional holographic models. For example, one can consider a near-extremal BTZ black hole \cite{Ghosh:2019rcj} in AdS$_3$ holography, and it is plausible that there could be a full 3D gravity version of our construction.\footnote{See \cite{Chua:2023ios} for a discussion about the bulk perturbative construction of the Hilbert space for 2-sided BTZ black holes analogous to our discussion in section \ref{sec:pureJT}. We believe this could serve as a starting point for analyzing the non-perturbative bulk Hilbert space in 3D gravity.} In the SUSY context, we could consider BPS black holes in $\mathcal{N} = 4$ SYM \cite{Gutowski:2004ez,Gutowski:2004yv,Chong:2005hr,Kinney:2005ej,Choi:2018hmj,Benini:2018ywd, Cabo-Bizet:2018ehj}.

\section*{Acknowledgments}

We are grateful to Douglas Stanford and Zhenbin Yang for countless discussions. We also thank Geoff Penington, Stephen Shenker, Scott Collier, Daniel Harlow and David Kolchmeyer  for useful comments. HL is supported by a Bloch Fellowship and by NSF Grant PHY-2310429. LVI was supported by the Simons Collaboration on Ultra-Quantum Matter, a Simons Foundation Grant with No.~651440, during part of this work. MM is supported in part by the STFC grant ST/X000761/1. AL is supported through the Packard Foundation as well as the Heising-Simons Foundation under grant no. 2023-4430.

\appendix

\section{The Hilbert space and eigenfunctions in the boundary particle formalism}\label{app:JTparticle}

\subsection{Hilbert space from the boundary particle formalism}

One way to construct the Hilbert space with matter uses the `boundary particle' formulation of JT gravity \cite{Yang:2018gdb,Kitaev:2018wpr}. Our starting point is the following expression ((6.74) in \cite{Yang:2018gdb}) for the disk-level correlation functions of JT coupled to matter (we will not be particularly careful with factors of 2 in the normalisation here):
\begin{align}\label{eq:Kcorr}
    &\langle \mathcal{O}_1(\tau_1)\cdots \mathcal{O}_n(\tau_n)\rangle_\mathrm{Disk} = \\
    &\; \int \left(\frac{1}{\mathrm{V}(SL(2,\mR))}\prod_{i=1}^n \frac{dx_i dz_i}{z_i^2}\right) \tilde{K}(\tau_{21};\mathbf{x}_1,\mathbf{x}_2)\cdots  \tilde{K}(\tau_{1n};\mathbf{x}_n,\mathbf{x}_1)  z_1^{\Delta_1}\cdots z_n^{\Delta_n} \langle \mathcal{O}_1(x_1)\cdots \mathcal{O}_n(x_n)\rangle_{QFT} \,. \nonumber
\end{align}
Here $\mathbf{x} = (x,z)$ denotes a position on the line $x\in \mR$ along with a `depth' $z$ into the hyperbolic plane, which is an infinitesimal version of the upper half-plane $y$ coordinate; $SL(2,\mR)$ acts as $(x,z)\mapsto \left(\frac{a x+b}{c x+d},\frac{z}{(c x+d)^2}\right)$. The volume of this group in the measure of the integral indicates that this symmetry should be gauge-fixed. The Euclidean times $\tau_1\cdots \tau_n$ are cyclically ordered around the disc, and the integral is over similarly cyclically ordered $x_i$'s (so $x_1<x_2<\cdots < x_n$ or a cyclic rearrangement of this; this is the opposite ordering convention to \cite{Yang:2018gdb}).

The propagators in this formula are given in terms of the half-disk path integral in \eqref{eq:ZHD} by
\begin{equation}\label{eq:Ktilde}
    \tilde{K}(\tau;\mathbf{x}_1,\mathbf{x}_2) = 2e^{-2\frac{z_1+z_2}{x_2-x_1}} e^{-\ell_{12}/2} Z_\HD(\tau;\ell_{12}),
\end{equation}
where $\ell_{12}$ is the renormalised distance between the points $\mathbf{x}_1,\mathbf{x}_2$ in the hyperbolic plane; this is the only $SL(2,\mR)$-invariant quantity formed from $\mathbf{x}_1,\mathbf{x}_2$, given by
\begin{equation}\label{eq:length}
    \ell_{12} = \log\frac{(x_2-x_1)^2}{z_1z_2}.
\end{equation}
The prefactor in $\tilde{K}$ is not $SL(2,\mR)$ invariant; it requires a choice of `gauge'. But once we include all the propagators the resulting cyclic combination
\begin{equation}
   \exp\left(-2 \sum_{i=1}^n \frac{z_i+z_{i+1}}{x_{i+1}-x_i}\right)
\end{equation}
is invariant. The exponent is proportional to $A-(n-2)\pi$, where $A$ is the area of the hyperbolic polygon with vertices $\mathbf{x}_i$. For the case $n=3$ of a triangle, we get the quantity defined in \eqref{eq:Iformula},
\begin{equation}\label{eq:I3}
    \exp\left(- 2\frac{z_1+z_2}{x_2-x_1}-2\frac{z_2+z_3}{x_3-x_2}-2\frac{z_3+z_1}{x_1-x_3}\right) = I(\ell_{12},\ell_{23},\ell_{31}).
\end{equation}

We would like a Hilbert space interpretation of \eqref{eq:Kcorr} as an inner product, which means splitting it into two pieces (a `bra' and `ket'), with a residual integral which computes the inner product. We split at times $\tau_L$ and $\tau_R$, and relabel insertions so that $n$ operators lie between $\tau_L$ and $\tau_R$; write $\tau_i$ for the time interval between the $i$th and $(i+1)$th operator, with $\tau_0$ the time between $\tau_0$ and the first insertion, and $\tau_n$ between the $n$th insertion and $\tau_R$. This leaves $n'$ insertions on the other side, which we label in the opposite order. To make the split, we use a composition relation (5.44) for the propagators:
\begin{equation}
   \tilde{K}(\tau_0'+\tau_0;\mathbf{x}_1',\mathbf{x}_1) =  \int \frac{dx_L dz_L}{z_L^2} \tilde{K}(\tau_0';\mathbf{x}_1',\mathbf{x}_L)\tilde{K}(\tau_0;\mathbf{x}_L,\mathbf{x}_1)\,,
\end{equation}
and similarly on the right between the $n$ and $n'$ operators. We will associate the $\tilde{K}(\tau_0;\mathbf{x}_L,\mathbf{x}_1)$ with the `ket', the $\tilde{K}(\tau_0';\mathbf{x}_1',\mathbf{x}_L)$ with the `bra', and the integral will become part of the inner product.

Having done this, we strip out the pieces of the correlator associated to the ket state, defining
\begin{equation}\label{eq:psiK}
    \begin{aligned}
     |\Psi(\mathbf{x}_L,\mathbf{x}_R)\rangle = e^{\frac{\ell}{2}} e^{2\frac{z_R+z_L}{x_R-x_L}}\int &\prod_{i=1}^n \frac{dx_i dz_i}{z_i^2} \tilde{K}(\tau_0;\mathbf{x}_L,\mathbf{x}_1)\tilde{K}(\tau_1;\mathbf{x}_1,\mathbf{x}_2)\cdots \tilde{K}(\tau_n;\mathbf{x}_n,\mathbf{x}_R) \\
     &\qquad z_1^{\Delta_1}\cdots z_n^{\Delta_n}  \mathcal{O}_1(x_1)\cdots \mathcal{O}_n(x_n)|\Omega\rangle \,,
\end{aligned}
\end{equation}
 The integral runs over $x_i$ with $x_L<x_1<x_2<\cdots <x_n<x_R$ (assuming $x_L<x_R$). The $e^{\frac{\ell}{2}}$ term is introduced to give us the usual flat measure on $\ell$ as we will see in a moment. The remainder of the prefactor makes the `area terms' $SL(2,\mR)$ invariant by completing the similar prefactors in $\tilde{K}$ into a cyclic combination as explained above. This state is a function of $\mathbf{x}_L,\mathbf{x}_R$ valued in the matter Hilbert space ($|\Omega\rangle \in \hilbM$ is the $SL(2)$-invariant vacuum). But it is only nontrivially a function of a single combination of $\mathbf{x}_L,\mathbf{x}_R$, namely the invariant length $\ell$ as defined in \eqref{eq:length} ($SL(2,\mR)$ acts in the usual way on $\mathbf{x}_L,\mathbf{x}_R$ and also on the matter Hilbert space). We do not lose any information by `gauge-fixing', for example taking $x_L=-1$, $x_R=+1$ and $z_L=z_R = 2e^{-\ell/2}$. The upshot is that \eqref{eq:psiK} is a function of $\ell$ valued in the matter QFT Hilbert space, which is the perturbative Hilbert space described in section \ref{sec:JT-with-matter}.

To combine this with another state to get a correlator we need to define the inner product. First,  the inner product on the matter Hilbert space  is similar to `radial quantisation'  for the Hilbert space on a sphere $S^{d-1}$ CFTs in $d\geq 2$ dimensions. For this, the Hermitian conjugate of an operator includes a reflection $x\mapsto \sigma(x)$, where $\sigma$ is an `inversion': the orientation-reversing $GL(2,\mR)$ map which fixes $x_L$ and $x_R$ and squares to the identity (e.g., for $x_L=-1,x_R=1$, we have $\sigma(x)=1/x$). Explicitly, this acts as $(\mathcal{O}(x))^\dag = |\sigma'(x)|^\Delta \mathcal{O}^\dag (\sigma(x))$. This leaves the part of the inner product which integrates over $\mathbf{x}_L,\mathbf{x}_R$, which is
\begin{equation}
    \int \frac{1}{\mathrm{V}(SL(2,\mR))}\frac{dx_L dz_L}{z_L^2}  \frac{dx_R dz_R}{z_R^2} \longrightarrow \int d\ell e^\ell \,,
\end{equation}
where the measure over $\ell$ results from gauge-fixing $x_L,x_R$ to any constants and $z_L=z_R$. The extra factor of $e^{\frac{\ell}{2}}$ in \eqref{eq:psiK} is designed to give the $e^\ell$ factor appearing here so we get the usual flat measure on length.

As a basic check, the $n=0$ case (with no insertions) gives us back the half-disk.

\subsection{One-particle Hilbert space}

We can now use this formalism to rewrite the asymptotic states $|\Delta;\tau_L,\tau_R\rangle$ introduced in section \ref{sec:JT-with-matter} as wavefunctions of the length valued in the single-particle Hilbert space of the matter sector. Since this sector is identified with a single irreducible representation of $\mathfrak{sl}(2)$, this is equivalent to getting the half-disk wavefunction.

Including the factor of $e^{S_0/2}$, our formula \eqref{eq:psiK} gives
\begin{equation}
     |\Delta;\tau_L,\tau_R\rangle = e^{S_0/2}e^{\ell/2} e^{2\frac{z_R+z_L}{x_R-x_L}}\int  \frac{dx dz}{z^2} \tilde{K}(\tau_L;\mathbf{x}_L,\mathbf{x})\tilde{K}(\tau_R;\mathbf{x},\mathbf{x}_R) 
      z^{\Delta}   \mathcal{O}^\Delta(x)|\Omega\rangle,
\end{equation}
where we recall that we are regarding the state as a wavefunction of $\ell$ valued in the matter Hilbert space. We can make this more explicit by inserting the expression \eqref{eq:Ktilde} for the propagators. We get the simplest expressions for the energy eigenstates $|\Delta;E_L,E_R\rangle$ from stripping off energy integrals in the propagators, getting
\begin{equation}
    |\Delta;E_L,E_R\rangle = e^{\ell/2}\int  \frac{dx dz}{z^2} I(\ell,\ell_L,\ell_R) e^{-(\ell_{L}+\ell_{R})/2}
       \phi_{E_L}(\ell_{L})\phi_{E_R}(\ell_{R})  z^{\Delta} \mathcal{O}^\Delta(x)|\Omega\rangle.
\end{equation}
We have written the propagators in terms of renormalised lengths $\ell_L,\ell_R$ as  in figure~\ref{fig:halfdiskmatter}, which are related to $x,z$ by
\begin{equation}
    x = \tanh\left(\frac{\ell_L-\ell_R}{4}\right), \quad z =\frac{e^{(\ell-\ell_L-\ell_R)/2}}{2\cosh^2\left(\frac{\ell_L-\ell_R}{4}\right)}.
\end{equation}
Writing the integration measure in terms of $\ell_{L,R}$ instead of $x,z$ gives us
\begin{equation}
    |\Delta;E_L,E_R\rangle = \frac{1}{2}\int d\ell_L d\ell_R I(\ell,\ell_L,\ell_R) \phi_{E_L}(\ell_L)\phi_{E_R}(\ell_R) z^\Delta \mathcal{O}^\Delta(x)|\Omega\rangle.
\end{equation}
Here if we rewrite the operator insertion $\mathcal{O}^\Delta$ in the $u$ coordinate (including a conformal transformation), we can write this as
\begin{equation}
    |\Delta;E_L,E_R\rangle =  \int du \,  \phi^{\Delta}_{E_L,E_R}(\ell,u)\mathcal{O}(u)|\Omega\rangle,
\end{equation}
where $\phi^{\Delta}_{E_L,E_R}(\ell,u)$ is precisely the eigenfunction \eqref{eq:phiELERlu} given in the main text.

\subsection{Evaluating the overlap of energy eigenfunctions}

We here compute the overlap of two energy eigenstates,
\begin{equation}
    \int dl du du' \left(2\cosh\left(\tfrac{u-u'}{2}\right)\right)^{-2\Delta} \phi^{\Delta}_{E_L',E_R'}(\ell,u')\phi^{\Delta}_{E_L,E_R}(\ell,u),
\end{equation}
which fixes the normalisation. The main idea is to use the knowledge that these must be orthogonal, so we only need to extract the coefficient of $\delta(E_L-E_L')\delta(E_R-E_R')$ which arises from the oscillatory tails of the integral (we will identify precisely the region which contributes in the following). For this we need the asymptotics
 \begin{equation}\label{eq:psiEasymp}
     \phi_E(\ell) \sim 2^{1-2i k}\Gamma(2ik)e^{ikl} + \mathrm{c.c} \quad\text{as } \ell\to\infty, \qquad E=\frac{k^2}{2}.
 \end{equation}

We first rewrite $\phi^{\Delta}_{E_L,E_R}(\ell,u)$ in terms of convenient variables $\tilde{\ell}_L = \frac{\ell}{2}+u$ and $\tilde{\ell}_R = \frac{\ell}{2}-u$, which we can interpret as the renormalised length from the particle's location to the boundary along the geodesic on either side. In the integral \eqref{eq:phiELERlu} we change integration variables to $s_{L,R}$ defined by $\ell_L = \tilde{\ell}_L+s_L$ and $\ell_R = \tilde{\ell}_R+s_R$, and the $\delta$-function fixes $s_L=s_R=s$ leaving us with
\begin{equation}\label{eq:psiLRwavefunction2}
    \phi_{E_L,E_R}(\ell,u) = 2\exp\left[-2(e^{-\tilde{\ell}_L}+e^{-\tilde{\ell}_R})\right] \int ds\,  e^{-\Delta s-2e^{-s}}  \phi_{E_L}(s+\tilde{\ell}_L)\phi_{E_R}(s+\tilde{\ell}_R).
\end{equation}
We want to extract the asymptotics as $\tilde{\ell}_{L,R}\to\infty$, so we take this limit (with the integration variable $s$ held fixed). The integrand is simplified using the asymptotics \eqref{eq:psiEasymp}, and the remaining $s$ integral yields a $\Gamma$-function:
\begin{equation}\label{eq:phiLRexpansion}
    \phi_{E_L,E_R}(\tilde{\ell}_L,\tilde{\ell}_R) \sim \sum_{\pm k_L,\pm k_R} 2^{3-\Delta-i k_L-i k_R}\Gamma(2ik_L)\Gamma(2ik_R)\Gamma(\Delta-i k_L-i k_R)e^{ik_L \tilde{\ell}_L+ik_R \tilde{\ell}_R} . 
\end{equation}
The sum  is over four terms for all possible sign flips of $k_{L,R}$.

The $\delta$-function contribution to the overlap of two states will come from the regions $\tilde{\ell}_{L,R}\to\infty$, so we would like to find the contribution from integrating plane waves
\begin{equation}
     \int d\ell  du du' \left(2\cosh\left(\tfrac{u-u'}{2}\right)\right)^{-2\Delta} e^{ik_L \tilde{\ell}_L+ik_R \tilde{\ell}_R-ik_L' \tilde{\ell}_L'+ik_R' \tilde{\ell}_R'},
\end{equation}
and we will subsequently add these with the coefficients found in \eqref{eq:phiLRexpansion}. On the right hand side, we have $\tilde{\ell}_{L,R} = \frac{1}{2}\ell \pm u$ and  $\tilde{\ell}_{L,R}' = \frac{1}{2}\ell \pm u'$. We first change variables to `averages' and `differences', $\hat{\ell}_L = \frac{1}{2}(\tilde{\ell}_L+\tilde{\ell}_L') =\frac{1}{2}(\ell + u+u')$, $\hat{\ell}_R =\frac{1}{2}(\tilde{\ell}_R+\tilde{\ell}_R') =\frac{1}{2}(\ell - u-u')$, $\delta =\tilde{\ell}_L-\tilde{\ell}_L' = \tilde{\ell}_R'-\tilde{\ell}_R= u -u'$, to get
\begin{equation}
     \int d\hat{\ell}_L d\hat{\ell}_R  d\delta \left(2\cosh\tfrac{\delta}{2}\right)^{-2\Delta} e^{i(k_L-k_L') \hat{\ell}_L+i(k_R-k_R') \hat{\ell}_R+i\frac{1}{2}(k_L+k_L'-k_R-k_R')\delta}.
\end{equation}
The $\hat{\ell}_{L,R}$ integrals give us $\delta$-functions\footnote{The coefficient of each $\delta(k-k')$ is $\pi$, since we are only picking up the part from the $\hat{\ell}\to+\infty$ tail. We also get a singular imaginary part proportional to the principal value of $\frac{1}{k-k'}$, but this cancels when we add up all the terms later.} setting $k_L=k_L'$ and $k_R=k_R'$.  The remaining $\delta$ integral is a Fourier transform that we can evaluate\footnote{It is easiest to check this by evaluating the inverse Fourier transform, which is a sum of residues in a half-plane coming from one of the $\Gamma$-functions.}, giving
\begin{equation}
     \pi\delta(k_L-k_L')\pi\delta(k_R-k_R') \frac{\Gamma(\Delta-i k_L+i k_R)\Gamma(\Delta+i k_L-i k_R)}{\Gamma(2\Delta)}.
\end{equation}
If we insert this result in the overlap integral, we have four equally contributing terms, with the final result
\begin{equation}
    \int dl du du' \left(2\cosh\left(\tfrac{u-u'}{2}\right)\right)^{-2\Delta} \phi^{\Delta}_{E_L',E_R'}(\ell,u')\phi^{\Delta}_{E_L,E_R}(\ell,u) =   \frac{\delta(E_L-E_L')}{\rho_0(E_L)}\frac{\delta(E_R-E_R')}{\rho_0(E_R)} \frac{\Gamma(\Delta \pm i k_L \pm i k_R)}{4^{\Delta-1}\Gamma(2\Delta)}. 
\end{equation}
This recovers the usual JT two-point function, as normalised  .

\section{Deriving the Casimir distribution from the Gravitational OPE}
\label{app:casimir}

Our goal in this section will be to derive the probability distribution in eq. \eqref{eqn:latetimedist} for the Casimir in the state $V_R(T)W_L(0)\ket{\beta/2}_{LR}$ by using the explicit form of the gravitational blocks in terms of Wilson polynomials. We begin with the exact expression for the Casimir distribution given by the expansion of the state overlaps into conformal blocks given in eq. \eqref{eqn:fulldistributionmain} namely 
\begin{align}\label{eqn:fulldistributionapp}
   & p(n) = \frac{4}{Z(\beta)\braket{VW}} \times \nonumber \\
&\int_0^{\infty} \prod_{j=1}^4 \left( ds_j \rho(s_j) e^{-\tau_j s_j^2}\right) \left(\Gamma_{12}^{\Delta_V}\Gamma_{23}^{\Delta_W}\Gamma_{34}^{\Delta_W}\Gamma_{41}^{\Delta_V}\right)^{1/2} P_n^{\Delta_V \Delta_W}(s_4;s_1, s_3) P_n^{\Delta_V \Delta_W}(s_2;s_1,s_3)
\end{align}
As explained in \cite{Jafferis:2022wez}, the blocks $P^{\Delta_V, \Delta_W}_n$ are related to the Wilson polynomials of degree $n$, $W_n$, by
\begin{align}
P_n(x;s_1,s_3) = \frac{1}{\sqrt{r_n} }\left(\Gamma(\Delta_V \pm is_1 \pm ix) \Gamma(\Delta_W \pm is_3 \pm ix)\right)^{1/2} W_n(x)
\end{align}
with $r_n$ given by
\begin{align}
r_n = n! \frac{(n-1 + 2(\Delta_V + \Delta_W))_n}{\Gamma(2n + 2\Delta_V + 2\Delta_W)} \times \left(\Gamma(2\Delta_V + n) \Gamma(2\Delta_W +n) \Gamma(\Delta_V + \Delta_W \pm is_1 \pm is_3 +n) \right).
\end{align}

We will only be interested in the distribution $p(n)$ at large Casimir values, which corresponds to large $n$. This is because at late time we expect the Casimir distribution to be dominated by large values. We then want to expand the Wilson polynomials at large $n$. These have an asymptotic expansion at large $n$ of the form \cite{Li_2019}
\begin{align}
W_n(x) =C_n \left( e^{2ix \log n} A(ix) + c.c.\right)
\end{align}
with $C_n$ defined to be
\begin{align}
C_n = (2\pi)^{3/2} e^{-3n} n^{3n+ 2(\Delta_W + \Delta_V) -3/2}.
\end{align}
and 
\begin{align}
A(z) = \frac{\Gamma(2z)}{\Gamma(\Delta_V \pm is_1 + z) \Gamma(\Delta_W \pm is_3 + z)}.
\end{align}
At large $n$, it turns out that
\begin{align}
C_n/\sqrt{r_n} \approx \sqrt{\frac{2}{n}}.
\end{align}
So the full integral we want to do for the distribution at a given large $n$ is 
\begin{align}\label{eqn:fulldistatlargen}
&\frac{8 \times 4^{2-\Delta_V-\Delta_W}}{n\Gamma(2\Delta_V) \Gamma(2\Delta_W)} \int \left(\prod_{i=1}^4 \rho(s_i) ds_i\right) \times e^{-(\beta/4) (s_2^2 + s_4^2) -\frac{\varepsilon}{2} s_1^2 -\frac{\varepsilon}{2} s_3^2 + i\frac{T}{2} (s_2^2 - s_4^2)} \times \nonumber \\
& \left(\Gamma(2is_4)\Gamma(-2i s_2) \Gamma(\Delta_V \pm is_1 - is_4) \Gamma(\Delta_V \pm is_1 + is_2) \Gamma(\Delta_W \pm is_3 - is_4)\Gamma(\Delta_W \pm is_3 + is_2) n^{-4is_-}\right. \nonumber \\
&+ c.c. \nonumber \\
&\left.+\Gamma(2is_4)\Gamma(2i s_2) \Gamma(\Delta_V \pm is_1 - is_4) \Gamma(\Delta_V \pm is_1- is_3) \Gamma(\Delta_W \pm is_3 - is_4)\Gamma(\Delta_W \pm is_3- is_2) n^{4is_+ } + c.c.\right),
\end{align}
with
\begin{align}
s_{\pm} = \frac{s_2 \pm s_4}{2}.
\end{align}
Here we have ignored the normalization factors in eq. \eqref{eqn:fulldistributionapp} for now. 

By examining the symmetry of this expression, we see that we can include the terms propotional to $n^{\pm 4i s_+}$ by considering just the terms propotional to $n^{\pm 4is_-}$ but letting the $s_2$ and $s_4$ integrals run from $-\infty$ to $+\infty$. Since we are interested in the leading contribution to this distribution in the limit that $\sqrt{\Delta_V \Delta_W}/\varepsilon \to \infty$, it will be helpful to make the re-scaling $s_{1,3} \to \sqrt{\frac{\Delta_{V,W}}{\varepsilon}} \sigma_{1,3}$. Then we have this contribution to the distribution
\begin{align}
&\frac{8 \times 4^{2-\Delta_V-\Delta_W}}{n\Gamma(2\Delta_V) \Gamma(2\Delta_W)} \int \left(\prod_i \rho(s_i)\right) \times e^{-(\beta/4) (s_2^2 + s_4^2) -\frac{\Delta_V}2\sigma_1^2 -\frac{\Delta_W}2 \sigma_3^2 + i\frac{T}2 (s_2^2 - s_4^2)} \times \Gamma(2is_4)\Gamma(-2i s_2)\nonumber \\
&  \Gamma \left(\Delta_V \pm i\sqrt{\frac{\Delta_V}{\varepsilon}} \sigma_1 - is_4 \right) \Gamma \left(\Delta_V \pm i\sqrt{\frac{\Delta_V }{\varepsilon}}\sigma_1 + i s_2 \right) \times \nonumber \\
& \Gamma \left(\Delta_W \pm i\sqrt{\frac{\Delta_W }{\varepsilon}}\sigma_3 - i s_4 \right)\Gamma \left(\Delta_W \pm i\sqrt{\frac{\Delta_W }{\varepsilon}}\sigma_3 + is_2\right) n^{-4is_-} +c.c.\ .
\end{align}

Defining the geometric mean of the dimensions $\Delta \equiv \sqrt{\Delta_V \Delta_W}$, we can expand this expression at large $\frac{\Delta }{\varepsilon}$. Using the fact that
\begin{align}
\Gamma(a \pm is ) = 2\pi |s|^{2a-1} e^{-\pi |s|},
\end{align}
at large $|s|$, we get 
\begin{align}\label{eqn:13}
&\frac{8 \times 4^{2-\Delta_V-\Delta_W} \times \Delta}{64n\Gamma(2\Delta_V) \Gamma(2\Delta_W) \varepsilon} \int d\sigma_1 d\sigma_3 ds_2 ds_4 \times \rho(s_2) \rho(s_4) \sigma_1 \sigma_3\times e^{-(\beta/4) (s_2^2 + s_4^2) -\frac{\Delta_V}2\sigma_1^2 -\frac{\Delta_W}2 \sigma_3^2 + i\frac{T}2 (s_2^2 - s_4^2)} \times \nonumber \\
& \Gamma(2i s_4)\Gamma(-2i s_2) \times \nonumber \\
& \exp\left( -4i s_- \log \frac{n}{ \frac{\Delta}{\varepsilon} \sigma_1 \sigma_3} +(4\Delta_V -2) \log \sqrt{\frac{\Delta_V }{\varepsilon}} \sigma_1 +  (4\Delta_W -2) \log \sqrt{\frac{\Delta_W }{\varepsilon}} \sigma_3 \right)
\end{align}
The first factor of $\Delta/\varepsilon$ out front comes from the measure when we integrate over $s_1$ and $s_3$. The second factor comes from the density of states after re-scaling to $\sigma$ variables. 

Now at large $T$ and $\Delta/\varepsilon$, we expect the integrals over $s_+$ and $s_-$ to be dominated by small $s_- \ll \Delta \ll s_+$ and so $s_- \log \sigma_1 \sigma_3$ is negligible compared to the linear in $\Delta$ terms in the exponent in eq. \eqref{eqn:13}. In the large $\Delta$ limit, one can do the $\sigma_1$ and $\sigma_3$ integrals by saddle point. One can easily check that they are dominated at $\sigma_{1,3} \approx 2$. Furthermore, the $\sigma_1$ and $\sigma_3$ integrals amount to a factor of $\bra{VW}\ket{VW}$. This cancels with the normalization factor in eq. \eqref{eqn:fulldistributionapp} and we find that the full normalized distribution is 
\begin{align}\label{eqn:totalprobdist}
&p(n) \nonumber \\
&= \frac{8}{n Z(\beta)}  \int_{-\infty}^{\infty} ds_2 ds_4 \rho(s_2) \rho(s_4)\exp \left(-\frac{\beta}4 (s_2^2 + s_4^2) + i\frac{T}2 (s_2^2 - s_4^2)  \right)   \times \nonumber \\
&\Gamma(2i s_4)\Gamma(-2is_2) \exp \left(i2 (s_2 - s_4) \log \frac{n}{4 \frac{\Delta}{\varepsilon}}\right)  + \mathcal{O}(\varepsilon/(\Delta)).
\end{align} 
Converting from $n$ to $C$, which at large $C$ is just $C \sim n^2$, we see that eq. \eqref{eqn:totalprobdist} agrees with expression eq. \eqref{eqn:latetimedist} in the main text, up to an order one re-scaling of $C_0$ which we did not fix.

\section{Details on the numerics}\label{app:Numerics}
\subsection{Rescaling from the semi-circle to the Schwarzian spectrum}
Let $X$ be a random variable that is drawn from the semi-circle law. Let $E= f(X)$, where $f$ is a deterministic function. Then 
\begin{align}
    \rho_\text{Sch}(E) d E = p (f(X)) f'(X) d X =  p(X) d X\\
 \frac{\pi}{2}\sqrt{1-X^2}   = \alpha\frac{E'(X)}{(2\pi)^2}  \sinh (2\pi\sqrt{E(X)})
\end{align}
Solving this differential equation gives
\begin{align}
 8 \pi   \alpha \left( X \sqrt{1-X^2} +  \frac{\pi}{2} - \cot^{-1} \frac{1+X}{\sqrt{1-X^2}}\right) = \frac{-1}{2\pi^2} \sinh (2 \pi \sqrt{E}) + \frac{\sqrt{E}}{\pi} \cosh (2 \pi \sqrt{E})
\end{align}
Here $\alpha$ is a parameter that sets the maximum energy cutoff for the spectrum.
We choose $\alpha = \frac{e^{12}}{16 \pi^2} \approx 1030.66$. This gives a JT spectrum with support from $E=0$ to $E=E_\text{max} \approx 3.58741$.

\subsection{Length and velocities}
\begin{figure}[H]
    \centering
    \includegraphics[width=0.9\columnwidth]{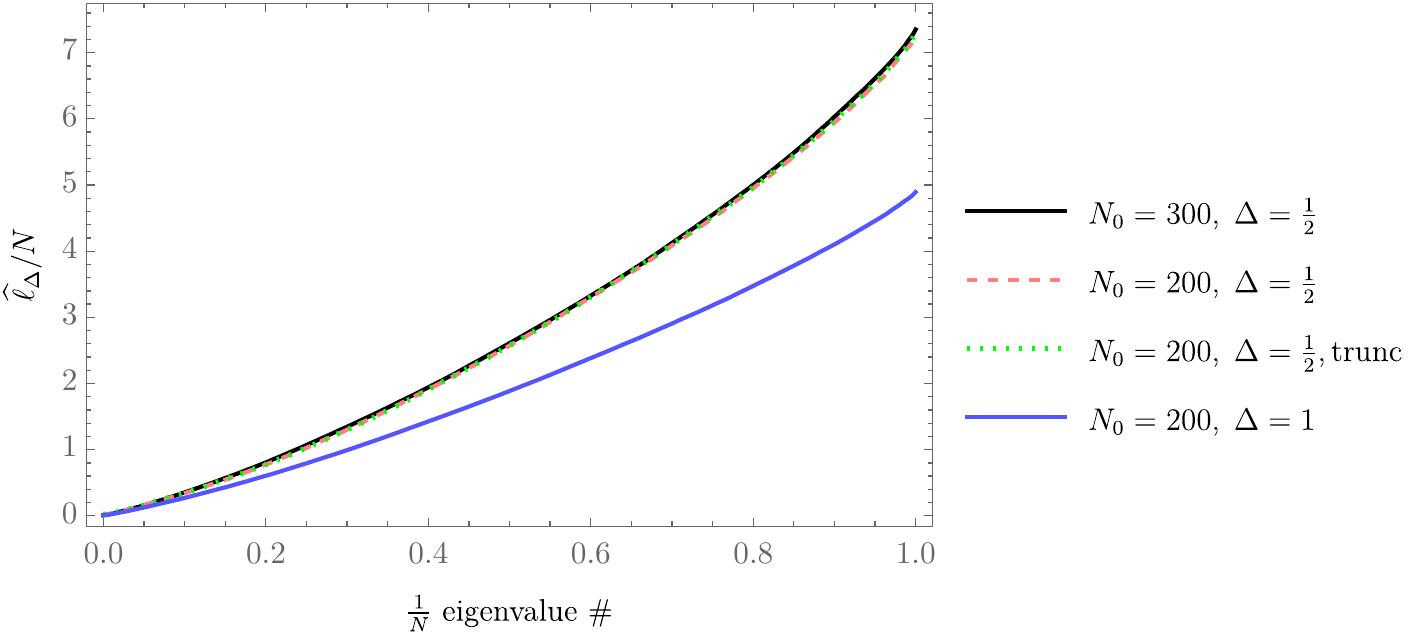}
        \caption{\label{fig:length_spectrum}
        Eigenvalue spectrum of the length operator $\hat{\ell}_\Delta$ for $\Delta = 1/2$ and $\Delta=1$. In contrast to the perturbative case where the length operator has an unbounded spectrum, once non-perturbative corrections are included, the spectrum becomes bounded with a maximum eigenvalue $\sim e^{S_0}$.  The green dotted line corresponds to the $N_0=200$ but truncated to $N=180$ by deleting the part of the spectrum corresponding to the largest 20 energy eigenvalues. This probes the sensitivity of the spectrum to the cutoff.  %
        }%
\end{figure}

\begin{figure}[h!]
    \centering
    \includegraphics[width=0.32\textwidth]{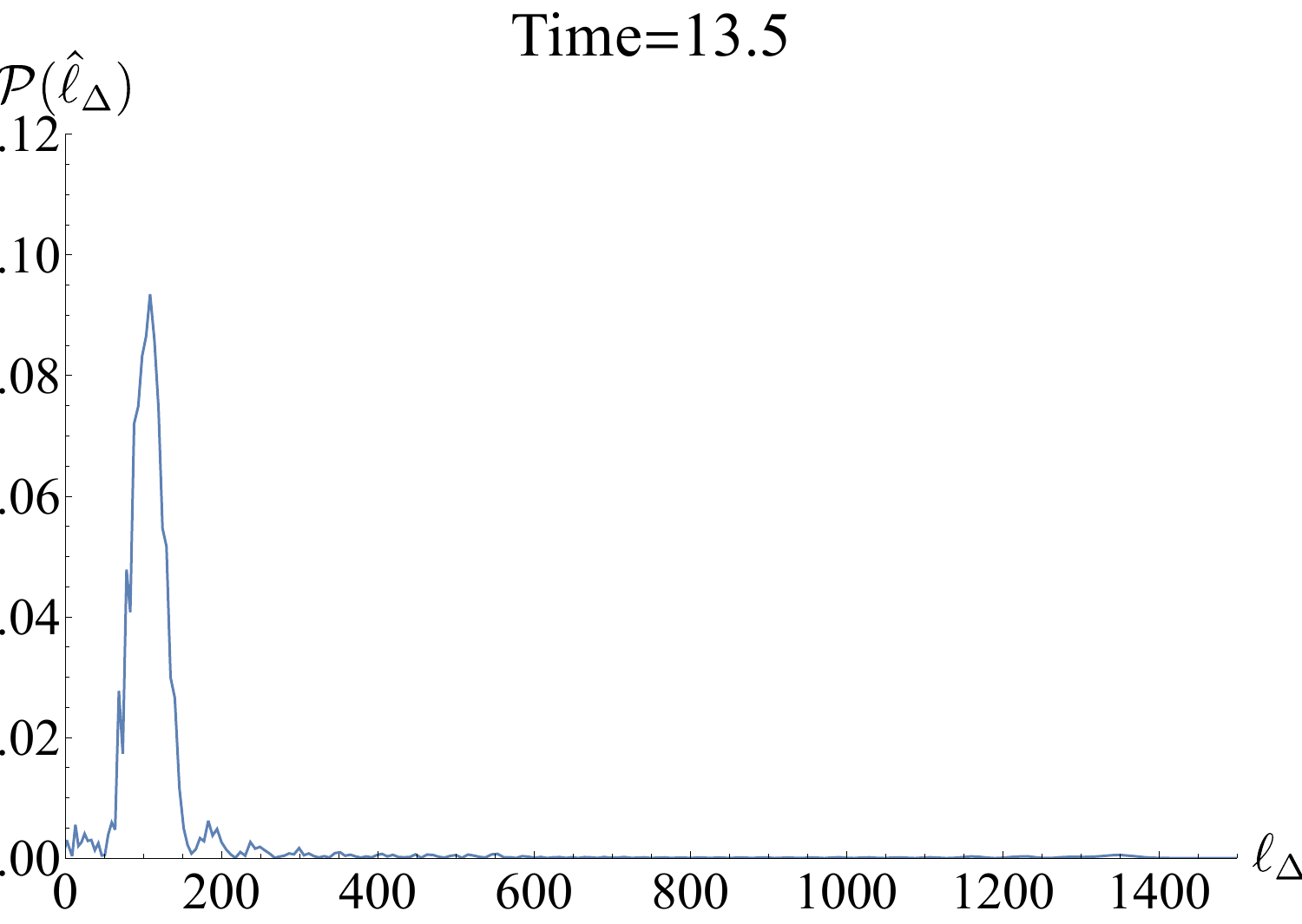}
      \includegraphics[width=0.32\textwidth]{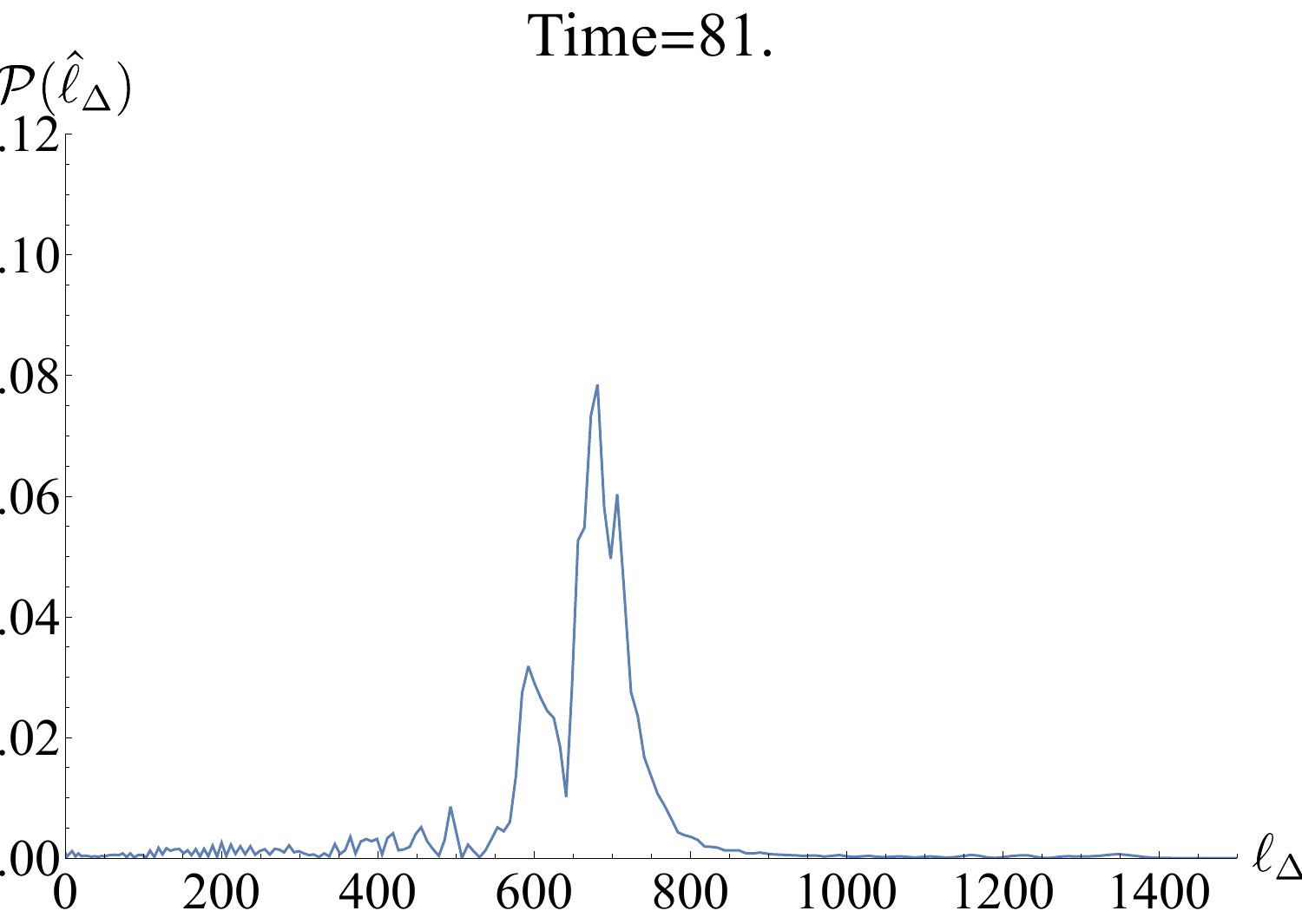}
        \includegraphics[width=0.32\textwidth]{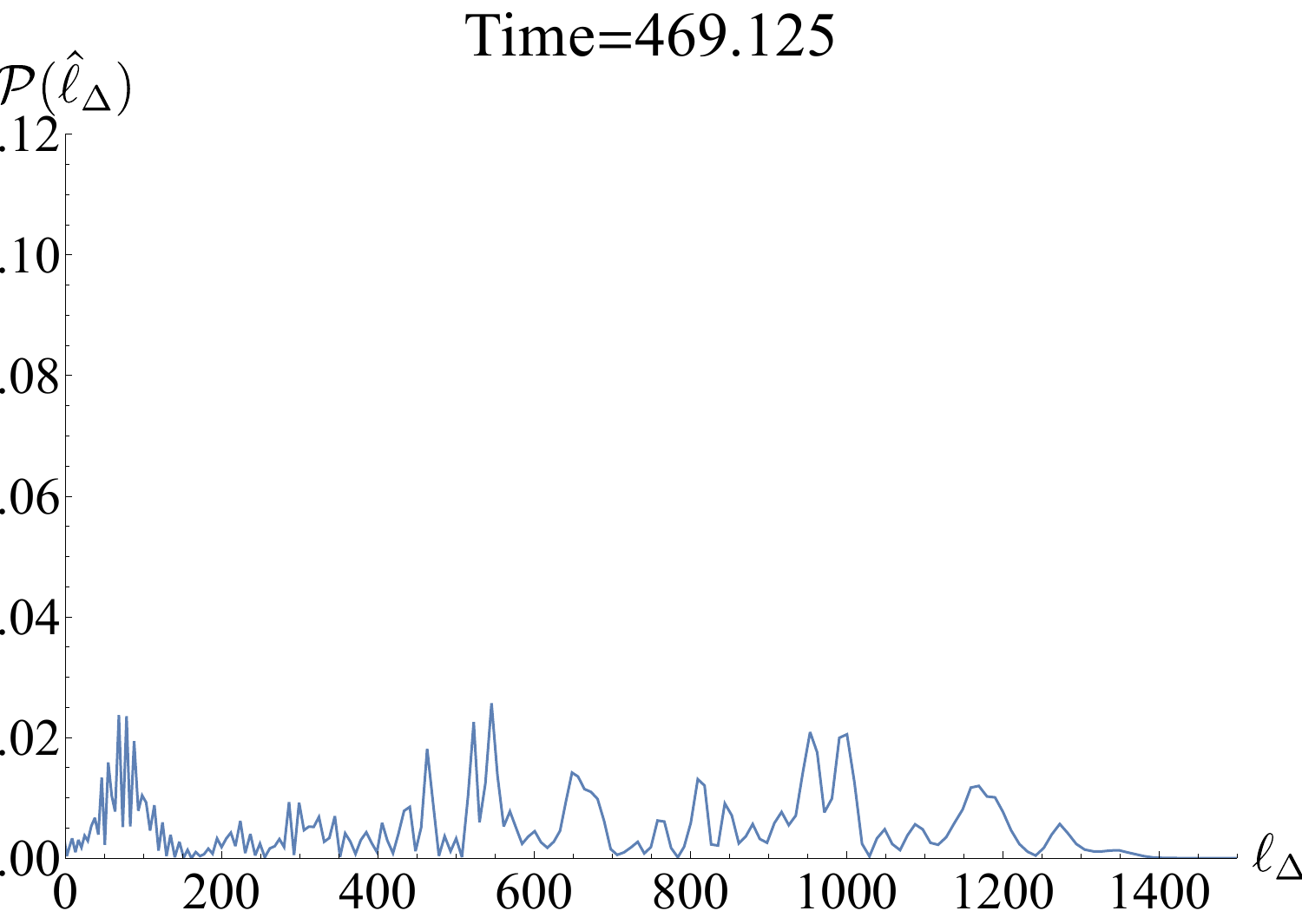}
            \caption{ \label{fig:probDistributionVelocity2}The probability of detecting a length eigenvalue at different times. The \textit{left plot} shows the probability at early times (with $t \ll e^{S_0}$) where the wavefunction is entirely peaked at small values of $\ell_\Delta$. The \textit{middle plot} shows the probability at intermediate times (with $t \lesssim e^{S_0}$) where the wavefunction now starts spreading over a larger number of eigenvalues $\ell_\Delta$.  The \textit{right plot} shows the probability at very late times (with $t > e^{S_0}$) where the wavefunction now has support on all lengths. A video of this probability distribution as a function of time can be found at \cite{video2}.}
\end{figure}

In this section we collect the results of further numerical experiments. In Figure \ref{fig:length_spectrum}, we plot the eigenvalue spectrum of the length operator and show its sensitivity to various choices of parameters. In Figure \ref{fig:probDistributionVelocity2} we plot the probability distribution for the length. At early times the wavefunction is relatively peaked around the semiclassical answer, whereas at late times the wavefunction is erratic.

\begin{figure}[H]
    \centering
    \includegraphics[width=0.9\columnwidth, trim = 0 0cm 0cm 0, clip=true]{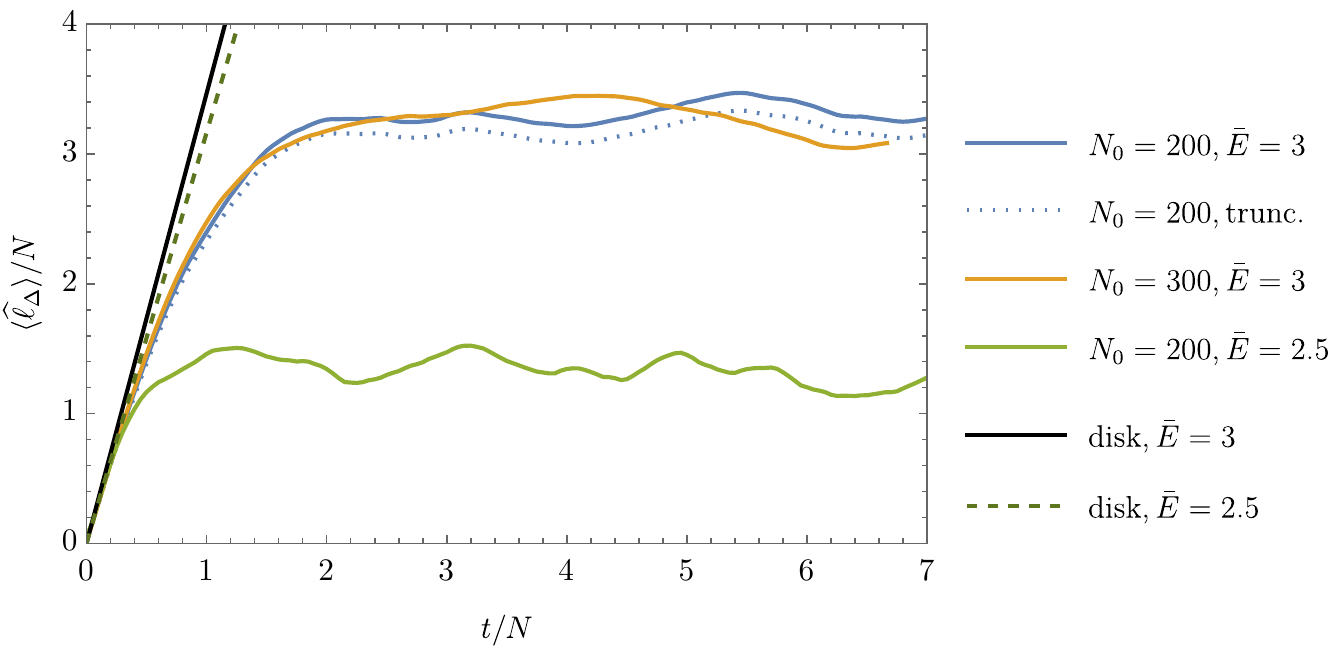}
        \caption{Numerical demonstration of the $\tau$ scaling limit.  We plot the average length as a function of $t/N \sim \tau = e^{-S_0} t$. This figure is a more complete version of Figure~\ref{fig:length_expectation_tau}.   note that the green curve with $\bar{E}=2.5$ plateaus at a smaller value since $e^{S(\bar{E})}$ is smaller than for $\bar{E}=3$. \label{fig:length_expectation_tau_complete}}
\end{figure}

\section{The dimension of the Hilbert space}
\label{sec:Hilbert-space-dimensions}

To check the over-redundancy of states in the $\ell$-basis, we can ask for the actual dimension of the Hilbert space. To be able to study the effective dimension of the Hilbert space, we will have to limit the maximum energy of the states that we study and smear the states $\ket{\ell}$ such that they are normalizable, instead of $\delta$-function normalizable. To achieve the latter step, we can define
\be 
\ket{\Psi_\ell} = \int d\tilde\ell\,  G_\ell(\tilde \ell) \ket{\tilde \ell}\,,
\ee
where $G_\ell(\tilde \ell)$ is some smearing function, peaked around $\tilde \ell= \ell$, defined to change the orthogonality property \eqref{eq:orth-lengths-to-energy} such that 
\be 
\int d\ell\, \left( \int d\tilde \ell G_\ell(\tilde \ell) \psi_{E_1}(\tilde \ell) \right) \left( \int d\tilde \ell' \, G_\ell(\tilde \ell') \psi_{E_2}(\tilde \ell') \right) \approx \frac{e^{-\frac{(E_1 - E_2)^2}{\epsilon^2}}}{\sqrt{\pi} \epsilon}\,,
\ee
such that as $\epsilon \to 0$ one recovers the exact orthogonality property \eqref{eq:orth-lengths-to-energy}. We can now define the density matrix, which is the naive resolution of the identity at the disk level 
\be 
\rho_\ell = \int d\ell \ket{\Psi_\ell} \bra{\Psi_\ell}\,,
\ee
but when including non-perturbative corrections, we can use its rank to determine the dimension of the Hilbert space. Nevertheless, as mentioned above, we should not expect to get a finite dimensional Hilbert space unless we truncate the energy of the states. Thus, instead, we will study the density matrix $\rho_{\ell, \, E_\text{max}}  = P_{E_\text{max}}\rho_{\ell}$, where $\bra{E_1}P_{E_\text{max}} \ket{E_2}= \delta(E_1-E_2) \Theta(E_\text{max} - E_1)$, projects to the subspace of states with energy smaller than $E_\text{max}$.\footnote{Here, we will solely consider $E_\text{max}\gg e^{-S_0}$.}

We will warm up to the computation of the rank of $\rho_\ell$ by computing a related quantity, the second R\'enyi entropy. Consider a mixed state projecting to a $d$ dimensional subspace
\beq
\rho=P/d,\quad \quad P^2=P.
\eeq
We have $\text{Tr}\rho=1$ and
\beq
\text{Tr}\rho^2=1/d.
\eeq
Given a mixed state, a proxy for the dimension of its support is
\beq
d_\text{eff}=[\text{Tr}\rho^2]^{-1}.
\eeq
In fact, this is a lower bound on the actual rank.\footnote{This is easy to show. Write:
\beq
\text{rank}\rho=\text{Tr}\Theta(\rho),
\eeq
where $\Theta$ is the Heaviside function. We have
\beq
1=\text{Tr}\rho=\text{Tr}[\Theta(\rho)\rho] \leq \sqrt{\text{Tr}\Theta(\rho)^2 \text{Tr}\rho^2}.
\eeq
Using $\Theta(\rho)^2=\Theta(\rho)$ we find
\beq
\text{rank}\rho \geq \frac{1}{\text{Tr}\rho^2}=d_\text{eff}.
\eeq}

We shall start by studying the effective dimension of this  $\rho_{\ell, \, E_\text{max}}$,
\be 
d_\text{eff}^{-1} = \frac{ \Tr \rho_{\ell, \, E_\text{max}}^2}{(\Tr \rho_{\ell, \, E_\text{max}})^2}\,.
\ee
Using the definition of the smearing function, as well as that of the projection operator, one finds 
\be 
 \overline{\Tr \rho_{\ell, \, E_\text{max}}^2} = \int_0^\infty d\ell d\ell' \, \overline{|\bra{\Psi_\ell} P_{E_\text{max}} \ket{\Psi_{{\ell}'}}|^2} = \int_0^{E_\text{max}} dE_1 dE_2 \,\overline{\rho(E_1) \rho(E_2) }\,\frac{e^{-2\frac{\left(E_1-E_2\right)^2}{\epsilon^2}}}{\pi \epsilon^2}\,.
\ee
On the other hand,
\be \overline{(\Tr \rho_{\ell, \, E_\text{max}})^2} = \int_0^\infty d\ell d\ell'\, \overline{\bra{\Psi_\ell} P_{E_\text{max}} \ket{\Psi_{{\ell}}}\bra{\Psi_{\ell'}} P_{E_\text{max}} \ket{\Psi_{{\ell'
}}}}=\int dE_1 dE_2\, \frac{\overline{\rho(E_1) \rho(E_2) } }{\pi \epsilon^2}\,.
\ee
Thus, we find that the effective dimension is given by the very simple formula
\be 
\label{eq:effective-dimension-explicit}
d_\text{eff}^{-1} = \frac{\int_0^{E_\text{max}} dE_1 dE_2\,\overline{\rho(E_1) \rho(E_2) }\, e^{-2\frac{\left(E_1-E_2\right)^2}{\epsilon^2}}  }{\int_0^{E_\text{max}} dE_1 dE_2\,\overline{\rho(E_1) \rho(E_2) }}
\ee
We want to study this wavefunction in the limit $\epsilon \to 0$ such that the wavefunctions $\ket{\Psi_\ell}$ are as close as possible to the original $\ket{\ell}$'s. In that limit, the integral in the numerator in \eqref{eq:effective-dimension-explicit} is dominated by the  universal form of the spectral correlator $\overline{\rho(E_1) \rho(E_2) }$ in the regime $E_1\to E_2$. This is given by, 
\be 
\label{eq:spectral-correlator-small-energy-differences}
\overline{\rho(E_1) \rho(E_2) }  &= e^{2S_0}\rho_\text{disk}(E_1)\rho_\text{disk}(E_2) + \delta(\Delta E) \rho_\text{disk}(\bar E) - \frac{1}{\pi^2(\Delta E)^2} \sin^2 \left[e^{S_0}\pi \rho_\text{disk}\left(\bar E\right) \Delta E\right]\,, \nn \\ \Delta E&\equiv {E_2-E_1}\,,\qquad \bar E = \frac{E_2-E_1}2\,.
\ee

In particular, consider the limit in which $\epsilon \ll e^{-S_0}$. In this limit the doubly non-perturbative piece cancels with the leading disconnected piece and only the contact term survives,
\be 
\label{eq:spectral-correlator-very-small-energy-differences}
\overline{\rho(E_1) \rho(E_2) } &\approx  \delta(\Delta E) e^{S_0}\rho_\text{disk}(\bar E)\,, \quad \text{for}\quad \Delta E \sim O(\epsilon)\,.
\ee
It thus follows that the numerator of \eqref{eq:effective-dimension-explicit} is given by 
\be
\int_0^{E_\text{max}} dE_1 dE_2\,\overline{\rho(E_1) \rho(E_2) }\, e^{-2\frac{\left(E_1-E_2\right)^2}{\epsilon^2}} = \int_0^{E_\text{max}} dE e^{S_0} \rho_\text{disk}(E) \equiv d(E_\text{max})\,.
\ee
where $d(E_\text{max})$ simply counts the number of eigenenergies in the interval $[0, E_\text{max}]$.  
The denominator is not strongly affected by any non-perturbative corrections and therefore can simply be written as, 
\be 
\int_0^{E_\text{max}} dE_1 dE_2\,\overline{\rho(E_1) \rho(E_2) } = \int_0^{E_\text{max}} dE_1 dE_2\,e^{2S_0}{\rho_\text{disk}(E_1) \rho_\text{disk}(E_2) } = d(E_\text{max})^2
\ee
From this, we thus find that
\be 
d_\text{eff} = d(E_\text{max})\,,
\ee
which is precisely the expected result. We can similarly compute higher moments of $\rho_{\ell, \, E_\text{max}}$. Following the same procedure, these are given by
\be 
\label{eq:moments-of-tr-rho-n}
\frac{\Tr \rho_{\ell, \, E_\text{max}}^n}{(\Tr \rho_{\ell, \, E_\text{max}})^n} = \frac{\int_0^{E_\text{max}} dE_1 \dots dE_n\,\overline{\rho(E_1) \dots \rho(E_n) }\, e^{-\frac{1}{\epsilon^2}\left[(E_1-E_2)^2 + (E_2-E_3)^2+ \dots + (E_n - E_1)^2\right] }  }{\int_0^{E_\text{max}}dE_1 \dots dE_n\,\overline{\rho(E_1) \dots \rho(E_n) }}
\ee
Once again, in the limit when $\epsilon \ll e^{-S_0}$, the spectral correlator is simply dominated by a single contact term
\be 
\overline{\rho(E_1) \dots \rho(E_n) } \approx e^{S_0} \rho_\text{disk}(E_1) \delta(E_1 - E_2)\dots \delta(E_n - E_1)\,.
\ee
Similarly, the denominator of \eqref{eq:moments-of-tr-rho-n} is dominated by the disconnected contributions. From this, we find that 
\be 
\frac{\Tr \rho_{\ell, \, E_\text{max}}^n}{(\Tr \rho_{\ell, \, E_\text{max}})^n}= \frac{1}{d(E_\text{max})^{n-1}}
\ee
This implies that $ \rho_{\ell, \, E_\text{max}}$ has an approximately flat ``entanglement spectrum'' (as $\epsilon \to 0$) and from a replica trick computation, that the rank of the density matrix is given by 
\be 
\text{Rank} \, \rho_{\ell, \, E_\text{max}} = \lim_{n \to 0}\frac{\Tr \rho_{\ell, \, E_\text{max}}^n}{(\Tr \rho_{\ell, \, E_\text{max}})^n} = d(E_\text{max})\,,
\ee
which is, once again, the expected result.

Note that this computation sheds light on the following question: naively the ``rectangle propagator,'' e.g.., the propagator $\bra{\ell'} e^{-\tau H_\text{Liouville} } \ket{\ell}_0$ can be used to construct a trace $\int d \ell \bra{\ell} e^{-\tau H_\text{Liouville} } \ket{\ell}_0 = \infty$. Here we see that this is not $\Tr \mathbf{1}$ because $\int d \ell \ketbra{\ell}$ is not a correct resolution of the identity non-perturbatively.%

\bibliographystyle{jhep}
\bibliography{biblio.bib}
\end{document}